\documentclass[11pt,thmsa, a4paper]{article}

\usepackage{tikz-qtree}
\usepackage{tikz}
\usepackage{amssymb}
\usepackage{amsfonts}
\usepackage{amsmath}
\usepackage{amsthm}
\usepackage{color}
\usepackage{graphicx}
\usepackage{geometry}
\usepackage{probsoln}%
\providecommand{\U}[1]{\protect\rule{.1in}{.1in}}
\renewcommand{\thesection}{\Roman{section}} 
\renewcommand{\thesubsection}{\Alph{subsection}}

\newtheorem{theorem}{Theorem}

\newtheorem{proposition}{Proposition}
\newtheorem{lemma}{Lemma}
\newtheorem{definition}{Definition}

\theoremstyle{definition}

\definecolor{indigo}{rgb}{0.0, 0.25, 0.42}
\definecolor{lillac}{rgb}{0.0, 0.15, 0.30}
\definecolor{amethyst}{rgb}{0.0, 0.18, 0.39}
\definecolor{amethyst}{rgb}{0.0, 0.0, 0.0}
\definecolor{byzred}{rgb}{0.66, 0.13, 0.24}

\newenvironment{super-boxer}
{\begin{center}
\begin{tabular}{||p{0.9\columnwidth}||}
\hline \hline \\
}
{
\\ \\ \hline \hline
\end{tabular}
\end{center}
}

\newenvironment{Hidden}{\onlysolution \color{indigo}
}
{
	\endonlysolution}
\showanswers
\hideanswers
\begin{document}

\title{Risk Aversion and Insurance Propensity\thanks{\lowercase{\uppercase{M}accheroni (\texttt{fabio.maccheroni@unibocconi.it}) and \uppercase{M}arinacci (\texttt{massimo.marinacci@unibocconi.it}) are
		with the \uppercase{D}epartment of \uppercase{D}ecision\ \uppercase{S}ciences,  \uppercase{U}niversit\`{a} \uppercase{B}occoni;
		\uppercase{W}ang (\texttt{ruodu.wang@uwaterloo.ca}) and \uppercase{W}u (\texttt{q35wu@uwaterloo.ca}) are with the \uppercase{D}epartment of \uppercase{S}tatistics and \uppercase{A}ctuarial \uppercase{S}cience,
		\uppercase{U}niversity of \uppercase{W}aterloo.
		\uppercase{A}ll authors contributed equally to this work.
		\uppercase{F}or helpful comments that substantially improved the paper, we thank the editors, three anonymous referees, numerous seminar and conference audiences, as well as \uppercase{A}lain \uppercase{C}hateauneuf, \uppercase{P}aul \uppercase{E}mbrechts, \uppercase{M}artin \uppercase{L}arsson, \uppercase{M}arcel \uppercase{N}utz, \uppercase{L}uciano \uppercase{P}omatto, \uppercase{M}arco \uppercase{S}carsini, 
		\uppercase{P}eter \uppercase{W}akker, \uppercase{J}ingni \uppercase{Y}ang, and \uppercase{B}in \uppercase{W}ang.
		\uppercase{T}his work was supported by the \uppercase{I}talian \uppercase{M}inistry of
		\uppercase{U}niversity and \uppercase{R}esearch [\uppercase{G}rant \uppercase{P}\uppercase{R}\uppercase{I}\uppercase{N}-2017\uppercase{C}\uppercase{Y}2\uppercase{N}\uppercase{C}\uppercase{A}] and the \uppercase{N}atural \uppercase{S}ciences and
		\uppercase{E}ngineering \uppercase{R}esearch \uppercase{C}ouncil of \uppercase{C}anada [\uppercase{G}rants \uppercase{R}\uppercase{G}\uppercase{P}\uppercase{I}\uppercase{N}-2018-03823 and
		\uppercase{R}\uppercase{G}\uppercase{P}\uppercase{A}\uppercase{S}-2018-522590].}
}}
\author{Fabio Maccheroni\quad Massimo Marinacci\quad Ruodu Wang\quad Qinyu Wu}
\date{\today}
\maketitle

\begin{abstract}
We provide a new foundation of risk aversion by showing that this attitude is fully captured by the propensity
to seize insurance opportunities. Our
foundation, which applies to all probabilistically sophisticated preferences,
well accords with the commonly held prudential interpretation of risk
aversion that dates back to the seminal works of Arrow (1963) and Pratt
(1964).
In our main results, we first characterize the Arrow-Pratt risk aversion in
terms of propensity to \textit{full} insurance and the stronger notion of
risk aversion of Rothschild and Stiglitz (1970) in terms of propensity to 
\textit{partial} insurance. We then extend the analysis to comparative risk
aversion by showing that the notion of Yaari (1969) corresponds to
comparative propensity to full insurance, while the stronger notion of Ross
(1981) corresponds to comparative propensity to partial insurance.

\end{abstract}

%\newpage

%\section{Introduction}

The seminal works of Arrow (1963), Pratt (1964), and Rothschild and Stiglitz
	(1970) provide different behavioral notions of risk aversion that under
	expected utility amount to the familiar concavity property of utility
	functions. All these papers relate their notions of risk aversion with
	insurance choices, arguably the most basic domain of application of any risk
	analysis. In this paper, we show how insurance choice behavior can be used to
\textit{define} risk attitudes. At a theoretical level, our results provide
compelling economic explanations of risk attitudes, thus clarifying their
economic appeal and normative status. At an empirical level, they show how
these attitudes may motivate the most common features of marketed insurance policies.

As both Arrow (1963) and Pratt (1964) observe, risk aversion can be characterized as
preference for full insurance over no insurance at an actuarially fair
premium. To formalize this claim, {consider an agent who, say because of real
	or financial assets held, faces a random wealth change $w$ that we call
	\emph{risk}.\footnote{{For instance, an agent who owns a car with market value
			$\nu$ faces a wealth change $w$ equal to $-\nu$ if the car is stolen, and to $0$
			otherwise. As it is typically the case, here the risk $w$ is a negative random
			variable, with $-w$ representing the agent loss. The `risk' terminology that we adopt comes from Pratt (1964) and is also used by Gollier (2001).}}}
{A full insurance for risk $w$ at premium $\pi$ is a contract with random
	payoff $-w-\pi$ that eliminates all uncertainty by replacing the random loss
	$-w$ with a fixed cost $\pi$. The actuarially fair premium is, by definition, the
	expected loss $\mathbb{E}\left[  -w\right]  $. Thus, the agent prefers to sign
	up an actuarially fair full insurance rather than facing the risk if%
	\begin{equation}
		w+\underset{\text{full insurance at fair premium }}{\underbrace{\left(
				-w-\mathbb{E}\left[  -w\right]  \right)  }}\ \succsim\ w+\underset{\text{no
				insurance }}{\underbrace{0}}\label{eq:pra}%
	\end{equation}
	that is, if $\mathbb{E}\left[  w\right]  \succsim w$ for all risks $w$. In
	words, it amounts to a preference for a sure amount over a random one with the
	same expectation. This is the classical Arrow-Pratt notion of risk aversion.}

While the concept of full insurance
	is
	natural and is, since a long time, common in the insurance practice, in reality no insurance comes at
	fair premium.  For instance, insurance companies have operating
	expenses that affect premia.\footnote{For a textbook treatment see, e.g., Dickson (2017), who
		writes `It is unlikely that an insurer who calculates premiums by this [fair
		premium] principle will remain in business very long'.} But then the connection of expression (\ref{eq:pra}) to actual insurance choices becomes a weak one -- we may want to define individuals who pick insurances at higher premia as risk
	averse, but little can be said of individuals who refuse them.

Our first contribution is to show that there is, indeed, a strong connection. We
provide an equivalent definition of risk aversion that is purely based on
insurance concepts and does not rely on expectations and fair premia. Consider an agent who,
at the same price $\pi$, can either buy full insurance $-w$ or make another
investment $h$ that has the same distribution of $-w$. Agent's payoff is
$w+\left(  -w-\pi\right)  $\ when full insurance is purchased, while it is
$w+\left(  h-\pi\right)  $ when {investment $h$ is purchased.} Arrow-Pratt's
risk aversion implies that
\begin{equation}
	w+\underset{\text{full insurance at price }\pi\text{ }}{\underbrace{\left(
			-w-\pi\right)  }}\succsim w+\underset{h\text{ distributed as }-w\text{ at
			price }\pi\text{ }}{\underbrace{\left(  h-\pi\right)  }}\label{eq:wra}%
\end{equation}
for all $w$ and all $\pi$. Indeed, the sure payoff $-\pi$ on the left-hand
side is the expectation of the random payoff $w+\left(  h-\pi\right)  $ on the
right-hand side (because $h$ is distributed as $-w$). Preference pattern
(\ref{eq:wra}) has a
clear meaning of \emph{propensity to full insurance}, does not make use of
expected values, and applies to actual insurance choices dealing with insurance premia that are not actuarially fair. In terms of dependence, a full insurance is perfectly
	negatively correlated with risk $w$ and therefore guarantees a constant
	payoff. In contrast, the equally distributed $h$ might well have a different
	correlation structure with risk $w$, as a simple example in Section
	\ref{sec:analysis}.\ref{sec:a-analysis} illustrates. Arrow-Pratt's risk aversion thus manifests
	itself into different attitudes towards identically distributed modifications
	of $-w$ based on their correlation with risk $w$, perfect negative correlation
	being favored because it eliminates wealth variability.

Our first main result, Theorem \ref{th:wra}, shows that this propensity to
full insurance is equivalent to Arrow-Pratt's risk aversion for every
transitive preference $\succsim$ over random payoffs that depends only on
payoffs' distributions. It thus applies to expected utility, where it provides a novel underpinning for concave utility, but goes well beyond it. In particular, it applies to all probabilistically
sophisticated preferences in the sense of Machina and Schmeidler
(1992)   such as cumulative prospect theory  with probability weighting (Tversky and Kahneman, 1992).\footnote{\label{PS} This class also includes the
	preferences introduced by Machina (1982), rank-dependent utility (Quiggin,
	1982, Yaari, 1987), betweenness preferences (Dekel, 1986, Chew, 1989, Gul,
	1991), multiplier preferences (Hansen and Sargent, 2008), quantile
	preferences (Rostek, 2010),
	and cautious expected utility (Cerreia-Vioglio, Dillenberger, and Ortoleva,
	2015).} 
    It also applies to preferences that do not satisfy stochastic
dominance, like the original prospect theory of Kahneman and Tversky (1979),
and that might not even be complete, like the mean-variance preferences of
Markowitz (1952).

One may then argue that many insurance contracts do not provide full coverage.
Some of them, like most health insurance policies, have a proportional form as
they reimburse only a fraction of the loss. Others, like many property
insurance policies, have a deductible-limit form as they impose a deductible
and a policy limit. The resulting notions of propensity to partial insurance
are natural extensions of (\ref{eq:wra}). For instance, \emph{propensity to
	proportional insurance} requires%
\begin{equation}
	w+\underset{\text{proportional insurance at price }\pi\text{ }}{\underbrace
		{\left(  -\alpha w-\pi\right)  }}\succsim w+\underset{h\text{ distributed as
		}-\alpha w\text{ at price }\pi\text{ }}{\underbrace{\left(  h-\pi\right)  }%
	}\label{eq:sra}%
\end{equation}
for all $w$, all $\pi$, and all
$\alpha\in\left(  0,1\right]  $.\footnote{The \emph{percentage excess}
	$1-\alpha$ is the fraction of the loss not covered by the insurance policy.
	When $\alpha=1$ proportional coverage corresponds to full coverage.
	As a consequence, propensity to proportional insurance is a stronger
	requirement than propensity to full insurance.} The definition of
\emph{propensity to deductible-limit insurance} is analogous.

Our second main result, Theorem \ref{th:sra}, shows that propensity to
proportional insurance and propensity to deductible-limit insurance are both
equivalent to the Rothschild-Stiglitz notion of risk aversion%
\begin{equation}
	\mathbb{E}\left[  \varphi\left(  f\right)  \right]  \geq\mathbb{E}\left[
	\varphi\left(  g\right)  \right]  \text{ for all concave }\varphi
	:\mathbb{R}\rightarrow\mathbb{R} \text{ implies } f\succsim g \label{eq:rsa}%
\end{equation}
Like the equivalence of Arrow-Pratt's risk aversion (\ref{eq:pra}) and
propensity to full insurance (\ref{eq:wra}), also the equivalence between
Rothschild-Stiglitz's risk aversion (\ref{eq:rsa}) and propensity to
proportional insurance (\ref{eq:sra}) is based on basic insurance concepts
that do not rely on expectations. At a theoretical level, our findings provide
	new definitions of risk aversion that are economically founded and may clarify its normatively appeal. 
 At an empirical level, they show how Rothschild-Stiglitz's risk
aversion may underlie two important market phenomena: (i) the prevalence of
proportional and deductible-limit policies in insurance markets, (ii) the fact
that insurance policyholders typically have both kinds of contracts in their portfolios.

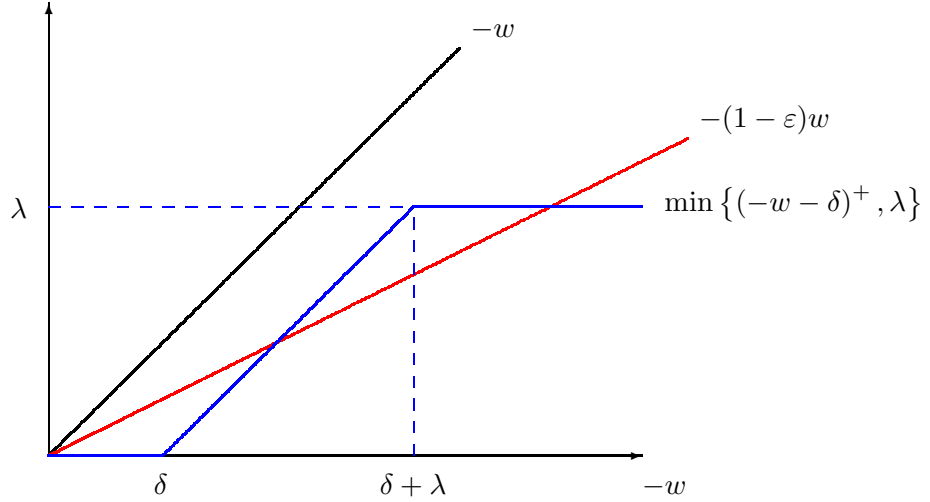
\begin{figure}[ptbh]
{ \setlength{\unitlength}{3cm}
\begin{center}
\begin{picture}(2.6,2.4)(0,0)
				
				%axes
				
				\put(0,0){\vector(0,1){2}}				
				\put(0,0){\vector(1,0){2.6}}
				
				%letters
				
				\put(0.46,-0.17){$\delta$}
				\put(1.45,-0.17){$\delta+\lambda$}
				\put(2.6,-0.17){$-w$}
				\put(-0.17,1.05){$\lambda$}
				\put(1.85,1.85){$-w$}
				\put(2.85,1.45){$-(1-\varepsilon) w$}
				\put(2.70,1.08){$\min\left\{\left(  -w-\delta\right)  ^{+} , \lambda \right\}$}
				
				%loss
				
				\linethickness{0.25mm}
				\qbezier(0,0)(1,1)(1.8,1.8)
				
				%proportional
				
				\color{red}
				\linethickness{0.25mm}
				\qbezier(0,0)(1,0.5)(2.8,1.4)
				
				%deductible-limit
				
				\color{blue}
				\linethickness{0.25mm}
				\qbezier(0.5,0)(1,0.5)(1.6,1.1)
				\qbezier(1.6,1.1)(2,1.1)(2.6,1.1)
				
				\qbezier(0,0)(.5,0)(.5,0)
				
				\linethickness{0.1mm}
				\multiput(0,1.1)(0.1,0.0){17}{\line(1,0){0.05}}
				\multiput(1.6,0)(0.0,0.1){11}{\line(0,1){0.05}}
				
			\end{picture}
\end{center}}
\caption{Proportional insurance (in red) and deductible-limit insurance (in
blue) for loss $-w$}%
\label{figa}
\end{figure}

In our analysis we also consider more general, yet standard, definitions of
partial insurance that only require coverage to increase with loss.\footnote{ They
	include proportional and deductible-limit insurances as special
	cases since their payoffs are, indeed, (weakly) increasing functions of the loss (see
	Figure \ref{figa}).} We show that the resulting notions of propensity to
partial insurance correspond again to Rothschild-Stiglitz risk aversion, thus
providing further support to this popular concept.

We then extend the analysis to comparative risk attitudes. We show that
comparative risk aversion in the sense of Yaari (1969) corresponds to
comparative propensity to full insurance, while the stronger notion of
comparative risk aversion due to Ross (1981) corresponds to comparative
propensity to partial insurance, in its various forms (proportional,
deductible-limit, and so on). These comparative results complete our analysis,
which thus provides a unified economic perspective on weak and strong notions
of absolute and comparative risk aversion in terms of insurance choices, as
Figure \ref{fig:GC2} shows in the conclusion.

Finally, we relate our results to the ones on correlation aversion of Epstein
and Tanny (1980), and to the ones on expected-value preferences of the
classical de Finetti (1931) and of the recent Pomatto, Strack, and Tamuz (2020).

\section{Preliminaries} \label{sect:pre}

\subsection{Risk setting}

We study an agent who has to choose, at time $0$, among actions that yield,
at time $1$, monetary payoffs that depend on uncertain contingencies outside
the agent control. Uncertainty resolves at time 1 and is represented by a
probability space $\left( S,\Sigma ,P\right) $, where $S$ is a space of
payoff-relevant states (the contingencies), $\Sigma $ is a $\sigma $-algebra
of events in $S$, and $P$ is the probability measure on $\Sigma $ that
governs states' realizations.

Each action corresponds to a
random variable 
$
f:S\rightarrow\mathbb{R}%
$
with $f\left(  s\right)  $ interpreted as the, positive or negative, monetary
payoff obtained in state $s$ when the action is taken.

The probability measure $P$ is given %, known to the agent,@@
and, in the tradition of Savage (1954), it is assumed throughout to be \emph{%
	adequate}, that is, either nonatomic on $\Sigma $ or uniform on a finite
partition that generates $\Sigma $. Nonatomicity is a standard divisibility
assumption requiring that, for each event A with $P\left( A\right) >0$,
there exists an event $B\subseteq A$ such that $0<P\left( B\right) <P\left(
A\right) $; it amounts to the existence of a random variable with continuous
distribution (e.g., a normal distribution).

We restrict our attention to random variables that admit all moments. We
call them \emph{random payoffs} and denote their collection by $\mathcal{F}$%
, with typical elements $f$, $g$, and~$h$. Formally, we denote by $\mathcal{L%
}^{0}$ the space of all random variables and by $\mathcal{L}^{\infty }$
the subspace of $\mathcal{L}^{0}$ that consists of all (almost surely)
bounded random variables. Moreover, for each $p\in \left[ 1,\infty \right) $
we denote by $\mathcal{L}^{p}$ the subspace of all elements $f$ of $\mathcal{L}%
^{0}$ with finite absolute $p$-th moment $\mathbb{E}[|f|^{p}]$. With this,
we consider 
\begin{equation*}
	\text{either }\mathcal{F}=\mathcal{L}^{\infty }\text{ or }\mathcal{F}=%
	\mathcal{M}^{\infty }
\end{equation*}%
where $\mathcal{M}^{\infty }=\bigcap_{p\in \mathbb{N}}\mathcal{L}^{p}$ is
the space of all random variables with finite moments of all orders (as
usual, $\mathbb{N}=\{0,1,...\}$ is the set of all natural numbers). The
space $\mathcal{M}^{\infty }$ contains $\mathcal{L}^{\infty }$, the usual
setting of decision theory under risk, yet it allows for random variables
that are commonly used in applications -- like normals, log-normals, and
gammas -- with distributions that admit all moments, but may have unbounded
support. All the results in the main text, with the exception of Proposition %
\ref{th:PST}, hold for both spaces. Also, in the main text, we
consider convergence of random payoffs in $\mathcal{F}$ with respect to all
integer $p$-norms, that is, $f_{n}\rightarrow f$ whenever $\mathbb{E}%
[|f_{n}-f|^{p}]\rightarrow 0$ for all $p\in \mathbb{N}$. In Appendix \ref{sect:prea} we
detail how this mode of convergence can be weakened.

Each random payoff $f$ induces a distribution $P_{f}=P\circ f^{-1}$ of
deterministic payoffs, called `lottery' in the decision theory jargon. In
particular, $P_{f}\left(  B\right)  $ is the probability that $f$ yields an
outcome in the Borel subset $B$ of the real line.

\begin{definition}
Two random payoffs $f$ and $g$ are \emph{equally distributed}, written
$f\overset{d}{=}g$, when $P_{f}=P_{g}$.
\end{definition}

Equally distributed random payoffs generate the same lottery, but   they may have different realizations in the same state, as the example in Section \ref{sec:analysis}.\ref{sec:a-analysis} illustrates.

\subsection{Risk preferences}

The agent preferences are represented by a binary relation $\succsim$ on the
space $\mathcal{F}$ of random payoffs. We read $f\succsim g$ as `the
agent prefers $f$ to $g$'. As usual, $\sim$ and $\succ$ denote the indifference and strict preference relations.

\begin{definition}
A binary relation $\succsim$ on $\mathcal{F}$ is a \emph{risk preference} when
it is both transitive and law invariant, that is, 
\[
f\overset{d}{=}g\Longrightarrow f\sim g
\]
for all random payoffs $f$ and $g$. 

\end{definition}

Besides the standard assumption of transitivity, the
definition of risk preference assumes law invariance, which requires the agent to be indifferent between
equally distributed random payoffs. The fact that only the lottery\ $P_{f}$
induced by $f$ matters to the agent is what characterizes choice under risk,
hence the name risk preferences. Law
invariance guarantees reflexivity, which is thus automatically satisfied by a risk preference.

As previously mentioned, risk preferences include all probabilistically sophisticated preferences, like the
	preferences introduced by Machina (1982), rank-dependent utility (Quiggin,
	1982, Yaari, 1987), betweenness preferences (Dekel, 1986, Chew, 1989, Gul,
	1991), cumulative prospect theory (Tversky and Kahneman, 1992), multiplier
	preferences (Hansen and Sargent, 2008), quantile preferences (Rostek, 2010), and cautious expected utility
	(Cerreia-Vioglio, Dillenberger, and Ortoleva, 2015). Risk preferences also include some classes of incomplete preferences, like the expected multi-utility of Dubra, Maccheroni, and Ok (2004), classes of preferences that do not satisfy stochastic dominance, like the original prospect
theory of Kahneman and Tversky (1979), and classes with both of these features, like the mean-variance preferences of Markowitz (1952) defined by $f\succsim_{\mathrm{MV}}g$ if and only
if $\mathbb{E}\left[  f\right]  \geq\mathbb{E}\left[  g\right]  $ and
$\mathbb{V}\left[  f\right]  \leq\mathbb{V}\left[  g\right]$.

\begin{definition}
\label{def:cont} A risk preference is \emph{continuous} when%
\[
f_{n}\succsim g_{n}\text{ for all }n\text{ }\implies\text{ }\lim_{n}%
f_{n}\succsim\lim_{n}g_{n}%
\]
for all convergent sequences $\left\{  f_{n}\right\}  $ and $\left\{
g_{n}\right\}  $ of random payoffs.
\end{definition}

This assumption is weaker than  continuity in 
distribution because our notion of convergence implies convergence in distribution.

\subsection{Classical risk attitudes}

As mentioned in the introduction, there are two classical approaches to risk
attitudes. One approach is due to Arrow (1963) and Pratt (1964). It is based
on the observation that a random payoff $f$ is `risky' when it is not
constant, that is, when $f\neq\mathbb{E}\left[  f\right]  $. This leads to the
definition of \emph{weak} risk attitudes.

\begin{definition}
A risk preference $\succsim$ is:

\begin{enumerate}
\item[(i)] \emph{weakly risk averse} when, for all random payoffs $f$,
\[
\mathbb{E}\left[  f\right]  \succsim f
\]

\item[(ii)] \emph{weakly risk propense} when, for all random payoffs $f$,%
\[
\mathbb{E}\left[  f\right]  \precsim f
\]

\item[(iii)] \emph{weakly risk neutral} when it is both weakly risk averse and propense.
\end{enumerate}
\end{definition}

The other approach is due to Rothschild and Stiglitz (1970). They show that
the relation $\geq_{\mathrm{cv}}$ on $\mathcal{F}$ defined by%
\[
f\geq_{\mathrm{cv}}g\iff\mathbb{E}\left[  \varphi\left(  f\right)  \right]
\geq\mathbb{E}\left[  \varphi\left(  g\right)  \right]  \text{ for all concave
}\varphi:\mathbb{R}\rightarrow\mathbb{R}%
\]
meaningfully captures the idea that `$f$ is less risky than $g$' (for example,
in terms of mean preserving spreads). This leads to the definition of
\emph{strong} risk attitudes.

\begin{definition}
A risk preference $\succsim$ is:

\begin{enumerate}
\item[(i)] \emph{strongly risk averse} when, for all random payoffs $f$ and
$g$,
\[
f\geq_{\mathrm{cv}}g\Longrightarrow f\succsim g
\]

\item[(ii)] \emph{strongly risk propense} when, for all random payoffs $f$ and
$g$,
\[
f\geq_{\mathrm{cv}}g\Longrightarrow f\precsim g
\]

\item[(iii)] \emph{strongly risk neutral} when it is both strongly risk averse
and propense.
\end{enumerate}
\end{definition}

Clearly, strong risk aversion (propensity) implies weak risk aversion
(propensity). As well-known, these two notions are equivalent for expected utility
preferences, but not in general.\footnote{Yaari (1987), Wakker (1994), Cohen (1995), and Schmidt and Zank (2008) study several notions of risk aversion for rank-dependent and cumulative prospect theory preferences. For instance, in the dual model of Yaari (1987) weak risk aversion corresponds to a probability weighting function that is dominated by the identity function, while strong risk aversion to a convex probability weighting function.} In contrast,
the strong and weak notions of  risk neutrality always coincide, so we can talk of `risk
neutrality' without further qualification.

\section{Absolute attitudes} \label{sec:analysis}

\subsection{Insurance contracts and attitudes} \label{sec:a-analysis}

As discussed in the introduction, our main objective is to characterize
classical risk attitudes in terms of insurance choices. To tackle this problem
we need to answer two questions:

\begin{itemize}
	\item Which random payoffs can be seen as insurances for a given risk $w$?
	
	\item How can we describe the attitudes towards insurance of an agent facing a risk  $w$?
\end{itemize}

Let us identify, as in Arrow (1974), insurance policies (also called contracts) with the random
payoffs detailing their state-contingent net payments to the agent.
 	Formally, an insurance policy that pays $h(s)$ in each state $s$ and has
	premium $\pi $ corresponds to the random payoff $f=h-\pi $.
We consider an agent who, \emph{before} an insurance
policy is chosen, faces a risk $w\in \mathcal{F}$, so a random loss $-w$%
. Therefore, \emph{after} the policy $f\in \mathcal{F}$ is chosen, the risk changes to $w+f$.

In our static analysis, the policy is chosen at time $0$ and uncertainty
resolves at time $1$. For our purposes, it is immaterial whether we interpret
the risk $w$ as the random wealth change over the considered
period (as we maintain throughout to ease exposition) or, rather, as the final random wealth of the agent. Indeed, in our setting the insurance problem of an agent with initial wealth $w_0 \in \mathbb{R}$ who confronts  risk $w\in\mathcal{F}$ is equivalent to that of an agent with initial wealth $0$ who confronts risk $w_0+w\in\mathcal{F}$. By purchasing insurance, agents seek
protection against payoff variability, which is unaffected by the
addition of constants. For instance, to fully insure risk $w$ at premium $\pi $, thus
receiving $-w$ at cost $\pi $, is equivalent to fully insure $w_0 +w$ at premium $\pi -w_0 $. Formally, this equivalence
corresponds to the accounting identity $-w-\pi =-\left( w_0 +w\right)
-(\pi -w_0 )$.\footnote{%
	Similar identities hold for partial insurances,
	which thus feature analogous equivalences.
	Appendix \ref{app:vs} discusses these properties and their behavioral
	implications in more detail.}
For this reason, in our analysis, random payoffs can take both positive and negative values.

Next we introduce a basic taxonomy of insurance policies.
\begin{definition}
\label{def:pirla}
Given any risk $w$, a random payoff $f$ is:

\begin{enumerate}
\item[(i)] a \emph{full insurance} for $w$, written $f\in\mathcal{I}%
^{\mathrm{fi}}(w)$, when%
\[
f=-w-\pi
\]
for some premium $\pi\in\mathbb{R}$;

\item[(ii)] a \emph{proportional insurance }for $w$, written $f\in
\mathcal{I}^{\mathrm{pr}}(w)$, when%
\[
f=-\left(  1-\varepsilon\right)  w-\pi
\]
for some premium $\pi\in\mathbb{R}$ and percentage excess $\varepsilon
\in\left[  0,1\right)  $;

\item[(iii)] a \emph{deductible-limit insurance }for $w$, written
$f\in\mathcal{I}^{\mathrm{dl}}(w)$, when%
\[
f= \min\left\{\left(  -w-\delta\right)  ^{+} , \lambda \right\} -\pi
\]
for some premium $\pi\in\mathbb{R}$, deductible $\delta\in\mathbb{R}$, and
limit $\lambda \in \left[  0,\infty\right)$.\footnote{As usual, $\left(  -w-\delta\right) ^{+}$ denotes the positive part of $-w-\delta$.}
\end{enumerate}
\end{definition}

Full insurances completely cover the agent position by neutralizing, at a cost, the uncertainty that the agent faces. Instead, proportional and deductible-limit
insurances provide only partial cover: they reimburse either a proportion
$1-\varepsilon$ of the loss or the part of the loss exceeding $\delta$, up to
$\lambda$. They are the most common and simplest kinds of insurance contracts.
Health insurance contracts are typically proportional, while property
insurance ones have a deductible-limit form.

Next we introduce attitudes towards insurance using the types of contracts
that we just presented. Recall that an agent facing risk $w$ who
purchases insurance $f$ ends up with $w+f$.

\begin{definition}
\label{def:iciente}A risk preference $\succsim$ is:

\begin{enumerate}
\item[(i)] \emph{propense to full insurance} when, for all $w,f,g\in
\mathcal{F}$ with $g\overset{d}{=}f$,%
\[
f\in\mathcal{I}^{\mathrm{fi}}\left(  w\right)  \implies w+f\succsim w+g
\]

\item[(ii)] \emph{propense to proportional insurance} when, for all
$w,f,g\in\mathcal{F}$ with $g\overset{d}{=}f$,%
\[
f\in\mathcal{I}^{\mathrm{pr}}(w)\implies w+f\succsim w+g
\]

\item[(iii)] \emph{propense to deductible-limit insurance} when, for all
$w,f,g\in\mathcal{F}$ with $g\overset{d}{=}f$,%
\[
f\in\mathcal{I}^{\mathrm{dl}}(w)\implies w+f\succsim w+g
\]

\end{enumerate}
\end{definition}

Intuitively, insurance has the benefit of being negatively correlated to the risk that the agent is facing: being insured flattens the agent payoff. Other contracts may have the same distribution, but they may not be correlated in the same way. Our whole point is that an insurance-propense agent prefers a contract that, being negatively correlated with the risk, reduces payoff variability.
	 In particular, an
	agent who is propense to full insurance favors a perfect negative correlation that
	altogether eliminates variability. When the agent also values a milder
	negative correlation that partially reduces variability, the
	stronger notion of propensity to partial insurance takes the stage.

These definitions address the initial questions of this section by relying on a common principle: once a kind of
insurance is defined for risk $w$ (Definition \ref{def:pirla}), propensity to insurance of
that kind means that the agent prefers to purchase these insurances $f$
over other equally distributed random payoffs $g$ (Definition \ref{def:iciente}). Equidistribution is a \emph{ceteris paribus} assumption
that disciplines comparisons by ensuring, for example, that neither of the
random payoffs at hand be statewise dominated (with dominance considerations
then confounding insurance motives).\footnote{%
	The notions presented in Definition \ref{def:iciente} are equivalent to the
	ones discussed in the introduction because the equidistribution relation $%
	\overset{d}{=}$ is invariant under the addition of constants. For instance,
	Definition \ref{def:iciente}-(i) is just a theoretically convenient
	rewriting of (\ref{eq:wra}) because, when $g\overset{d}{=}f=-w-\pi \in 
	\mathcal{I}^{\mathrm{fi}}(w)$, by setting $h=g+\pi $ we have $h\overset{d}{%
		=}-w$ as well as $w+f=w+\left( -w-\pi \right) $ and $w+g=w+\left( h-\pi
	\right) $. See also Appendix \ref{rem:anal}.}

For a simple
illustration, consider two important extreme weather events like `excess
rainfall' and `drought'. Wine grapes are an example of crop much more
vulnerable to excess rainfall than to drought, while the opposite is true
for rice. If the two extreme events are equally likely, the rain insurance $f
$ and the drought insurance $g$ paying, respectively, $1$ in case of excess rainfall and $0$
otherwise, and $1$ in case of drought and $0$ otherwise, have the same
distribution. For a viticulturist growing wine grapes -- with, say, revenue $w_{%
	\text{grapes}}$ equal to $0$ in case of excessive rainfall and to $1$
otherwise -- the rain insurance $f$ is a full insurance policy, while the
equally distributed drought insurance $g$ is not. Representing these in a table, we have, 
\begin{equation*}
	\renewcommand{\arraystretch}{1.25}%
	\begin{array}{cccc}
		\multicolumn{1}{c|}{} & \multicolumn{1}{c|}{\text{excess rainfall}} & 
		\multicolumn{1}{c|}{\text{drought}} & \text{other weather conditions} \\ \hline
		\multicolumn{1}{c|}{w_{\text{grapes}}+f} & \multicolumn{1}{c|}{1} & 
		\multicolumn{1}{c|}{1} & \multicolumn{1}{c}{1} \\[3pt]\hline
		\multicolumn{1}{c|}{w_{\text{grapes}}+g} & \multicolumn{1}{c|}{0} & 
		\multicolumn{1}{c|}{2} & \multicolumn{1}{c}{1}%
	\end{array}%
\end{equation*}%
In contrast, for a rice farmer -- with, say, revenue $w_{\text{rice}}$ equal
to $0$ in case of drought and to $1$ otherwise -- it is the drought insurance $g$
that becomes a full insurance policy, while the equally distributed rain insurance $f$ is
not. Indeed,%
\begin{equation*}
	\renewcommand{\arraystretch}{1.25}%
	\begin{array}{cccc}
		\multicolumn{1}{c|}{} & \multicolumn{1}{c|}{\text{excess rainfall}} & 
		\multicolumn{1}{c|}{\text{drought}} & \text{other weather conditions} \\ \hline
		\multicolumn{1}{c|}{w_{\text{rice}}+f} & \multicolumn{1}{c|}{2} & 
		\multicolumn{1}{c|}{0} & \multicolumn{1}{c}{1} \\[3pt]\hline
		\multicolumn{1}{c|}{w_{\text{rice}}+g} & \multicolumn{1}{c|}{1} & 
		\multicolumn{1}{c|}{1} & \multicolumn{1}{c}{1}%
	\end{array}%
\end{equation*}%
In conclusion, equally distributed random payoffs can be vastly different
when viewed as possible insurance policies for a given agent, depending on
their correlation with the risk that the agent is facing. Agents' behavior will differ accordingly: when both farmers are propense to full insurance, the viticulturist will prefer the acquisition of $f$, the rice farmer that of $g$.

There is a natural hierarchy among the insurance
attitudes introduced in  Definition \ref{def:iciente}. When $w$ is bounded, it can be shown that,
\begin{equation}
	\mathcal{I}^{\mathrm{fi}}\left(  w\right)  =\mathcal{I}^{\mathrm{pr}}%
	(w)\cap\mathcal{I}^{\mathrm{dl}}(w) \label{eq:fi-ga}%
\end{equation}
Thus, propensity to full insurance is weaker than propensity to either
proportional or deductible-limit insurance. The next results show that this hierarchy in insurance
attitudes corresponds to the hierarchy in risk attitudes.

\subsection{Weak risk aversion} \label{sec:wr} 

We start with weak risk aversion.

\begin{theorem}
\label{th:wra}The following properties are equivalent for a risk preference:

\begin{itemize}
\item[(i)] weak risk aversion;

\item[(ii)] propensity to full insurance.
\end{itemize}
\end{theorem}

This theorem provides a novel foundation of weak
risk aversion for all risk preferences. It shows how the traditional notion of `preference for the
expectation of a random payoff over the random payoff itself' emerges from a
minimal requirement of propensity to insurance. It is minimal because only the
purchase of full insurance is required to be preferred over the purchase of
other equally distributed random payoffs; in other words, because weak risk
aversion is silent about attitudes towards partial insurance. Furthermore, by
making no use of expectations, the equivalence presented in Theorem \ref{th:wra} also
addresses the normative critique of weak risk aversion that hinges on the
seemingly \emph{ad hoc} use of expectations over other possible statistics
(such as the median).

Theorem \ref{th:wra} relies on a novel result in probability theory of some
independent interest.

\begin{lemma}
\label{lm:deus}The following properties are equivalent for $f\in\mathcal{F}$:

\begin{itemize}
\item[(i)] $\mathbb{E}\left[  f\right]  =0$;

\item[(ii)] there exist $h,h^{\prime}\in\mathcal{F}$ such that $h\overset
{d}{=}h^{\prime}$ and $f\overset{d}{=}h-h^{\prime}$.
\end{itemize}
\end{lemma}

The nontrivial part is that (i) implies (ii). Yet, a simple explanation is
possible in the finite uniform case when $S=\left\{ 1,2,...,n\right\} $ and $%
P\left( s\right) =1/n$ for all $s\in S$. In this case,%
\begin{equation*}
	\mathbb{E}\left[ f\right] =0\iff \sum_{s=1}^{n}f\left( s\right) =0
\end{equation*}%
Define the random payoffs $h$ and $h^{\prime }$ by $h\left( s\right)
=\sum_{i=1}^{s}f\left( i\right) $ and $h^{\prime }\left( s\right)
=\sum_{i=1}^{s-1}f\left( i\right) $ for all $s\in S$, with the convention $%
h^{\prime }\left( 1\right) =0$. Diagrammatically,

\begin{equation*}
	\renewcommand{\arraystretch}{1.25}%
	\begin{array}{cccccc}
		\multicolumn{1}{c|}{} & \multicolumn{1}{c|}{1} & \multicolumn{1}{c|}{2} & 
		\multicolumn{1}{c|}{\cdots } & \multicolumn{1}{c|}{n-1} & n \\ \hline
		\multicolumn{1}{l|}{h} & \multicolumn{1}{l|}{f(1)} & \multicolumn{1}{l|}{
			f(1)+f(2)} & \multicolumn{1}{l|}{\cdots } & \multicolumn{1}{l|}{
			\sum_{i=1}^{n-1}f(i)} & \multicolumn{1}{l}{\sum_{i=1}^{n}f(i)=0} \\%
		[3pt]\hline
		\multicolumn{1}{l|}{h^{\prime }} & \multicolumn{1}{l|}{0} & 
		\multicolumn{1}{l|}{f(1)} & \multicolumn{1}{l|}{\cdots } & 
		\multicolumn{1}{l|}{\sum_{i=1}^{n-2}f(i)} & \multicolumn{1}{l}{
			\sum_{i=1}^{n-1}f(i)}%
	\end{array}%
\end{equation*}%
Therefore, $f\left( s\right) =h\left( s\right) -h^{\prime }\left( s\right) $
for all $s\in S$, that is, $f=h-h^{\prime }$. Moreover, it is easy to see that $%
h\overset{d}{=}h^{\prime }$ since all states are equally probable. Thus, $h$
and $h^{\prime }$ are the sought-after random payoffs showing that (i)
implies (ii).

%Although it seems natural to take a limit from this finite uniform case to deal with the general nonatomic case, 
%the above simple construction fails to deliver the desired result because the cumulants we built 
%above may lose boundedness or integrability when passing to the limit. Sophisticated rearrangement techniques and convergence arguments are then needed to prove the lemma.

The general nonatomic case cannot be directly tackled through a limit
argument building upon the finite uniform case because the cumulant random
payoffs that we constructed above may lose boundedness or integrability when
passing to the limit. {Different techniques are needed.} Interestingly, the
probabilistic Lemma \ref{lm:deus} and the decision-theoretic Theorem \ref%
{th:wra} turn out to be mathematically equivalent, as detailed in Appendix %
\ref{sec:sued}. If one were able to prove directly Theorem \ref{th:wra}
(something that eluded us), the lemma would follow.

To see how Lemma \ref{lm:deus} implies Theorem \ref{th:wra}, assume
propensity to full insurance. For each random payoff $f$, by the lemma there
exist two equally distributed risks $w$ and $w^{\prime }$ such that $f-%
\mathbb{E}\left[ f\right] \overset{d}{=}w-w^{\prime }$. Now, $-w+\mathbb{E}%
\left[ f\right] $ is a full insurance for risk $w$ that is equally
distributed with $-w^{\prime }+\mathbb{E}\left[ f\right] $. Propensity to
full insurance then implies   
\begin{equation*}
	w+\left( -w+\mathbb{E}\left[ f\right] \right) \succsim w+\left( -w^{\prime }+%
	\mathbb{E}\left[ f\right] \right)
\end{equation*}%
Hence, $\mathbb{E}\left[ f\right] \succsim w-w^{\prime }+\mathbb{E}\left[ f%
\right] \overset{d}{=}f$ and so, by the law invariance of $\succsim $, we
have $w-w^{\prime }+\mathbb{E}\left[ f\right] \sim f$. By transitivity, we
conclude that $\mathbb{E}\left[ f\right] \succsim f$, i.e., weak risk
aversion holds, as desired. The easy converse was explained in the introduction.
\subsection{Strong risk aversion}

We now move to strong risk aversion.

\begin{theorem}
\label{th:sra}The following properties are equivalent for a continuous risk preference:

\begin{itemize}
\item[(i)] strong risk aversion;

\item[(ii)] propensity to proportional insurance;

\item[(iii)] propensity to deductible-limit insurance.
\end{itemize}
\end{theorem}

This result shares the same features as the previous one in terms of
economic scope, technical accessibility, and normative soundness. Furthermore, it
justifies the use of the concave order instead of other dispersion orders
(such as the one of Bickel and Lehmann, 1976) to define strong risk aversion.

 Theorem \ref{th:sra} also has clear empirical relevance because proportional and
deductible-limit insurances are the most commonly held and legally  disciplined
insurance policies. On the one hand, it shows that the
strong risk aversion of agents may motivate the demand for these two types of
insurance contracts. On the other hand, the prevalence of these two contracts
in the insurance practice may support the hypothesis of strong risk aversion
of most policyholders. Moreover, the \textit{a priori} non-obvious equivalence between propensity to
proportional insurance (ii) and to deductible-limit insurance (iii) is consistent with the fact that
policyholders often have both kinds of contracts in their insurance portfolios.\footnote{Different forms of insurance policies address various issues in the insurance market. For example, deductible-limit insurance reduces labor costs in damage assessment for auto insurance, where damage verification is costly and moral hazard is a concern. Proportional insurance, on the other hand, is common in health insurance, where claim assessment is simpler.}

Finally, by expressing both concepts in the same language (that of
insurance), Theorems \ref{th:wra} and \ref{th:sra} jointly provide a new
perspective on the well-known fact that weak risk aversion is implied by
strong risk aversion. When offered equally distributed payoffs, a weakly
risk-averse agent only favors full insurance, whereas a strongly risk-averse
one also favors some forms of partial insurance. 
%\footnote{As we already mentioned, we only know that weakly
	%	risk-averse agents will take full insurance in the comparison, but we do not
	%	know if they will take partial insurance.}
In the next section we show that strongly risk-averse agents actually favor
any kind of partial insurance.

\subsection{More insurances}

To further develop our analysis, and make it more realistic, we consider more
general forms of partial insurance. A first principle of insurance
theory is that an insurance policy pays more when the incurred loss is
larger. There are two   similar ways to formalize this principle,
depending on whether we require the insurance payment to be a function of the
realized loss. We regroup them in the following definition.

\begin{definition}
Given any risk $w$, a random payoff $f$ is:

\begin{enumerate}
\item[(iv)] an \emph{indemnity-schedule insurance} for $w$, written
$f\in\mathcal{I}^{\mathrm{is}}(w)$, when%
\[
f=I\left(  -w\right)
\]
for some real-valued increasing map $I$ defined on the image of $-w$;
\item[(v)] a \emph{contingency-schedule insurance }for $w$, written
$f\in\mathcal{I}^{\mathrm{cs}}(w)$, when%
\[
-w\left(  s\right)  >-w\left(  s^{\prime}\right)  \implies f\left(  s\right)
\geq f\left(  s^{\prime}\right)
\]
for almost all states $s$ and $s^{\prime}$.\footnote{That is, almost surely with respect to the product probability measure $P
	\times P$.}
\end{enumerate}
\end{definition}

Again, these two notions have a common basic meaning: greater losses cannot lead to
smaller insurance payments. This is best seen by writing condition (iv), due to
Arrow (1963), as
\[
-w\left(  s\right)  \geq-w\left(  s^{\prime}\right)  \implies f\left(
s\right)  \geq f\left(  s^{\prime}\right)
\]
for all states $s$ and $s^{\prime}$. Thus, (iv) is obtained by (v) under the
additional requirement that equal losses must lead to equal insurance payments.

With this, (v) is the most general notion of insurance that we consider.\footnote{Proportional insurances with state-dependent percentage excesses might not be contingency-schedule insurances. This is also the case for deductible limit insurances with state-dependent deductibles.}
It
embodies a strong form of positive correlation between insurance $f$ and
loss $-w$, known as comonotonicity (see Schmeidler, 1989). This property is what ultimately characterizes insurances, among
all possible random payoffs, for an agent confronting risk  $w$.
We can now enrich relation (\ref{eq:fi-ga}) by adding the inclusions:%
\begin{equation*}
	\mathcal{I}^{\mathrm{pr}}\left( w\right) \cup \mathcal{I}^{\mathrm{dl}%
	}\left( w\right) \subseteq \mathcal{I}^{\mathrm{is}}(w)\subseteq \mathcal{I}^{%
		\mathrm{cs}}\left( w\right) 
\end{equation*}

The next definition is based on a different notion: rather than defining
insurance for $w$, it describes different degrees of coverage for the loss
$-w$ provided by two different policies $f$ and $g$. Yet, as it will be seen
momentarily, this concept naturally connects to the previous ones.

\begin{definition} 
Given any risk $w$, a random payoff $f$ is a \emph{better hedge} for
$w$ than a random payoff $g$, written $f\trianglerighteq_{w}g$, when $f\overset{d}{=}g$
and%
\[
P\left(  f\leq \tau \mid w\leq \lambda \right)  \leq P\left(  g\leq \tau \mid w\leq \lambda \right)
\]
for all payments $\tau\in\mathbb{R}$ and risk levels $\lambda\in\mathbb{R}$.
\end{definition}

This means that $f$ first-order stochastically dominates $g$ on the left tails
of $w$. In the language of Epstein and Tanny (1980, p.~18), $f\trianglerighteq_{w}g$ if
and only if $f$ is \emph{less correlated} (or \emph{less concordant}) with $w$
than $g$. The next proposition connects the concepts of insurance and hedge.

\begin{proposition}
\label{prop:counter}Given any risk $w$, a random payoff $f$ is a
contingency-schedule insurance for $w$ if and only if it is a best hedge for
$w$, that is,%
\[
\mathcal{I}^{\mathrm{cs}}\left(  w\right)  =\left\{  f\in\mathcal{F}:f\trianglerighteq_{w}g\text{ for all }g\overset{d}{=}f\right\}
\]

\end{proposition}

In other words, contingency-schedule insurances for $w$ are the policies that
are less correlated to $w$ within any given distribution class. 
Next we introduce the definitions of propensity to insurance and to hedging relevant here, which are
completely analogous to the ones given before.

\begin{definition}
A risk preference $\succsim$ is:

\begin{enumerate}
\item[(iv)] \emph{propense to indemnity-schedule insurance} when, for all
$w,f,g\in\mathcal{F}$ with $g\overset{d}{=}f$,%
\[
f\in\mathcal{I}^{\mathrm{is}}(w)\implies w+f\succsim w+g
\]

\item[(v)] \emph{propense to contingency-schedule insurance} when, for all
$w,f,g\in\mathcal{F}$ with $g\overset{d}{=}f$,%
\[
f\in\mathcal{I}^{\mathrm{cs}}\left(  w\right)  \implies w+f\succsim w+g
\]

\item[(vi)] \emph{propense to hedging} when, for all $w,f,g\in\mathcal{F}$
with $g\overset{d}{=}f$,%
\[
f\trianglerighteq_{w}g\implies w+f\succsim w+g
\]

\end{enumerate}
\end{definition}

We are now ready for an omnibus result on the equivalence of strong risk aversion and propensity to partial insurance.

\begin{theorem}
\label{th:omni}The following properties are equivalent for a continuous risk preference:

\begin{itemize}
\item[(i)] strong risk aversion;

\item[(ii)] propensity to proportional insurance;

\item[(iii)] propensity to deductible-limit insurance;

\item[(iv)] propensity to indemnity-schedule insurance;

\item[(v)] propensity to contingency-schedule insurance;

\item[(vi)] propensity to hedging.
\end{itemize}
\end{theorem}

Some implications easily  follow from our earlier analysis, others are less
obvious. Some attitudes, like (ii) and (iii), seem mild, easy to understand,
and normatively compelling. Others, like (i) and (vi), seem instead more demanding and
theoretically sophisticated. Be that as it may, they are all equivalent.
In particular, as points (ii)-(v) embody different forms of propensity to partial insurance, we can summarize this result as:
\[
\text{strong risk aversion} \iff \text{propensity to partial insurance} \iff \text{propensity to hedging}
\]

To the best of our knowledge, the only precursor of this result is the
equivalence between (i) and (vi) for expected utility preferences that can be
derived from the findings of Epstein and Tanny (1980). Their results connect risk aversion and hedging propensity for expected utility preferences, but remain silent
about insurance choice behavior, which is the lens that we adopt here to analyze risk aversion.

Finally, let us recall that both  Theorems \ref{th:wra} and \ref{th:omni} (which subsumes Theorem \ref{th:sra}) are valid for all preferences on
$\mathcal{F}$ that are transitive, law invariant, and continuous. %\footnote{These theorems can also be extended to preferences on $\mathcal{L}^{p}$, with $p\in\left[
%	1,\infty\right)  $, and on $\mathcal{F}_{0}$, the space of simple random payoffs (see Section \ref{sect:ext}). Remarkably, in these spaces continuity is
%	often an automatic consequence of monotonicity, a very
%	natural decision-theoretic assumption (see, e.g., Cheridito and Li, 2009).}
Therefore, the applicability of our results goes well beyond expected utility.
This makes the present analysis relevant for popular models of risk
behavior in psychology (such as the prospect theory of Kahneman and Tversky, 1979)
and allows us to account for robustness concerns in economics and finance (as
captured, e.g., by the multiplier preferences of Hansen and Sargent, 2008, or by the expected shortfall criterion of Artzner, Delbaen, Eber, and Heath, 1999). Our analysis shows that insurance propensity characterizes risk aversion irrespective of whether preferences abide to the expected utility model or violate it.

%Finally, for all $p \in\left[  1,\infty\right)  $, Theorem \ref{th:omni}
%extends to risk preferences on $\mathcal{L}^{p}=\mathcal{L}^{p}\left(
%S,\Sigma,P\right)  $ that are continuous with respect to convergence in the
%$p$-th moment (see Appendix \ref{proof:thomni}).

\section{Neutrality}

The definitions of \emph{aversion} to the different kinds of insurance and to
hedging are obtained from those of \emph{propensity} by replacing $\succsim$
with $\precsim$. As usual, \emph{neutrality} is then defined as simultaneous
propensity and aversion. With this, the counterparts of Theorems \ref{th:wra}, \ref{th:sra}, and \ref{th:omni} hold as expected. In particular, all
definitions of insurance neutrality coincide both with risk neutrality and
with hedging neutrality.

The concept of neutrality is important because it serves as a benchmark to connect the absolute attitudes that we studied in the previous section and the comparative ones that we will analyze in the next section. With this motivation we go a bit deeper in its study. To this end, we
introduce two more notions. 

\begin{definition}
A risk preference $\succsim$ is:

\begin{itemize}
\item \emph{monotone} when, for all $w\in\mathcal{F}$ and all $\varepsilon \in \left(  0,\infty\right)$,%
\[
w + \varepsilon\succ w
\]

\item \emph{dependence neutral} when, for all $w,f,g\in\mathcal{F}$,%
\[
g\overset{d}{=}f\implies w+f\sim w+g
\]

\end{itemize}
\end{definition}

Monotonicity just requires that the addition of a sure positive payoff is
always preferred, a natural assumption when monetary outcomes are
considered. %\footnote{For some of the results that follow, a weaker version of
%monotonicity, just requiring that larger sure payoffs are preferred to smaller
%ones, is sufficient. This is detailed in {\color{black}Appendix \ref{app:weak}}.}
Dependence neutrality
means that preferences are unaffected by the possible correlation between
risk $w$ and two identically distributed investments $f$ and $g$. It
strengthens the requirement of law invariance, which corresponds to $w=0$, to situations where risk is present.

\begin{proposition}
\label{prop:rn}The following conditions are equivalent for a risk preference
$\succsim$:

\begin{enumerate}
\item[(i)] risk neutrality;

\item[(ii)] neutrality to full insurance;

\item[(iii)] neutrality to hedging;

\item[(iv)] dependence neutrality.
\end{enumerate}

\noindent Moreover, $\succsim$ is monotone and satisfies any of the equivalent
conditions above if and only if
\begin{equation}
f\succsim g\iff\mathbb{E}\left[  f\right]  \geq\mathbb{E}\left[  g\right]
\label{eq:ev}%
\end{equation}
for all random payoffs $f$ and $g$.
\end{proposition}

This proposition characterizes risk neutrality and makes explicit its relation
with expected-value preferences.
%\footnote{Observe that the equivalence of (i), (ii), and (iii) implies that risk neutrality is also equivalent to neutrality to proportional insurance, deductible-limit insurance,
%indemnity-schedule insurance, and contingency-schedule insurance.}
Its most innovative contribution is the characterization of these preferences based on dependence neutrality, which shows how law invariance irrespective of the outstanding risk leads to expected value maximization.  

 A fundamental feature of these preferences is
their consistency with first-order stochastic dominance, $\geq_{\mathrm{fsd}}%
$. This consistency is crucial in the existing characterizations of
expected-value preferences, in particular the classic one of de Finetti (1931)
and the more recent one of Pomatto, Strack, and Tamuz (2020). In our result{\color{black},} consistency
with $\geq_{\mathrm{fsd}}$ is implicit because, in the derivation, it follows
from monotonicity and dependence neutrality. Yet, to better connect {\color{black}the}
approaches, next we provide a characterization of expected-value preferences
that makes explicit the role of first-order stochastic dominance.

\begin{proposition}
\label{th:PST}Let $\mathcal{F}=\mathcal{M}^{\infty}$ and $P$ be nonatomic. The
following conditions are equivalent for a monotone risk preference $\succsim$:

\begin{enumerate}
\item[(i)] $\succsim$ admits an expected-value representation (\ref{eq:ev});

\item[(ii)] for all $w,f,g\in\mathcal{F}$,
\[
f\geq_{\mathrm{fsd}}g\implies w+f\succsim w+g
\]

\item[(iii)] for all $w,f,g\in\mathcal{F}$,%
\[
f\succsim g\implies w+f\succsim w+g
\]

\item[(iv)] $\succsim$ is complete and
\[
f\succ g\Longrightarrow w+\tilde{f}>_{\mathrm{fsd}}w+\tilde{g}%
\]
for some $w,\tilde{f},\tilde{g}\in\mathcal{F}$ such that $f\overset{d}%
{=}\tilde{f}$, $g\overset{d}{=}\tilde{g}$ and $w$ is independent of both
$\tilde{f}$ and $\tilde{g}$.
\end{enumerate}
\end{proposition}

The equivalence of conditions (i) and (ii) is the sought-after
characterization of expected-value preferences in terms of first-order
stochastic dominance. For perspective, {\color{black} Proposition \ref{th:PST}} also reports the earlier
characterizations of de Finetti (1931), which in preferential form corresponds
to the equivalence of points (i) and (iii), and of Pomatto, Strack, and Tamuz (2020),
which corresponds to the equivalence of points (i) and (iv).

In comparing condition (ii) with (iii), it is important to contrast the
objective premise $f\geq_{\mathrm{fsd}}g$ of the implication in (ii) with the
subjective premise $f\succsim g$ of the one in (iii).
%Implicit in the analysis
%of de Finetti is the consistency requirement that $f\geq_{\mathrm{fsd}}g$
%implies $f\succsim g$. This is why our assumption (ii) is weaker and so our
%result improves de Finetti's one.
In comparing (ii) with (iv), it is important to observe that the former is not
the contrapositive of the latter. Indeed, the equivalence between (i), (ii)
and (iii) continues to hold when $\mathcal{F}=\mathcal{L}^{\infty}$, while
expected-value preferences on $\mathcal{L}^{\infty}$ fail to satisfy (iv).

\section{Comparative attitudes} \label{sect:compa}

We have shown how absolute risk attitudes -- both strong and weak -- can be
characterized in terms of insurance behavior, without recurring to the concept
of expectation, and how this leads to novel insights on old and recent results
about risk preferences. It is then natural to wonder whether the same exercise
can be performed for comparative attitudes.

\subsection{Classical comparative risk attitudes}

As it is the case for absolute risk attitudes, also comparative attitudes have
a weak and {\color{black}a }strong form. According to Yaari (1969), agent $\mathrm{B}$ (Bob) is
\emph{weakly more risk averse than} agent $\mathrm{A}$ (Ann) if whenever Ann
prefers a sure payoff to a random one, so does Bob. Formally,%
\[
\gamma\succsim_{\mathrm{A}}f\implies\gamma\succsim_{\mathrm{B}}f
\]
for all $f\in\mathcal{F}$ and $\gamma\in\mathbb{R}$. Ross (1981) introduces a stronger notion: $\mathrm{B}$ is \emph{strongly
more risk averse than} $\mathrm{A}$ if
\[
\left.
\begin{array}
[c]{l}%
f\geq_{\mathrm{cv}}g \smallskip \\
g\sim_{\mathrm{A}}f-\rho_{\mathrm{A}} \smallskip \\
g\sim_{\mathrm{B}}f-\rho_{\mathrm{B}}%
\end{array}
\right\}  \implies\rho_{\mathrm{B}}\geq\rho_{\mathrm{A}}%
\]
for all $f,g\in\mathcal{F}$ and $\rho_{\mathrm{A}},\rho_{\mathrm{B}}%
\in\mathbb{R}$. The interpretation becomes transparent once one observes that
$\rho_{\mathrm{A}}$ (resp.~$\rho_{\mathrm{B}}$) is the amount of money
Ann (resp.~Bob) is willing to pay to replace $g$ with
the less risky $f$. For the ease of exposition, next we introduce a class of
risk preferences for which this amount always exists.

\begin{definition}
A risk preference $\succsim$ is \emph{secular} when, for all $f,g\in
\mathcal{F}$, there exists $\rho\in\mathbb{R}$ such that 
$
g\sim f-\rho
$.

\end{definition}

When $\succsim$ is monotone, $\rho$ is the largest scalar $r$ such that $f-r \succsim g$, that is, the highest amount of money that the agent is willing to pay to trade $g$ with $f$. Equivalently, $-\rho$ is the smallest compensation for which the agent accepts this trade. Secularity, implicit in Ross (1981), thus requires that the agent is willing to
trade any random payoff with another one for some suitable compensation. Briefly, `every risk has its price' (see
Gollier, 2001). %\footnote{Like for monotonicity, for some of the results that
%follow, a weaker version of secularity, just requiring that each random payoff
%admits a certainty equivalent, is sufficient. Again, this is detailed in {\color{black}Appendix \ref{app:weak}}.}
This notion allows us to extend the observation of Ross, who studies the
monotone and strictly concave expected utility case, that his
definition is stronger than the one of Yaari.

\begin{lemma}
\label{prop:Ya}The following conditions are equivalent for two monotone and
secular risk preferences $\succsim_{\mathrm{A}}$ and $\succsim_{\mathrm{B}}$:

\begin{enumerate}
\item[(i)] $\mathrm{B}$ is weakly more risk averse than $\mathrm{A}$;

\item[(ii)] for all $f,g\in\mathcal{F}$ and $\rho_{\mathrm{A}},\rho
_{\mathrm{B}}\in\mathbb{R}$,%
\[
\left.
\begin{array}
[c]{l}%
f=\mathbb{E}\left[  g\right] \smallskip \\
g\sim_{\mathrm{A}}f-\rho_{\mathrm{A}} \smallskip \\
g\sim_{\mathrm{B}}f-\rho_{\mathrm{B}}%
\end{array}
\right\}  \implies\rho_{\mathrm{B}}\geq\rho_{\mathrm{A}}%
\]

\end{enumerate}

\noindent In particular, if $\mathrm{B}$ is strongly more risk averse than
$\mathrm{A}$, then $\mathrm{B}$ is weakly more risk averse than $\mathrm{A}$.
\end{lemma}

This lemma also shows how Yaari's and Ross' notions are the comparative counterparts of
the ones of Arrow-Pratt and Rothschild-Stiglitz. Indeed, in both the absolute and
comparative cases, the weak notion corresponds to preference for expectation,
the strong one to preference for less risky payoffs in general. The
parallel does not stop here: the absolute risk  aversion notions can be obtained from
the comparative ones by assuming agent $\mathrm{A}$ to be risk neutral, as
next we show. The result is known for the definition of Yaari (we
report it for the sake of completeness), while it seems novel for the one of Ross.

\begin{lemma}
\label{prop:Ro}Let $\succsim_{\mathrm{A}}$ and $\succsim_{\mathrm{B}}$ be
monotone and secular risk preferences. If $\mathrm{A}$ is risk neutral, then:

\begin{enumerate}
\item $\mathrm{B}$ is weakly more risk averse than $\mathrm{A}$ if and only if
$\mathrm{B}$ is weakly risk averse.

\item $\mathrm{B}$ is strongly more risk averse than $\mathrm{A}$ if and only
if $\mathrm{B}$ is strongly risk averse.
\end{enumerate}
\end{lemma}

To further elaborate, observe that when a risk preference is monotone and secular, given any $g$ and $f$ in $\mathcal{F}$ the sure amount $\rho =\rho\left(  g,f\right) $ such that $g\sim f-\rho$ exists and is unique. So, the function
\begin{equation}
	\left(  g,f\right)  \mapsto\rho\left(  g,f\right)  \label{eq:secular-map}%
\end{equation}
is well defined.
Intuitively, the greater
$\rho\left(  g,f\right)  $ is, the more $f$ is preferred over $g$. With this,
we can interpret the function
(\ref{eq:secular-map}) as a measure of the strength of preference. In view of Lemma \ref{prop:Ya}, this function permits to reformulate the
comparative notions of Yaari and Ross as follows:
%\footnote{As usual in comparative analyses, here 	preferences are not required to be risk averse or risk propense to start with.}

\begin{itemize}
\item $\mathrm{B}$ is weakly more risk averse than $\mathrm{A}$ when $
f=\mathbb{E}\left[  g\right] $ implies $\rho_{\mathrm{B}}\left(  g,f\right)
\geq\rho_{\mathrm{A}}\left(  g,f\right)
$
for all $f,g\in\mathcal{F}$.

\item $\mathrm{B}$ is strongly more risk averse than $\mathrm{A}$ when 
$
f\geq_{\mathrm{cv}}g$ implies $\rho_{\mathrm{B}}\left(  g,f\right)  \geq
\rho_{\mathrm{A}}\left(  g,f\right)
$
for all $f,g\in\mathcal{F}$.
\end{itemize}

The difference in the definitions is now evident. Not only $\mathbb{E}\left[
g\right]  \geq_{\mathrm{cv}}g$, but we also have $\mathbb{E}\left[  g\right]
\geq_{\mathrm{cv}}h$ for all $h\geq_{\mathrm{cv}}g$. In words, $\mathbb{E}%
\left[  g\right]  $ is the least risky among the random payoffs that are less
risky than $g$. The sure payoff $\mathbb{E}\left[  g\right]  $
completely eliminates the risk involved in $g$, while a generic payoff
$h\geq_{\mathrm{cv}}g$ only reduces it. Thus, Yaari compares the strength of
preferences only when risk is eliminated, while Ross compares it
also when risk is just reduced.

%Last but not least, the preferences considered in this section are not required to be risk averse or risk loving to start with, both the weak and the strong notions are purely comparative.

\subsection{Comparative insurance propensity}

In light of the previous analysis, the formalization of the concept of
comparative propensity to full insurance {\color{black}is} now natural:

\begin{definition}
\label{def:compi}Let $\succsim_{\mathrm{A}}$ and $\succsim_{\mathrm{B}}$ be
monotone and secular risk preferences. We say that $\mathrm{B}$ is \emph{more
propense to full insurance than} $\mathrm{A}$ when, for all $w,f,g\in
\mathcal{F}$ with $g\overset{d}{=}f$,%
\[
f\in\mathcal{I}^{\mathrm{fi}}\left(  w\right)  \implies\rho_{\mathrm{B}%
}\left(  w+g,w+f\right)  \geq\rho_{\mathrm{A}}\left(  w+g,w+f\right)
\]

\end{definition}

In words, Bob is `more willing to pay than'
Ann in order to achieve full insurance for the risk $w$ that he faces. We can now state the comparative version of
Theorem \ref{th:wra}.

\begin{theorem}
\label{th:comp-wra}The following properties are equivalent for two monotone
and secular risk preferences $\succsim_{\mathrm{A}}$ and $\succsim
_{\mathrm{B}}$:

\begin{itemize}
\item[(i)] $\mathrm{B}$ is weakly more risk averse than $\mathrm{A}$;

\item[(ii)] $\mathrm{B}$ is more propense to full insurance than $\mathrm{A}$.
\end{itemize}
\end{theorem}

To move to strong comparative attitudes, first observe that the definitions of
comparative propensity to proportional insurance, deductible-limit insurance,
indemnity-schedule insurance, and contingency-schedule insurance can be
obtained by replacing $\mathcal{I}^{\mathrm{fi}}\left(  w\right)  $ with
$\mathcal{I}^{\mathrm{pr}}\left(  w\right)  $, $\mathcal{I}^{\mathrm{dl}%
}\left(  w\right)  $, $\mathcal{I}^{\mathrm{is}}(w)$, and $\mathcal{I}%
^{\mathrm{cs}}\left(  w\right)  $ in Definition \ref{def:compi}. Also the
comparative version of propensity to hedging yields no surprises.

\begin{definition}
Let $\succsim_{\mathrm{A}}$ and $\succsim_{\mathrm{B}}$ be monotone and
secular risk preferences. We say that $\mathrm{B}$ is \emph{more propense to
hedging than} $\mathrm{A}$ when, for all $w,f,g\in\mathcal{F}$ with
$g\overset{d}{=}f$,%
\[
f\trianglerighteq_{w}g\implies\rho_{\mathrm{B}}\left(  w+g,w+f\right)  \geq\rho
_{\mathrm{A}}\left(  w+g,w+f\right)
\]

\end{definition}

We can now state the comparative version of Theorem \ref{th:omni}.

\begin{theorem}
\label{th:omni-comp}The following properties are equivalent for two
continuous, monotone, and secular risk preferences $\succsim_{\mathrm{A}}$ and
$\succsim_{\mathrm{B}}$:

\begin{itemize}
\item[(i)] $\mathrm{B}$ is strongly more risk averse than $\mathrm{A}$;

\item[(ii)] $\mathrm{B}$ is more propense to proportional insurance than
$\mathrm{A}$;

\item[(iii)] $\mathrm{B}$ is more propense to deductible-limit insurance than
$\mathrm{A}$;

\item[(iv)] $\mathrm{B}$ is more propense to indemnity-schedule insurance than
$\mathrm{A}$;

\item[(v)] $\mathrm{B}$ is more propense to contingency-schedule insurance
than $\mathrm{A}$;

\item[(vi)] $\mathrm{B}$ is more propense to hedging than $\mathrm{A}$.
\end{itemize}
\end{theorem}

The interpretations and implications of these comparative results are similar
to the absolute ones we discussed in Section \ref{sec:analysis}.
In particular,
\[
\text{stronger risk aversion} \iff \text{higher propensity to partial insurance} \iff \text{higher propensity to hedging}
\]

In view of the fact that $f$ is a better hedge than $g$ for $w$ if and only if
$f$ is more correlated than $g$ with $-w$ (in the sense of Epstein and Tanny,
1980), these equivalences confirm the classical intuition that agents are more
risk averse if and only if they exhibit a stronger preference for insurance
contracts that are more correlated with losses.

\section{Conclusion}

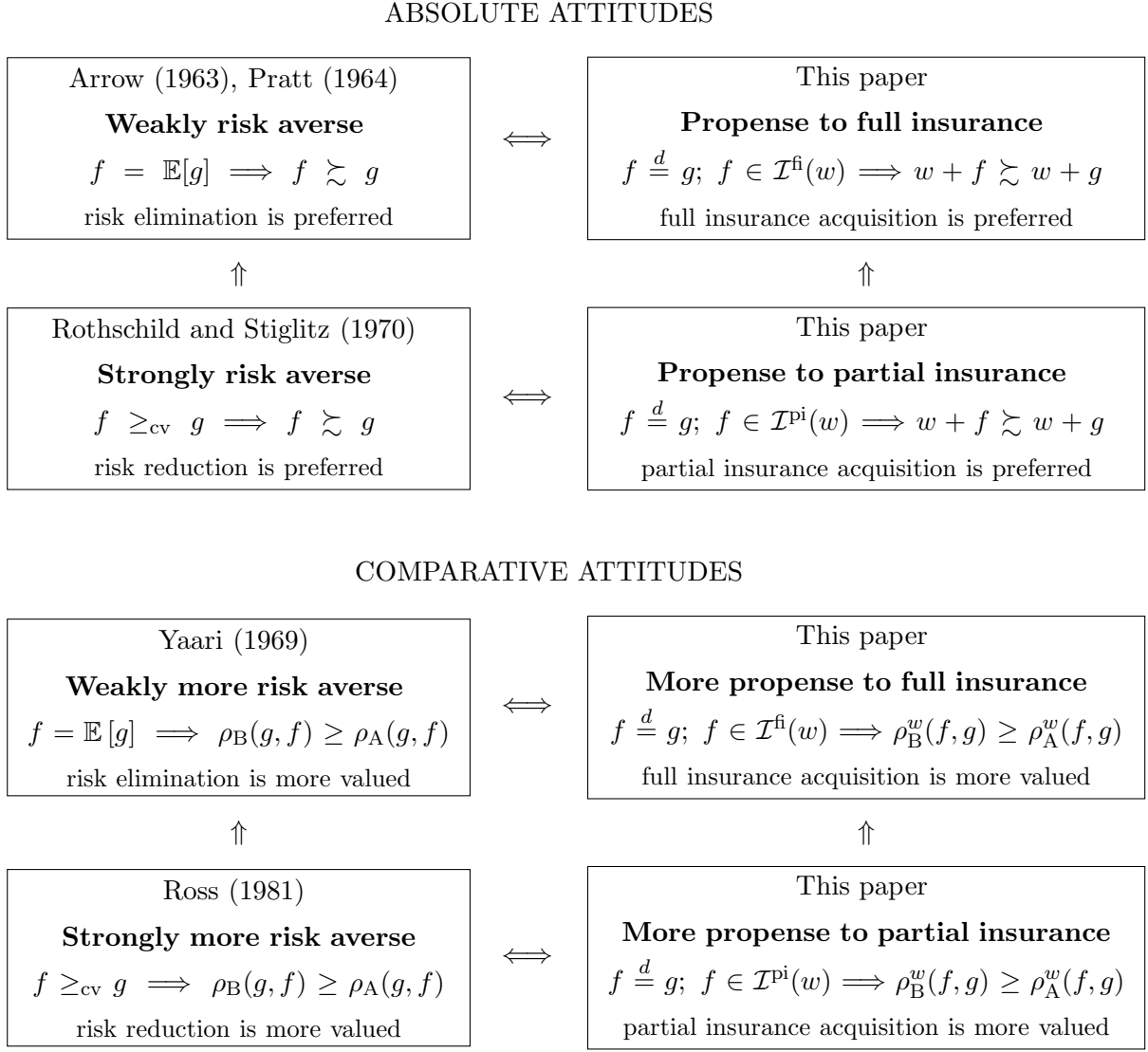
\begin{figure}[ht!]

\tikzstyle{bag} = [text width=16em, text centered]
\tikzstyle{bag2} = [text width=19.4em, text centered]
\tikzstyle{end} = []
\begin{center}
\begin{tikzpicture}[sloped]
\node  (c1) at (4.3,5.4) {ABSOLUTE ATTITUDES};
        \node[draw] (b1) at ( 0,3.5) [bag] {Arrow (1963), Pratt (1964) \vspace{0.3em}\\ \textbf{Weakly risk averse}  \vspace{0.3em} \\
        $f= \mathbb E[g] \Longrightarrow f \succsim g$  \vspace{0.3em} \\ \small   risk elimination is preferred };
        \node[draw] (a2) at ( 8.7 ,0) [bag2]   {This paper  \vspace{0.3em} \\ \textbf{Propense to partial insurance}   \vspace{0.3em} \\
        $f\overset{d}{=}g;  ~ f \in \mathcal I^{\mathrm{pi}}(w)   \Longrightarrow w+ f \succsim w+ g $  \vspace{0.3em}
        \\ \small partial insurance acquisition is preferred} ;
                \node[draw]  (a1) at ( 0,0) [bag]  {Rothschild and Stiglitz  (1970)  \vspace{0.3em} \\ \textbf{Strongly risk averse} \vspace{0.3em}  \\
        $f \ge_{\rm cv} g \Longrightarrow f \succsim g$  \vspace{0.3em} \\ \small risk reduction is preferred};
        \node[draw] (b2) at ( 8.7 ,3.5) [bag2]  {This paper \vspace{0.3em} \\ \textbf{Propense to full insurance}  \vspace{0.3em} \\
        $f\overset{d}{=}g;  ~ f \in \mathcal I^{\mathrm{fi}}(w)  \Longrightarrow w+ f \succsim w+ g$
         \vspace{0.3em} \\  \small full insurance acquisition is preferred}; 
            \node (e1) at ( 0,1.75) [bag] {$\Uparrow$}; 
           \node (e2) at ( 8.7 ,1.75) [bag] {$\Uparrow$};
                      \node (e3) at ( 4,3.6) [bag] {$\Longleftrightarrow$};
                           \node (e4) at ( 4,0) [bag] {$\Longleftrightarrow$};   
            \end{tikzpicture} 
            \end{center}
%\caption{Summary of absolute attitudes}
%\label{fig:GC}
%\end{figure}
%\vspace{-1em}
%\begin{figure}[ht!]
\tikzstyle{bag} = [text width=16em, text centered]
\tikzstyle{bag2} = [text width=19.4em, text centered]
\tikzstyle{end} = []
\begin{center}
            \begin{tikzpicture}[sloped]
\node  (c1) at (4.3,5.4) {COMPARATIVE ATTITUDES};
        \node[draw] (b1) at ( 0,3.5) [bag] {Yaari (1969)  \vspace{0.3em} \\ \textbf{Weakly more risk averse}  \vspace{0.3em} \\ $f=\mathbb{E}\left[  g\right]  \implies\rho_{\mathrm{B}}(  g,f)
\geq\rho_{\mathrm{A}}(  g,f)
$\vspace{0.3em}
   \\   \small risk elimination is more valued};
        \node[draw] (a2) at ( 8.7,0) [bag2]   {This paper \vspace{0.3em} \\ \textbf{More propense to partial insurance}\vspace{0.3em} \\
             $f \overset{d}{=} g;  ~ f \in \mathcal I^{\mathrm{pi}}(w)     \Longrightarrow \rho_{\mathrm{B}}^w(g,f ) \ge  \rho_{\mathrm{A}}^w (g,f )$\vspace{0.3em} \\\small {partial insurance acquisition is more valued}}  ;
                \node[draw]  (a1) at ( 0,0) [bag]  {Ross (1981) \vspace{0.3em}  \\  \textbf{Strongly more risk averse}\vspace{0.3em} \\ $f\geq_{\mathrm{cv}}g\implies\rho_{\mathrm{B}}(  g,f)  \geq
\rho_{\mathrm{A}}(  g,f)$\vspace{0.3em} 
         \\  \small risk reduction  is more valued};
        \node[draw] (b2) at ( 8.7,3.5) [bag2]  {This paper \vspace{0.3em} \\  \textbf{More propense to full insurance}\vspace{0.3em} \\ $f \overset{d}{=} g;  ~ f \in \mathcal I^{\mathrm{fi}}(w)     \Longrightarrow
        	 \rho_{\mathrm{B}}^w(g,f ) \ge  \rho_{\mathrm{A}}^w(g,f )$\vspace{0.3em}  \\\small full insurance acquisition is more valued};
            \node (e1) at ( 0,1.75) [bag] {$\Uparrow$}; 
           \node (e2) at ( 8.7,1.75) [bag] {$\Uparrow$};
                      \node (e3) at ( 4,3.5) [bag] {$\Longleftrightarrow$};
                           \node (e4) at ( 4,0) [bag] {$\Longleftrightarrow$};   
            \end{tikzpicture}
            
\end{center}
\caption{Summary tables, where $\rho^w(g,f )$ stands for $\rho(w+g, w+f)$}
\label{fig:GC2}
\end{figure}

We have shown how the classic, weak and strong, absolute and comparative
notions of risk aversion can be completely characterized through insurance
choice behavior. Our analysis thus provides a unified economic perspective
on these all-important attitudes. Figure \ref{fig:GC2}
summarizes. In the tables, the superscript `$\mathrm{pi}$'\ (partial insurance) stands for any one of `$%
\mathrm{pr},\mathrm{dl},\mathrm{is},\mathrm{cs}$',  with the set $\mathcal{I}^{\mathrm{pi}}(w)$ describing the corresponding class of partial insurance
contracts for $w$.

In sum, our unified analysis of the classical notions of
risk aversion in the expectation-free language of insurance contracts roots
these concepts into basic economic objects, thus improving their economic
appeal. It also makes it possible to talk of risk attitudes for random
variables with an infinite first moment, like those with some Pareto or Cauchy distributions, something that the traditional expectational analysis is unable
to do. The study of these extended notions is an object of future study.

A secondary contribution of our analysis is to highlight the potential
advantages of a state-space approach, based on random variables, for studying
risk attitudes and decisions under risk more broadly. Lotteries may fully describe random variables that are considered in isolation. This is the case, for instance, in the preferential rankings over pairs of random variables -- with either of them being possibly chosen -- that underlie the classical von Neumann-Morgenstern
axiomatization of expected utility. However, insurance inherently involves
multiple interacting random variables, where correlations play a central role.
A state-space framework then provides a natural way to address interdependence. In contrast, a purely lottery-type distributional analysis would require more
intricate tools, such as multivariate distributions or copulas, which our
approach does not need.

Finally we remark that our results can be
readily extended to $\mathcal{L}^{p}$ spaces featuring any scalar
$p\in\left[  1,\infty\right)  $, integer or not, as well as to the space
$\mathcal{F}_{0}$ of simple random payoffs. Moreover, the conclusions of our theorems continue to hold under weaker definitions of propensity to insurance when premium calculation principles are explicitly specified.  See  Appendices \ref{sec-gen-th1} and \ref{app:arente} for these extensions.

\appendix

\renewcommand{\thesection}{Appendix \Alph{section}}

\renewcommand{\thesubsection}{\Alph{section}.\arabic{subsection}}

\section{Outlines of the proofs of Theorems \protect\ref{th:omni}, \protect
	\ref{th:comp-wra}, and \protect\ref{th:omni-comp}}

 The proofs of Theorem \ref{th:wra} and Lemma \ref{lm:deus} are outlined in
Section \ref{sec:analysis}.\ref{sec:wr}. As Theorem \ref{th:sra} is implied by Theorem \ref%
{th:omni}, here we will consider Theorems \ref{th:omni}, \ref%
{th:comp-wra}, and \ref{th:omni-comp}.

\bigskip

\noindent \textbf{Proof sketch of Theorem \ref{th:omni}.} Property (i) implies the other
properties because, for any $w,f,g\in \mathcal{F}$, we have $w+f\geq _{%
	\mathrm{cv}}w+g$ if $f\overset{d}{=}g$ and either $f\in \mathcal{I}^{\mathrm{%
		pi}}(w)$ -- where `pi' denotes any of `pr, dl, is, cs' -- or the condition $%
f\trianglerighteq _{w}g$ is satisfied. This is a standard result in
stochastic orders (see M\"{u}ller and Stoyan, 2002). Property (vi) implies
(v) by Proposition \ref{prop:counter}. Since $\mathcal{I}^{\mathrm{pr}}(w)$
and $\mathcal{I}^{\mathrm{dl}}(w)$ are subsets of $\mathcal{I}^{\mathrm{is}%
}(w)\subseteq \mathcal{I}^{\mathrm{cs}}(w)$ for all $w\in \mathcal{F}$, it
is easy to see that (v) implies (iv), and that (iv) implies both (ii) and
(iii).

The most challenging parts are that (ii) implies (i) and that (iii) implies
(i). We first focus on the latter implication. Let us start with the finite
uniform case, where $S=\{1,\dots ,n\}$ and $P(s)=1/n$ for all $s\in S$. The
proof is based on constructing {mean preserving spreads}. 
%, and $g$ is a mean preserving spread of $f$ if there exist $\delta\ge 0$ and two distinct cells $i,j\in S$, with $f(i)\le f(j)$ such that $g=f-\delta 1_{\{i\}}+\delta 1_{\{j\}}$.
The first and most critical step is to verify that (iii) implies that $%
f\succsim g$ when $g$ is a mean preserving spread of $f$. For such $f$ and $g
$, our purpose is to construct $\tilde{f},\tilde{g},\tilde{w}\in \mathcal{F}$
such that 
\begin{equation}
	\label{eq-outline-th3-1}
	\tilde{f}\overset{d}{=}\tilde{g}\text{, }\tilde{f}\in \mathcal{I}^{\mathrm{dl%
	}}(\tilde{w})\text{, }f=\tilde{w}+\tilde{f}\text{ and }g=\tilde{w}+\tilde{g}
\end{equation}%
Indeed, this yields $f\succsim g$ by (iii). An explicit construction of $%
\tilde{f}$, $\tilde{g}$, and $\tilde{w}$ is provided in the proof of Lemma %
\ref{lm-MPSdl} in Appendix \ref{sec:mps}. Further, it is well-known that
when $f\geq _{\mathrm{cv}}g$ there exists a sequence $h_{0},,h_{1},\dots
,h_{m}$ such that $f=h_{0}$, $g=h_{m}$ and each $h_{k+1}$ is a mean
preserving spread of $h_{k}$. Therefore, transitivity and the existence of $%
\tilde{f},\tilde{g},\tilde{w}\in \mathcal{F}$ in (\ref{eq-outline-th3-1})
yield (i). The general nonatomic case can be directly tackled through a
limiting argument building upon the finite uniform case. 
%Then, (i) holds by transitivity as it is well-established that if $f\ge_{\rm cv}g$, then there exists a sequence $h_0,,h_1,\dots,h_m$ such that $f=h_0$, $g=h_m$ and each $h_{k+1}$ is a mean preserving spread of $h_k$. 
%Note that the risk preference satisfies transitivity. This ensures that to prove (iii) $\Rightarrow$ (i) it suffices to confirm that if \(g\) is a mean preserving spread of \(f\), then there exist \(\tilde{f}, \tilde{g}, \tilde{w} \in \mathcal{F}\) such that
%\begin{align}\label{eq-outline-th3-1}
%\tilde{f}\overset{d}{=}\tilde{g}\text{, }\tilde{f}\in\mathcal{I}^{\mathrm{dl}%
	%}(\tilde{w})\text{, }f=\tilde{w}+\tilde{f}\text{ and }g=\tilde{w}+\tilde{g}%
%\end{align}
%and this yields $f\succsim g$ by (iii).
%An explicit construction of \(\tilde{f}, \tilde{g}, \tilde{w}\) is provided in the proof of Lemma \ref{lm-MPSdl} in Appendix \ref{sec:mps}. 
%The general nonatomic case can be directly tackled through a limit argument building upon the finite uniform case. 

The proof that (ii) implies (i) is similar, with `dl' replaced by `pr' in %
\eqref{eq-outline-th3-1}. The additional technical complexity is that $%
\tilde{f}$, $\tilde{g}$, and $\tilde{w}$ may not exist for all pairs $f,g$
with a mean preserving relationship. They do exist, however for a
sufficiently large subset of such pairs, from which all pairs $f,g$ with a
mean preserving relationship can be approximated as limits of sequences
within this subset. This, combined with continuity, confirms that (ii)
implies (i). %\hfill $\square$
\bigskip 

\noindent \textbf{Proof sketch of Theorem \ref{th:comp-wra}.} It is easy to see that (i)
implies (ii) by noting that $w+f$ is constant when $f\in \mathcal{I}^{%
	\mathrm{fi}}(w)$. Conversely, Lemma \ref{lm:deus} plays an important role.
To be specific, let $\gamma \succsim _{\mathrm{A}}h$ with $h\in \mathcal{F}$
and $\gamma \in \mathbb{R}$. Lemma \ref{lm:deus} implies that there exist $%
w,w^{\prime }\in \mathcal{F}$ such that $w\overset{d}{=}w^{\prime }$ and $h-%
\mathbb{E}[h]\overset{d}{=}w-w^{\prime }$. Thus, we can construct two random
payoffs $f=-w+\mathbb{E}\left[ h\right] $ and $g=-w^{\prime }+\mathbb{E}%
\left[ h\right] $ satisfying $f\overset{d}{=}g$, $f\in \mathcal{I}^{\mathrm{%
		fi}}(w)$, $w+g\overset{d}{=}h$ and $w+f$ constant. Set 
\begin{equation*}
	\eta _{\mathrm{A}}=\left( w+f\right) -\rho _{\mathrm{A}}\left(
	w+g,w+f\right) \quad \text{and\quad }\eta _{\mathrm{B}}=\left( w+f\right)
	-\rho _{\mathrm{B}}\left( w+g,w+f\right) 
\end{equation*}%
They are both constant. It follows from (ii) that $\rho _{\mathrm{A}}\left(
w+g,w+f\right) \leq \rho _{\mathrm{B}}\left( w+g,w+f\right) $, which implies 
$\eta _{\mathrm{A}}\geq \eta _{\mathrm{B}}$. Note that 
\begin{equation*}
	\eta _{\mathrm{A}}\sim _{\mathrm{A}}{w+g}\overset{d}{=}h\precsim _{\mathrm{A}%
	}\gamma \quad \text{and\quad }\eta _{\mathrm{B}}\sim _{\mathrm{B}}w+g\overset%
	{d}{=}h
\end{equation*}%
By law invariance, $\eta _{\mathrm{A}}\sim _{\mathrm{A}}h$ and $\eta _{%
	\mathrm{B}}\sim _{\mathrm{B}}h$. Monotonicity of $\succsim _{\mathrm{A}}$
yields $\gamma \geq \eta _{\mathrm{A}}$. As $\eta _{\mathrm{A}}\geq \eta _{%
	\mathrm{B}}$, we get $\gamma \geq \eta _{\mathrm{B}}$. Further, we have $%
\gamma \succsim _{\mathrm{B}}\eta _{\mathrm{B}}\sim _{\mathrm{B}}h$, where
the `$\succsim _{\mathrm{B}}$' step is due to the monotonicity of $\succsim
_{\mathrm{B}}$. Transitivity shows that \textrm{B} is weakly more risk
averse than \textrm{A}. %\hfill $\square$
\bigskip 

\noindent \textbf{Proof sketch of Theorem \ref{th:omni-comp}.} Using arguments similar to
those in the proof of Theorem \ref{th:omni}, we can demonstrate that
property (i) implies the other properties and establish the sequence of
implications from (vi) to (v), from (v) to (iv), and from (iv) to both (ii)
and (iii).

We now address the more challenging implications (ii) $\Rightarrow $ (i) and
(iii) $\Rightarrow $ (i). We focus on the latter implication as the former
is similar but, like in Theorem \ref{th:omni}, involves additional technical
complexities requiring some standard convergence arguments. As in the proof
of Theorem \ref{th:omni}, the technique of mean preserving spreads is
central. In the finite uniform case, we recall that if $g$ is a mean
preserving spread of $f$, there exist $\tilde{f},\tilde{g},\tilde{w}\in 
\mathcal{F}$ such that \eqref{eq-outline-th3-1} holds. By (iii), we have $%
\rho _{\mathrm{B}}(g,f)\geq \rho _{\mathrm{A}}(g,f)$. To extend the result
for all $f,g\in \mathcal{F}$ with $f\geq _{\mathrm{cv}}g$, we can assume
that $f=h_{0}$, $g=h_{m}$, and $h_{k+1}$ is a mean preserving spread of $%
h_{k}$ for $k=0,\dots ,m-1$. In this step, we use some standard analysis to
prove by induction that for every $x\in \mathbb{R}$ and for each $%
j=1,2,\dots ,m$, 
\begin{equation*}
	\rho _{\mathrm{B}}(h_{j}-x,h_{0}-x)\geq \rho _{\mathrm{A}}(h_{j}-x,h_{0}-x)
\end{equation*}%
This establishes a result stronger than $\rho _{\mathrm{B}}(g,f)\geq \rho _{%
	\mathrm{A}}(g,f)$. The extension to the general nonatomic case can be
directly tackled through a limit argument. In particular, by noting that the
risk preference in Theorem \ref{th:omni-comp} is continuous, we demonstrate
that $\rho :\mathcal{F}\times \mathcal{F}\rightarrow \mathbb{R}$ is
(jointly) sequentially continuous by Lemma \ref{lem:strength} in Appendix %
\ref{app:compa}. %\hfill $\square$

\section{Proofs and related analysis}

\subsection{Preamble} \label{sect:prea}

Recall that $\left(  S,\Sigma,P\right)  $ is an adequate probability space.
We denote by $\mathcal{L}^{0}=\mathcal{L}^{0}\left(  S,\Sigma,P\right)  $ the
space of all measurable functions $f:S\rightarrow\mathbb{R}$, by
$\mathcal{L}^{\infty}=\mathcal{L}^{\infty}\left(  S,\Sigma,P\right)  $ the
space of all almost surely (a.s.)~bounded elements of $\mathcal{L}^{0}$, and
by $\mathcal{L}^{p}=\mathcal{L}^{p}\left(  S,\Sigma,P\right)  $ the space of
all elements of $\mathcal{L}^{0}$ which admit finite absolute $p$-th moment
(for $p\in\left(  0,\infty\right)  $). For $p\in\left[  1,\infty\right]  $,
$\left\Vert \cdot\right\Vert _{p}$ is the usual (semi-)norm of $\mathcal{L}%
^{p}$. By \emph{convergence in }$\mathcal{L}^{p}$, we mean convergence in this
norm. By \emph{convergence in }$\mathcal{M}^{\infty}=\bigcap_{p\in\mathbb{N}%
}\mathcal{L}^{p}$, we mean convergence in all of the $\left\Vert
\cdot\right\Vert _{p}$ norms (for $p\in\mathbb{N}$). By \emph{bounded
a.s.~convergence in }$\mathcal{L}^{\infty}$, we mean almost sure convergence
of a sequence which is bounded in $\left\Vert \cdot\right\Vert _{\infty}$
norm. By the Dominated Convergence Theorem, bounded a.s.~convergence implies
convergence in $\mathcal{M}^{\infty}$.

When we say that a risk preference $\succsim$ is continuous on $\mathcal{F}$, we consider bounded a.s.~convergence of sequences if $\mathcal{F} = \mathcal{L}^{\infty}$, and convergence of sequences in  $\mathcal{M}^{\infty}$ otherwise.

We denote by $L^{p}=L^{p}\left(  S,\Sigma,P\right)  $ the quotient of
$\mathcal{L}^{p}=\mathcal{L}^{p}\left(  S,\Sigma,P\right)  $ when almost
surely equal measurable functions are identified (e.g.~Pollard, 2002).
Analogously, $M^{\infty}=M^{\infty}\left(  S,\Sigma,P\right)  $ is the
quotient of $\mathcal{M}^{\infty}=\mathcal{M}^{\infty}\left(  S,\Sigma
,P\right)  $.

Given any $f\in\mathcal{L}^{0}$, the cumulative distribution function
$F:\mathbb{R}\rightarrow\left[  0,1\right]  $ of $f$ is defined by $F\left(
x\right)  =P\left(  f\leq x\right)  $ for all $x\in\mathbb{R}$. The function
$F$ is increasing  and right-continuous, with $\lim_{x\rightarrow-\infty
}F\left(  x\right)  =0$ and $\lim_{x\rightarrow\infty}F\left(  x\right)  =1$.
Its left-continuous inverse $F^{-1}:\left(  0,1\right)  \rightarrow\mathbb{R}$
is defined by $F^{-1}\left(  t\right)  =\inf\left\{  x\in\mathbb{R}:F\left(
x\right)  \geq t\right\}  $, also denoted by $q_{f}^{-}\left(  t\right)  $ or
$F_{f}^{-1}\left(  t\right)  $ when its dependence on $f$ needs to be
emphasized. The function $F^{-1}$ is always increasing, and it belongs to
$\mathcal{L}^{p}\left(  \lambda\right)  $ if and only if $f\in\mathcal{L}%
^{p}\left(  P\right)  $ (for all $p\in\left[  0,\infty\right]  $), where as
usual $\lambda$ is the Lebesgue measure on $\left(  0,1\right)  $.

Let $\mathcal{U}$ be the collection of all $v\in\mathcal{L}^{0}$ having
a uniform distribution on $\left(  0,1\right)  $, i.e.~$P\left(  v\leq
t\right)  =t$ for all $t\in\left(  0,1\right)  $. It is without loss to assume
$v\left(  S\right)  =\left(  0,1\right)  $ for all $v
\in\mathcal{U}$. For each $f\in\mathcal{L}^{0}$, $f_{v}\in
\mathcal{L}^{0}$ is defined by $f_{v}=q_{f}^{-}\circ v$.

\begin{lemma}
\label{lm:fs}Let $P$ be nonatomic and $f\in L^{0}$. Then:

\begin{enumerate}
\item[(i)] for each $v\in\mathcal{U}$, it holds $f_{v}%
\overset{d}{=}f$;

\item[(ii)] there exists $v\in\mathcal{U}$ such that $f_{v}=f$ a.s.
\end{enumerate}
\end{lemma}

\noindent\textbf{Proof.} See, e.g., Lemmas A.23 and A.32 of F\"{o}llmer and
Schied (2016).\hfill$\blacksquare\bigskip$

In what follows, for each $n\in\mathbb{N}$ we denote by%
\[
\Psi_{n}=\left\{  \left(  \frac{0}{2^{n}},\frac{1}{2^{n}}\right]  ,\left(
\frac{1}{2^{n}},\frac{2}{2^{n}}\right]  ,\dots,\left(  \frac{2^{n}-1}{2^{n}%
},\frac{2^{n}}{2^{n}}\right)  \right\}
\]
the partition of $\left(  0,1\right)  $ into segments of equal length $2^{-n}%
$. If $P$ is nonatomic and $v\in\mathcal{U}$, for each $n\in\mathbb{N}%
$,%
\[
\Pi_{n}^{v}=v^{-1}\left(  \Psi_{n}\right)
\]
is a partition of $S$ in $\Sigma$ such that $P(E)=1/2^{n}$ for all $E\in
\Pi_{n}^{v}$. By setting $\Sigma_{n}^{v}=\sigma\left(  \Pi
_{n}^{v}\right)  =v^{-1}\left(  \sigma\left(  \Psi_{n}\right)
\right)  $ for all $n\in\mathbb{N}$, we have a filtration $\left\{  \Sigma
_{n}^{v}\right\}  _{n\in\mathbb{N}}$ in $\Sigma$. As usual,
$\Sigma_{\infty}^{v}=\sigma\left(  \bigcup_{n\in\mathbb{N}}\Sigma
_{n}^{v}\right)  $.

\begin{lemma}
\label{lm:sti} Let $P$ be nonatomic and $p\in\lbrack1,\infty]$. For each
$v\in\mathcal{U}$,%
\[
\Sigma_{\infty}^{v}=\sigma\left(  v\right)
\]
and, for each $f\in\mathcal{L}^{p}$ (resp.~$f\in\mathcal{M}^{\infty}$),%
\[
\mathbb{E}\left[  f_{v}\mid\Sigma_{n}^{v}\right]  \rightarrow
f_{v}%
\]
almost surely, in $\mathcal{L}^{p}$ if $p<\infty$, and in bounded a.s.~convergence if $p=\infty$ (resp.~in $\mathcal{M}^{\infty}$). In particular, by
choosing $v$ such that $f=f_{v}$ a.s., it follows that%
\[
\mathbb{E}\left[  f\mid\Sigma_{n}^{v}\right]  \rightarrow f
\]
in the above senses. Moreover, for each $v\in\mathcal{U}$ and each
$f\in\mathcal{L}^{p}$,%
\[
q_{\mathbb{E}\left(  f_{v}\mid\Sigma_{n}^{v}\right)  }%
^{-}=\mathbb{E}_{\lambda}\left[  q_{f}^{-}\mid\sigma\left(  \Psi_{n}\right)
\right]  \qquad\lambda\text{-a.s.}%
\]
for all $n\in\mathbb{N}$.
\end{lemma}

\noindent\textbf{Proof.} Note that the $\sigma$-algebra $\sigma\left(
%TCIMACRO{\dbigcup _{n\in\mathbb{N}}}%
%BeginExpansion
{\displaystyle\bigcup_{n\in\mathbb{N}}}
%EndExpansion
\Psi_{n}\right)  $ is the Borel $\sigma$-algebra $\mathcal{B}\left(
0,1\right)  $ on $\left(  0,1\right)  $ because $%
%TCIMACRO{\dbigcup _{n\in\mathbb{N}}}%
%BeginExpansion
{\displaystyle\bigcup_{n\in\mathbb{N}}}
%EndExpansion
\Psi_{n}\ $is countable and separates the points of $\left(  0,1\right)  $
(see, e.g., Mackey, 1957, Theorem 3.3). Then,%
\begin{align*}
\sigma\left(  v\right)   &  =v^{-1}\left(  \mathcal{B}\left(
0,1\right)  \right)  =v^{-1}\left(  \sigma\left(
%TCIMACRO{\dbigcup _{n\in\mathbb{N}}}%
%BeginExpansion
{\displaystyle\bigcup_{n\in\mathbb{N}}}
%EndExpansion
\Psi_{n}\right)  \right)  =\sigma\left(  v^{-1}\left(
%TCIMACRO{\dbigcup _{n\in\mathbb{N}}}%
%BeginExpansion
{\displaystyle\bigcup_{n\in\mathbb{N}}}
%EndExpansion
\Psi_{n}\right)  \right)  =\sigma\left(
%TCIMACRO{\dbigcup _{n\in\mathbb{N}}}%
%BeginExpansion
{\displaystyle\bigcup_{n\in\mathbb{N}}}
%EndExpansion
v^{-1}\left(  \Psi_{n}\right)  \right) \\
&  =\sigma\left(
%TCIMACRO{\dbigcup _{n\in\mathbb{N}}}%
%BeginExpansion
{\displaystyle\bigcup_{n\in\mathbb{N}}}
%EndExpansion
\Pi_{n}^{v}\right)  =\sigma\left(
%TCIMACRO{\dbigcup _{n\in\mathbb{N}}}%
%BeginExpansion
{\displaystyle\bigcup_{n\in\mathbb{N}}}
%EndExpansion
\Sigma_{n}^{v}\right)  =\Sigma_{\infty}^{v}%
\end{align*}
By the Martingale Convergence Theorem on $\mathcal{L}^{1}$ and on
$\mathcal{L}^{p}$, $p\in(1,\infty)$ (see Theorems 4.2.11 and 4.4.6 of Durrett,
2019, respectively),
%Theorem 35.6 of Billingsley,
%1995),%
\[
\mathbb{E}\left[  f_{v}\mid\Sigma_{n}^{v}\right]  \rightarrow
\mathbb{E}\left[  f_{v}\mid\Sigma_{\infty}^{v}\right]
\]
both almost surely, and in $\mathcal{L}^{p}$ if $p<\infty$ (in particular,
if $f\in\mathcal{M}^{\infty}$, convergence in $\mathcal{M}^{\infty}$ follows).
In case $p=\infty$, we have bounded a.s.~convergence because $f$ is a.s.
bounded. But $\Sigma_{\infty}^{v}=\sigma\left(  v\right)  $ and
$f_{v}$ is $\sigma\left(  v\right)  $-measurable, and so, almost
surely%
\[
f_{v}=\mathbb{E}\left[  f_{v}\mid\sigma\left(  v\right)
\right]  =\mathbb{E}\left[  f_{v}\mid\Sigma_{\infty}^{v}\right]
\]
This proves the first part of the statement.

For each $n\in\mathbb{N}$. Define $G_{n}=\mathbb{E}_{\lambda}\left[  q_{f}%
^{-}\mid\sigma\left(  \Psi_{n}\right)  \right]  $ on $\left(  0,1\right)  $,
and observe that $G_{n}$ is an increasing function. Moreover, by the change of
variable formula (Lemma \ref{lm:change} below),%
\[
\mathbb{E}_{P}\left[  f_{v}\mid\Sigma_{n}^{v}\right]
=\mathbb{E}_{P}\left[  q_{f}^{-}\circ v\mid v^{-1}\left(
\sigma\left(  \Psi_{n}\right)  \right)  \right]  =\mathbb{E}_{\lambda}\left[
q_{f}^{-}\mid\sigma\left(  \Psi_{n}\right)  \right]  \circ v=G_{n}%
\circ v
\]
almost surely. By Lemma A.27 of F\"{o}llmer and Schied (2016), we then have,
$\lambda$-a.s.,%
\[
q_{\mathbb{E}\left[  f_{ v}\mid\Sigma_{n}^{ v}\right]  }%
^{-}=q_{G_{n}\circ v}^{-}=G_{n}\circ q_{ v}^{-}=G_{n}%
\]
as desired.\hfill$\blacksquare$\bigskip

We close with two technical results.

\begin{lemma}
\label{lm:change}Let $\left(  X,\Sigma_{X},P\right)  $ be a probability space,
$\left(  Y,\Sigma_{Y}\right)  $ be a measurable space, $T:X\rightarrow Y$ be a
measurable function, $g:Y\rightarrow\mathbb{R}$ be a measurable function such
that $g\circ T$ is $P$-summable. Then $g$ is $P\circ T^{-1}$-summable and, for
every sub-$\sigma$-algebra $\mathcal{A}$ of $\Sigma_{Y}$,
\[
\mathbb{E}_{P}\left[  g\circ T\mid T^{-1}\left(  \mathcal{A}\right)  \right]
=\mathbb{E}_{P\circ T^{-1}}\left[  g\mid\mathcal{A}\right]  \circ T
\]
Moreover, for all $A\in\Sigma_{Y}$, it holds $\mathbb{E}_{P}\left[  g\circ
T\mid T^{-1}\left(  A\right)  \right]  =\mathbb{E}_{P\circ T^{-1}}\left[
g\mid A\right]  $.
\end{lemma}

\noindent\textbf{Proof.} The proof is standard.\hfill$\blacksquare$\bigskip

\begin{lemma}
\label{lm:com-cv}Let $f,g,f^{\prime},g^{\prime}\in\mathcal{L}^{1}$. If
$P\left(  f\leq x,g\leq y\right)  \leq P\left(  f^{\prime}\leq x,g^{\prime
}\leq y\right)  $ for all $x,y\in\mathbb{R}$, then $f+g\geq_{\mathrm{cv}%
}f^{\prime}+g^{\prime}$.
\end{lemma}

\noindent\textbf{Proof.} See, e.g., Theorem 3.8.2 of M\"{u}ller and Stoyan
(2002).\hfill$\blacksquare$\bigskip

\subsection{On equivalent definitions of insurance propensity} \label{rem:anal}

The definitions of propensity to full (resp.~proportional) insurance that we
provide in the introduction are equivalent to those appearing in Section
\ref{sec:analysis}.
Indeed, $f\in\mathcal{I}^{\mathrm{pr}}(w)$ if and only if $f=-\left(
1-\varepsilon\right)  w-\pi$ for some $\pi\in\mathbb{R}$ and some
$\varepsilon\in\left[  0,1\right)  $. Thus, propensity to proportional
insurance, as defined by point (ii) of Definition \ref{def:iciente}, requires
\[
w-\left(  1-\varepsilon\right)  w-\pi\succsim w+g
\]
for all $w\in\mathcal{F}$, $\varepsilon\in\left[  0,1\right)  $, $\pi
\in\mathbb{R}$ and $g\overset{d}{=}-\left(  1-\varepsilon\right)  w-\pi$, that
is,
\[
w-\left(  1-\varepsilon\right)  w-\pi\succsim w+h-\pi
\]
for all $w\in\mathcal{F}$, $\varepsilon\in\left[  0,1\right)  $, $\pi
\in\mathbb{R}$ and$\ h\overset{d}{=}-\left(  1-\varepsilon\right)  w$. The
latter is the definition of propensity to proportional insurance proposed in
the introduction.

The case of full insurance is obtained by considering only the case
$\varepsilon=0$.

\subsection{On mean preserving spreads} \label{sec:mps}

In this section, we assume that $\Sigma$ is generated by a partition
$\mathcal{S}$ of equiprobable events (called cells), and we fix a risk
preference $\succsim$ on $\mathcal{F}$.

\begin{definition}
Given $f,g\in\mathcal{F}$, we say that $g$ is a \emph{mean preserving spread}
of $f$ when there exist $\delta\geq0$ and two distinct cells
$S_{1}$ and $S_{2}$ in $\mathcal{S}$, with $f\left(  S_{1}\right)  \leq
f\left(  S_{2}\right)  $ such that\footnote{Clearly, $f$ is constant on cells,
so $f\left(  S_{i}\right)  $ is the constant value of $f$ on $S_{i}%
$,\thinspace\ for $i=1,2$.}%
\[
g=f-\delta1_{S_{1}}+\delta1_{S_{2}}%
\]

\end{definition}

\begin{lemma}
\label{lm-MPSpr} \label{lm-finiteNL}Let $f,g\in\mathcal{F}$ be such that $g$
is a mean preserving spread of $f$ satisfying $g=f-\delta1_{S_{1}}%
+\delta1_{S_{2}}$ with $f(S_{1})<f(S_{2})$ and $\delta>0$. Then there exist
$\tilde{f},\tilde{g},\tilde{w}\in\mathcal{F}$ such that
\begin{equation}
\tilde{f}\overset{d}{=}\tilde{g},\text{\ }\tilde{f}=\eta\tilde{w}\text{ with
}\eta\in\left(  -1,0\right)  \text{, and }f=\tilde{w}+\tilde{f}\text{ and
}g=\tilde{w}+\tilde{g} \label{eq-NL}%
\end{equation}
in particular $\tilde{f}\in\mathcal{I}^{\mathrm{pr}}\left(  \tilde{w}\right)
$, and so $f\succsim g$ if the risk preference $\succsim$ is propense to
proportional insurance.
\end{lemma}

\noindent\textbf{Proof.} Denote by $m_{i}=f(S_{i})$, $i=1,2$. Let
$a=(m_{1}-m_{2})/\delta-1<-1$, and define
\[
\tilde{f}=f/(a+1),~\tilde{g}=\tilde{f}1_{S\setminus\{S_{1},S_{2}\}}+\tilde
{f}(S_{1})1_{S_{2}}+\tilde{f}(S_{2})1_{S_{1}},~\tilde{w}=a\tilde{f}.
\]
We aim to show that $\tilde{f},\tilde{g},\tilde{w}$ satisfy all conditions in
(\ref{eq-NL}). It is straightforward to see $\tilde{f}\overset{d}{=}\tilde{g}%
$. Moreover, it is easy to verify that%
\[
\tilde{f}=\frac{1}{a}\tilde{w}  \quad\text{with}\quad  \frac{1}{a}\in\left(  -1,0\right)
\]
because $a<-1$, and
\[
\tilde{w}+\tilde{f}=a\tilde{f}+\tilde{f}=\left(  a+1\right)  \frac{f}{a+1}=f
\]
and%
\begin{align*}
&\tilde{w}+\tilde{g}    =a\tilde{f}+\tilde{f}1_{S\setminus\{S_{1},S_{2}%
\}}+\tilde{f}(S_{1})1_{S_{2}}+\tilde{f}(S_{2})1_{S_{1}}\\
&  =a\tilde{f}1_{S\setminus\{S_{1},S_{2}\}}+a\tilde{f}(S_{2})1_{S_{2}}%
+a\tilde{f}(S_{1})1_{S_{1}}+\tilde{f}1_{S\setminus\{S_{1},S_{2}\}}+\tilde
{f}(S_{1})1_{S_{2}}+\tilde{f}(S_{2})1_{S_{1}}\\
&  =f1_{S\setminus\{S_{1},S_{2}\}}+(a\tilde{f}(S_{2})+\tilde{f}(S_{1}%
))1_{S_{2}}+(a\tilde{f}(S_{1})+\tilde{f}(S_{2}))1_{S_{1}}\\
&  =f1_{S\setminus\{S_{1},S_{2}\}}+\left(  a\frac{f(S_{2})}{a+1}+\frac
{f(S_{1})}{a+1}\right)  1_{S_{2}}+\left(  a\frac{f(S_{1})}{a+1}+\frac
{f(S_{2})}{a+1}\right)  1_{S_{1}}\\
&  =f1_{S\setminus\{S_{1},S_{2}\}} +\underset{=m_{2}+\delta=f\left(
S_{2}\right)  +\delta}{\underbrace{\left(  \left(  \frac{m_{1}-m_{2}}{\delta
}-1\right)  \frac{m_{2}}{\frac{m_{1}-m_{2}}{\delta}}+\frac{m_{1}}{\frac
{m_{1}-m_{2}}{\delta}}\right)  }}1_{S_{2}}%\\
%&  \hspace*{2.41cm}
+\underset{=m_{1}-\delta=f\left(  S_{1}\right)  -\delta
}{\underbrace{\left(  \left(  \frac{m_{1}-m_{2}}{\delta}-1\right)  \frac
{m_{1}}{\frac{m_{1}-m_{2}}{\delta}}+\frac{m_{2}}{\frac{m_{1}-m_{2}}{\delta}%
}\right)  }}1_{S_{1}}\\
&  =g
\end{align*}
as desired.\hfill$\blacksquare$

\begin{lemma}
Let $f,g\in\mathcal{F}$ be such that $g$ is a mean preserving spread of $f$.
If the risk preference $\succsim$ is continuous and propense to proportional
insurance, then $f\succsim g$.
\end{lemma}

\noindent\textbf{Proof.} Let $f,g\in\mathcal{F}$ be such that $g$ is a mean
preserving spread of $f$. Then there exist $\delta\geq0$ and two distinct
cells $S_{1}$ and $S_{2}$ in $\mathcal{S}$, with $f\left(  S_{1}\right)  \leq
f\left(  S_{2}\right)  $ such that%
\[
g=f-\delta1_{S_{1}}+\delta1_{S_{2}}%
\]
If $\delta=0$, then $f=g$ and reflexivity of $\succsim$ yields $f\succsim g$.
If $\delta>0$ and $f(S_{1})<f(S_{2})$, the previous lemma yields $f\succsim
g$. If $\delta>0$ and $f(S_{1})=f(S_{2})$, define $f_{\varepsilon
}=f-\varepsilon1_{S_{1}}+\varepsilon1_{S_{2}}$ with $\varepsilon\in(0,\delta
)$. Note that 
%$
%f_{\varepsilon}     \left(  S_{1}\right)  <f_{\varepsilon}\left(  S_{2}\right)$
%and 
\begin{align*}
f_{\varepsilon}  &  =f-\varepsilon1_{S_{1}}+\varepsilon1_{S_{2}}%
=f1_{S\setminus\{S_{1},S_{2}\}}+\left(  f(S_{1})-\varepsilon\right)  1_{S_{1}%
}+\left(  f(S_{2})+\varepsilon\right)  1_{S_{2}}\\
g  &  =f-\delta1_{S_{1}}+\delta1_{S_{2}}=f-\left(  \varepsilon+\left(
\delta-\varepsilon\right)  \right)  1_{S_{1}}+\left(  \varepsilon+\left(
\delta-\varepsilon\right)  \right)  1_{S_{2}} =f_{\varepsilon}-\left(  \delta-\varepsilon\right)  1_{S_{1}}+\left(
\delta-\varepsilon\right)  1_{S_{2}}%
\end{align*}
Thus $g$ is a mean preserving spread of $f_{\varepsilon}$ with $f_{\varepsilon
}(S_{1})<f_{\varepsilon}(S_{2})$ and $\delta-\varepsilon>0$. By the previous
argument $f_{\varepsilon}\succsim g$ for all $\epsilon\in(0,\delta)$. Letting
$\varepsilon_{n}=\delta/2^{n}\rightarrow0$, we have $f_{\varepsilon_{n}%
}\rightarrow f$, and continuity implies $f\succsim g$. \hfill$\blacksquare$

\begin{lemma}
\label{lm-MPSdl} Let $f,g\in\mathcal{F}$ be such that $g$ is a mean preserving
spread of $f$. Then there exist $\tilde{f},\tilde{g},\tilde{w}\in\mathcal{F}$
such that
\[
\tilde{f}\overset{d}{=}\tilde{g}\text{, }\tilde{f}\in\mathcal{I}^{\mathrm{dl}%
}(\tilde{w})\text{, }f=\tilde{w}+\tilde{f}\text{ and }g=\tilde{w}+\tilde{g}%
\]
and so $f\succsim g$ if the risk preference $\succsim$ is propense to
deductible-limit insurance.
\end{lemma}

\noindent\textbf{Proof.} For a mean preserving spread $g$ of $f$, we can
write
\[
g=f-2\delta1_{S_{1}}+2\delta1_{S_{2}}%
\]
where $\delta\geq0$ and $f(S_{1})\leq f(S_{2})$. Define the events%
\[
E_{1}=\left\{  f\leq f(S_{1})\right\}  \setminus S_{1} \qquad E_{2}=\left\{
f(S_{1})<f<f(S_{2})\right\}  \qquad E_{3}=\left\{  f\geq f(S_{2})\right\}
\setminus S_{2}
\]
The events $S_{1}$, $S_{2}$, $E_{1}$, $E_{2}$, and $E_{3}$ form a measurable
partition of $S$. Define $\tilde{f},\tilde{g},{\tilde{w}}$ by the following
table:
\[%
\begin{tabular}
[c]{c|c|c|c|c|c}
& $E_{1}$ & $S_{1}$ & $E_{2}$ & $S_{2}$ & $E_{3}$\\\hline
$\tilde{f}$ & $\delta$ & $\delta$ & $\delta$ & $-\delta$ & $-\delta$\\
$\tilde{g}$ & $\delta$ & $-\delta$ & $\delta$ & $\delta$ & $-\delta$\\
${\tilde{w}}$ & $f-\delta$ & $f-\delta$ & $f-\delta$ & $f+\delta$ & $f+\delta$%
\end{tabular}
\
\]
Write $\xi=-f(S_{2})-\delta$. One can check $\tilde{f}\overset{d}{=}\tilde{g}$
and
\[
\tilde{f}=(-\tilde{w}-\xi)^{+}\wedge(2\delta)-\delta
\]
in fact

\begin{itemize}
\item if $s\in E_{1}\cup S_{1}\cup E_{2}$, then $f\left(  s\right)  \leq
f\left(  S_{2}\right)  $, and%
\[
-\tilde{w}\left(  s\right)  -\xi=-f\left(  s\right)  +\delta+f(S_{2}%
)+\delta=f(S_{2})-f\left(  s\right)  +2\delta\geq2\delta\geq0
\]
so
\[
(-\tilde{w}\left(  s\right)  -\xi)^{+}=f(S_{2})-f\left(  s\right)
+2\delta\geq2\delta
\]
and
\[
(-\tilde{w}\left(  s\right)  -\xi)^{+}\wedge(2\delta)-\delta=2\delta
-\delta=\delta=\tilde{f}\left(  s\right)
\]

\item else $s\in S_{2}\cup E_{3}$, then $f\left(  s\right)  \geq f\left(
S_{2}\right)  $, and%
\[
-\tilde{w}\left(  s\right)  -\xi=-f\left(  s\right)  -\delta+f(S_{2}%
)+\delta=f(S_{2})-f\left(  s\right)  \leq0
\]
so%
\[
(-\tilde{w}\left(  s\right)  -\xi)^{+}=0
\]
and%
\[
(-\tilde{w}\left(  s\right)  -\xi)^{+}\wedge(2\delta)-\delta=-\delta=\tilde
{f}\left(  s\right)
\]

\end{itemize}

This implies $\tilde{f}\in\mathcal{I}^{\mathrm{dl}}(\tilde{w})$. On the other
hand, it is easy to see $\tilde{w}+\tilde{f}=f$ and $\tilde{w}+\tilde{g}=g$,
as wanted. \hfill$\blacksquare$

\begin{lemma}
\label{lem:cv}If the risk preference $\succsim$ is continuous, and 
propense to either proportional or deductible-limit insurance,
then
\[
f\geq_{\mathrm{cv}}g\implies f\succsim g
\]

\end{lemma}

\noindent\textbf{Proof.} If $f\geq_{\mathrm{cv}}g$ in $\mathcal{F}$, then there
exists a sequence $h_{0},h_{1},\dots,h_{m}$ such that $f=h_{0}$, $g=h_{m}$ and
each $h_{k+1}$ is either a mean preserving spread of $h_{k}$ or it is obtained
by $h_{k}$ through the permutation of the values that $h_{k}$ takes on two
cells. In the first case, $h_{k}\succsim h_{k+1}$ by what we just proved. In
the second, $h_{k}\sim h_{k+1}$ because $\succsim$ is law invariant. By the
transitivity of $\succsim$, we conclude that $f\succsim g$.\hfill
$\blacksquare$

\subsection{A deus ex machina}

In what follows, for $f\in L^{\infty}$, let $u_{f}$ be the essential supremum
of $f$ and $\ell_{f}$ be the essential infimum of $f$, defined by $u_{f} =
\inf\left\{  x\in\mathbb{R}: {P} (f \le x ) = 1\right\}  $ and $\ell_{f} =
\sup\left\{  x\in\mathbb{R}: {P} (f \ge x ) = 1\right\}  $.

\begin{theorem}
\label{th:deus} Let $k\geq1$ and $f\in L^{k}$. Then $\mathbb{E}\left[
f\right]  =0$ if and only if there exist $g,g^{\prime}\in L^{k-1}$ such that
$g\overset{d}{=}g^{\prime}$ and $g-g^{\prime}\overset{d}{=}f$. Moreover,

\begin{enumerate}
\item[(i)] if $f\in L^{\infty}$, it is possible to choose $g,g^{\prime}\in
L^{\infty}$ so that $\ell_{f}\leq g,g^{\prime}\leq u_{f}$;

\item[(ii)] if $f\in M^{\infty}$, it is possible to choose $g,g^{\prime}\in
M^{\infty}$;

\item[(iii)] if the probability space is finite, it is possible to choose $g$
and $g^{\prime}$ so that $g-g^{\prime}=f$.
\end{enumerate}
\end{theorem}

To prove Theorem \ref{th:deus}, we first note that the \textquotedblleft
if\textquotedblright\ direction can be verified in a straightforward manner.
Suppose that $f\overset{d}{=}g-g^{\prime}$ for some $g\overset{d}{=}g^{\prime
}$. If $g,g^{\prime}\in L^{1}$ then it is obvious that $\mathbb{E}\left[
g-g^{\prime}\right]  =0$. In general, Simons (1977) showed that $\mathbb{E}%
\left[  g-g^{\prime}\right]  =0$ even if $g,g^{\prime}$ are not in $L^{1}$, as
long as the mean $\mathbb{E}\left[  g-g^{\prime}\right]  $ is well defined,
justified by $f\in L^{k}$. Therefore, $\mathbb{E}\left[  f\right]
=\mathbb{E}\left[  g-g^{\prime}\right]  =0$.

Next, we focus on the more important \textquotedblleft only
if\textquotedblright\ direction of Theorem \ref{th:deus}. For this, we first
prove the case of $L^{\infty}$, and then the case of $L^{k}$, which is much
more technically involved. \bigskip

\noindent\textbf{Proof of Theorem \ref{th:deus} on finite spaces.}
We begin with a finite state space $S=\left\{  1,\dots,n\right\}  $ of
equiprobable states. Let $f:S\rightarrow\mathbb{R}$ have mean $0$, and set
$x_{i}=f\left(  i\right)  $ for each $i=1,\dots,n$. If $f=0$, there is nothing
to prove. Otherwise choose $j_{1}\in\left\{  1,\dots,n\right\}  $ such that
$x_{j_{1}}>0$. Now
\[
\min\left\{  x_{1},\dots,x_{n}\right\}  \leq\sum_{i=1}^{1}x_{j_{i}}\leq
\max\left\{  x_{1},\dots,x_{n}\right\}
\]
Assume for some $1\leq k<n$ to have found distinct $j_{1},j_{2},\dots,j_{k}%
\in\left\{  1,\dots,n\right\}  $ such that
\[
\min\left\{  x_{1},\dots,x_{n}\right\}  \leq\sum_{i=1}^{m}x_{j_{i}}\leq
\max\left\{  x_{1},\dots,x_{n}\right\}  \qquad\forall m=1,\dots,k
\]
We next show that there is $j_{k+1}\in J_{k+1}:=\left\{  1,\dots,n\right\}
\setminus\left\{  j_{1}, \dots,j_{k}\right\}  $ such that
\[
\min\left\{  x_{1},\dots,x_{n}\right\}  \leq\sum_{i=1}^{m}x_{j_{i}}\leq
\max\left\{  x_{1},\dots,x_{n}\right\}  \qquad\forall m=1,\dots,k,k+1
\]

\begin{enumerate}
\item If $x_{j}=0$ for some $j\in J_{k+1}$, set $j_{k+1}=j$.

\item If $\sum_{i=1}^{k}x_{j_{i}}=0$, arbitrarily choose $j_{k+1}\in J_{k+1}$.

\item Else $x_{j}\neq0$ for all $j\in J_{k+1}$ and $\sum_{i=1}^{k}x_{j_{i}
}\neq0$;

\begin{enumerate}
\item if $\sum_{i=1}^{k}x_{j_{i}}>0$, it cannot be the case that $x_{j} \geq0$
for all elements of $J_{k+1}=\left\{  1,\dots,n\right\}  \setminus\left\{
j_{1},\dots,j_{k}\right\}  $, otherwise we would have
\[
0<\sum_{i=1}^{k}x_{j_{i}} \leq\sum_{i=1}^{k}x_{j_{i}}+\sum_{j\in J_{k+1}}
x_{j}=\sum_{j=1}^{n}x_{j}=0
\]
then it is possible to choose $j_{k+1}\in J_{k+1}$ such that $x_{j_{k+1}}<0$,
and
\[
\min\left\{  x_{1},\dots,x_{n}\right\}  \leq x_{j_{k+1}}<\sum_{i=1}
^{k}x_{j_{i}}+x_{j_{k+1}}<\sum_{i=1}^{k}x_{j_{i}}\leq\max\left\{  x_{1}
,\dots,x_{n}\right\}
\]

\item else $\sum_{i=1}^{k}x_{j_{i}}<0$, it cannot be the case that $x_{j}
\leq0$ for all elements of $J_{k+1}=\left\{  1,\dots,n\right\}  \setminus
\left\{  j_{1},\dots,j_{k}\right\}  $, otherwise we would have
\[
0>\sum_{i=1}^{k}x_{j_{i}} \geq\sum_{i=1}^{k}x_{j_{i}}+\sum_{j\in J_{k+1}}
x_{j}=\sum_{j=1}^{n}x_{j}=0
\]
then it is possible to choose $j_{k+1}\in J_{k+1}$ such that $x_{j_{k+1}}>0$,
and
\[
\min\left\{  x_{1},\dots,x_{n}\right\}  \leq\sum_{i=1}^{k}x_{j_{i}}<\sum
_{i=1}^{k}x_{j_{i}}+x_{j_{k+1}}<x_{j_{k+1}}\leq\max\left\{  x_{1},\dots
,x_{n}\right\}
\]

\end{enumerate}
\end{enumerate}

In exactly $n$ steps this produces a rearrangement $(x_{j_{1}},\dots,x_{j_{n}%
})$ of $(x_{1},\dots,x_{n})$, which by construction satisfies
\begin{equation}
\min\left\{  x_{1},\dots,x_{n}\right\}  \leq\sum_{i=1}^{m}x_{j_{i}}\leq
\max\left\{  x_{1},\dots,x_{n}\right\}  \qquad\forall m=1,\dots,n
\label{eq:bds}%
\end{equation}

Define $g,g^{\prime}:S\rightarrow\mathbb{R}$ by $g\left(  j_{k}\right)
=\sum_{i=1}^{k}x_{j_{i}}$ and $g^{\prime}\left(  j_{k}\right)  =\sum
_{i=1}^{k-1}x_{j_{i}}$ for each $1\leq k\leq n$, with the convention
$g^{\prime}\left(  j_{1}\right)  =0$. Diagram $g$ and $g^{\prime}$ as follows:%
\[%
\begin{array}
[c]{cccccc}
& \medskip j_{1} & j_{2} & \cdot\cdot\cdot & j_{n-1} & j_{n}\\
g & \medskip x_{j_{1}} & x_{j_{1}}+x_{j_{2}} & \cdot\cdot\cdot & \sum
_{i=1}^{n-1}x_{j_{i}} & 0=\sum_{i=1}^{n}x_{j_{i}}\\
g^{\prime} & 0 & x_{j_{1}} & \cdot\cdot\cdot & \sum_{i=1}^{n-2}x_{j_{i}} &
\sum_{i=1}^{n-1}x_{j_{i}}%
\end{array}
\]
We have that $g\left(  j_{i}\right)  -g^{\prime}\left(  j_{i}\right)
=f\left(  j_{i}\right)  $ for all $i=1,\dots,n$ and hence%
\[
f=g-g^{\prime}%
\]
and since states are equally probable, $g\overset{d}{=}g^{\prime}$. In view of
(\ref{eq:bds}), we conclude that
\begin{equation}
f=g-g^{\prime}\quad\text{;\quad}g\overset{d}{=}g^{\prime}\quad\text{and\quad
}\min_{S}f\leq g,g^{\prime}\leq\max_{S}f \label{eq:bds-bis}%
\end{equation}
This proves the statement for a finite state space.\bigskip

\noindent\textbf{Proof of Theorem \ref{th:deus} on }$L$\textbf{$^{\infty}$.}
Now, let $S$ be an infinite state space. Let $f\in L^{\infty}$. Choose
$ v\in\mathcal{U}$ such that $f=f_{ v}$, by Lemma \ref{lm:sti},
\[
f_{n}:=\mathbb{E}\left[  f\mid\Sigma_{n}^{ v}\right]  \rightarrow f
\]
both almost surely and in $L^{1}$. Moreover, for all $n\in\mathbb{N}$,
\[
\ell_{f}\leq f\leq u_{f}%
\]
implies%
\[
\ell_{f}=\mathbb{E}\left[  \ell_{f}\mid\Sigma_{n}^{ v}\right]  \leq
f_{n}\leq\mathbb{E}\left[  u_{f}\mid\Sigma_{n}^{ v}\right]  =u_{f}%
\]
%Thus, $\ell_f  \le f_n \leq u_f $.
In view of (\ref{eq:bds-bis}), by choosing the standard versions of the
$f_{n}$, given by%
\[
f_{n}\left(  s\right)  =\frac{1}{2^{n}}\int_{E}f\mathrm{d}P\qquad\forall s\in
E\in\Pi_{n}^{ v}%
\]
there exist two sequences $\{g_{n}\}$ and $\{g_{n}^{\prime}\}$ such that, for
each $n\in\mathbb{N}$,%
\[
g_{n}\overset{d}{=}g_{n}^{\prime}\quad\text{,\quad}\ell_{f}\leq g_{n}%
,g_{n}^{\prime}\leq u_{f}\quad\text{and\quad}g_{n}-g_{n}^{\prime}=f_{n}%
\]
Since $f_{n}\in L^{\infty}$, we have $g_{n},g_{n}^{\prime}\in L^{\infty}$ for
all $n\in\mathbb{N}$. Moreover, by the almost sure convergence of
$f_{n}\ $to $f$, it follows that%
\begin{equation}
g_{n}-g_{n}^{\prime}\overset{d}{\rightarrow}f \label{eq:mg-cvg-bis}%
\end{equation}
Denote by $\mu_{n}$ the joint distribution of $(g_{n},g_{n}^{\prime})$. The
sequence $\{\mu_{n}\}$ is tight since is supported in the compact square
\[
C=[\ell_{f},u_{f}]\times\lbrack\ell_{f},u_{f}]
\]
of $\mathbb{R}^{2}$. By Prohorov's Theorem, there exists a subsequence
$\{\mu_{n_{k}}\}$ that converges weakly to a probability measure $\mu$ on
$\mathbb{R}^{2}$ with support in $C$. As $P$ is nonatomic, by a version of
Skorokhod's Theorem there exists a random vector $(g,g^{\prime}):S\rightarrow
\mathbb{R}^{2}$ with joint distribution $\mu$.\footnote{See, e.g., Theorem 3.1
of Berti, Pratelli, and Rigo (2007).} By the Continuous Mapping
Theorem,\footnote{See, e.g., Theorem 4.27 of Kallenberg (2002).}%
\[
g_{n_{k}}\overset{d}{\rightarrow}g\quad\text{,\quad}g_{n_{k}}^{\prime}%
\overset{d}{\rightarrow}g^{\prime}\quad\text{and\quad}g_{n_{k}}-g_{n_{k}%
}^{\prime}\overset{d}{\rightarrow}g-g^{\prime}%
\]
Since $g_{n_{k}}\overset{d}{=}g_{n_{k}}^{\prime}$ for all $k\geq1$, we have
$g\overset{d}{=}g^{\prime}$. By (\ref{eq:mg-cvg-bis}), we also have $g_{n_{k}%
}-g_{n_{k}}^{\prime}\overset{d}{\rightarrow}f$ and so $g-g^{\prime}\overset
{d}{=}f$. Note that $\ell_{f}\leq g,g^{\prime}\leq u_{f}$ since $\mu$ is
supported in $C$.\hfill$\blacksquare$ \bigskip

\noindent\textbf{Preparation for the proof on }$L^{k}$\textbf{.}
We first present some preliminaries. For $f\in L^{1}$, denote by $\mu
_{f}=P\circ f^{-1}$. Recall that the left quantile $q_{f}^{-}:(0,1)\to
\mathbb{R}$ is defined as $q_{f}^{-}\left(  p\right)  =\inf\left\{
x\in\mathbb{R}:P\left(  f\le x\right)  \geq p\right\}  $.

\begin{figure}[ptb]
\centering	
\begin{tikzpicture}
		\draw[-] (0,-3) -- (0,0) -- (6,0);
		\node at (-0.2,-0.2) {\footnotesize $0$};
		\node at (5.3,-0.2) {\footnotesize $1$};
		
		\draw [black] (0,0) .. controls (1,-1.8) and (1.2,-2.3).. (2,-2.55);
		\draw [black] (3,-2.55) .. controls (3.8,-2.3) and (4.2,-1.8).. (5.2,0);
		\draw [black] (2,-2.55)--(3,-2.55);

		\node at (-0.2,-2.55) {\footnotesize $c$};
		\draw[gray,dotted] (0,-2.55) -- (2,-2.55);
		
		\node at (-0.2,-1.5) {\footnotesize $r$};
		\draw[gray,dotted] (0,-1.5) -- (4.3,-1.5);
		
		\node[above] at (2,0) {\footnotesize $a^-$};
		\draw[gray,dotted] (2,0) -- (2,-2.55);
		
		\node[above] at (3,0) {\footnotesize $a^+$};
		\draw[gray,dotted] (3,0) -- (3,-2.55);
		
		\node[above] at (0.85,0) {\footnotesize $A(r)$};
		\draw[gray,dotted] (0.85,0) -- (0.85,-1.5);
		
		\node[above] at (4.3,0) {\footnotesize $B(r)$};
		\draw[gray,dotted] (4.3,0) -- (4.3,-1.5);
	\end{tikzpicture}
~ \begin{tikzpicture}
		\draw[-] (0,3) -- (0,0) -- (3,0);
		\draw [black] (-4,0)--(0,0);
		\node at (0,-0.2) {\footnotesize $0$};
		\node at (-0.2,2.75) {\footnotesize $1$};
		\node[below] at (-2.5,0)  {\footnotesize $c$};
		
		\node at (-2.5,-0.016) {$\circ$};
		\node at (-2.5,0.8) {$\bullet$};
		\draw [black] (-2.5,0.8) .. controls (-1.5,1.8) and (-1,2.3).. (0,2.6);
		\draw [black] (0,2.6)--(2.5,2.6);
		
		\node at (0.8,0.8) {\footnotesize $a^+-a^-$};
		\draw[gray,dotted] (-2.5,0.8) -- (0,0.8);
		
	\end{tikzpicture}
%			PLACEHOLDER \newline for\newline A TIKZ PICTURE\newline that \newline WILL BE
%			ADDED
\caption{The functions $H$ (left panel) and $K$ (right panel).}%
\label{fig-HK}%
\end{figure}
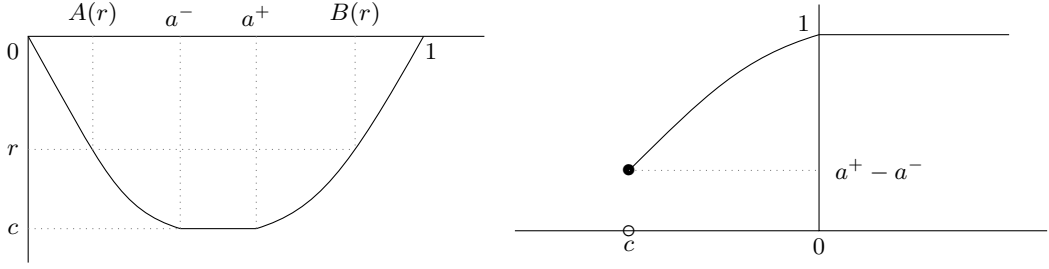

Let $f\in L^{1}$ with $\mathbb{E}\left[  f\right]  =0$. We assume that $f$ is
not constantly $0$. Define
\[
H(t)=\int_{0}^{t}q_{f}^{-}\mathrm{d}\lambda\qquad\forall t\in\lbrack0,1]
\]
and denote by $a^{-}=\mu_{f}((-\infty,0))$ and $a^{+}=\mu_{f}((-\infty,0])$.
It is easy to see that $H$ is strictly decreasing on $[0,a^{-}]$ and strictly
increasing on $[a^{+},1]$, $H(0)=H(1)=0$, and the minimum value of $H$ is
given by $c:=H(a^{-})=H(a^{+})<0$, which is attained by any point in
$[a^{-},a^{+}]$. Moreover, $H$ is convex because $q_{f}$ is increasing, and
hence $H$ is almost everywhere differentiable on $[0,1]$. For $r\in\lbrack
c,0]$, define
\begin{equation}
\label{eq-AB}A(r)=\inf\{t\in\lbrack0,1]:H(t)=r\}\quad\mathrm{and}\quad
B(r)=\sup\{t\in\lbrack0,1]:H(t)=r\}
\end{equation}
Obviously, $A(r)\in\lbrack0,a^{-}]$, $B(r)\in\lbrack a^{+},1]$, $A(c)=a^{-}$,
$B(c)=a^{+}$ and $H\circ A(r)=H\circ B(r)=r$. Moreover, $A(r)$ and $B(r)$ are
also both continuous and strictly monotone as $H$ is so on $[0,a^{-}]$ and
$[a^{+},1]$. Define
\begin{equation}
\label{eq-K}K(r)=%
\begin{cases}
1\quad & r>0\\
B(r)-A(r)\quad & c\leq r\leq0\\
0\quad & r<c
\end{cases}
\end{equation}
One can check that $K$ is right-continuous and increasing, with $\lim
_{x\uparrow c}K(x)=0$, $\lim_{x\downarrow c}K(x)=K(c)=a^{+}-a^{-}$ and
$K(0)=1$. Hence, $K$ is a distribution function on $[c,0]$. More precisely,
$K$ is continuous and strictly increasing on $(c,0]$ and has probability mass
$a^{+}-a^{-}$ at $c$. The functions $H$ and $K$ are plotted in Figure
\ref{fig-HK}.

Define the function $\Phi$ on $[c,0)$ by
\begin{align}
\label{eq-Phi}\Phi= q_{f}^{-}\circ B -q_{f}^{-}\circ A
\end{align}
It is easy to see that $\Phi$ is increasing. Note that $q_{f}^{-}\circ A (r)
\le0 \le q_{f}^{-}\circ B(r)$ for $r\in[c,0]$ with strict inequalities on
$(c,0)$. It holds that $\Phi\ge\max\{|q_{f}^{-}\circ B|,|q_{f}^{-}\circ A|\}$.
The functions $A, B, H, K, \Phi$ have been studied by Wang and Wang (2015) in
a different context. An important technical tool that we will use is Lemma 2.4
of Wang and Wang (2015), which says that $\Phi(w)$ where $w\overset{d}{\sim}K$
is in $L^{k-1}$ (the distribution of $\Phi(w)$ is denoted by $\tilde F$ in
that paper). That is,
\begin{align}
\int_{c}^{0}(\Phi(r))^{k-1}{\mathrm{d}}K(r)<\infty\label{eq:WW15}%
\end{align}

%Next, we aim to construct an increasing sequence of $\{s_n\}\subseteq[c,0]$ such that $q_f^-(B(s_{n+1}))-q_f^-(A(s_{n+1}))\le 2( q_f^-(B(s_n)+)-q_f^-(A(s_n))+)$ for all $n\in\mathbb{N}$ and $\lim_{n\to\infty} s_n=0$.
%let
%$\{A(x_n)\}\subseteq(0,a^-)$ and $\{B(y_n)\}\subseteq (a^+,1)$ be the set of all non differentiable points of $H$ such that $A(x_1)>A(x_2)>\cdots$ and $B(y_1)<B(y_2)<\cdots$, and denote by $A(x_0)=a^-$ and $B(y_0)=a^+$ which implies $x_0=y_0=c$. By the definition of the function $H$, one can check that $q_f^-$ is continuous on $(A(x_{n+1}),A(x_{n})]$ and $(B(y_n),B(y_{n+1})]$ for all $n\ge 0$. Hence, we can construct an increasing sequence of $\{s_n\}$ such that $q_f^-$ is continuous on $(A(s_{n+1}),A(s_{n})]$ and $(B(s_n),B(s_{n+1})]$ and $\lim_{n\to\infty}s_n=0$. More precisely, let $s_0=c$ and $s_1=\min\{x_1,y_1\}$. Then, $s_n$ can be obtained in the following way. When $s_n=\min\{x_{k},y_m\}$, if  $s_n=x_k$, then $s_{n+1}=\min\{x_{k+1},y_m\}$, otherwise $s_{n+1}=\min\{x_k,y_{m+1}\}$.
%If $\{A(x_n)\}$ and $\{B(x_n)\}$ are both infinite,  then the
\bigskip

\noindent\textbf{Proof of Theorem \ref{th:deus} on }$L^{k}$\textbf{.}
Consider any $f\in L^{k}$ with $k\geq1$ which satisfies $\mathbb{E}\left[
f\right]  =0$. The case of $f=0$ is trivial. We assume that $f$ is not
constantly $0$. Choose $v\in\mathcal{U}$ such that $f=f_{v}$. Recall the
functions $A$, $B$ defined by \eqref{eq-AB} and $K$ by \eqref{eq-K}. We have
that $K=B-A$ is continuous and strictly increasing on $(c,0]$ and has
probability mass $a^{+}-a^{-}$ at $c$.
%Note that $A$ and $B$ defined by \eqref{eq-AB} are both continuous and mildly monotone on $[c,0]$ with $A(c)=a^-$, $B(c)=a^+$ and $A(0)=1-B(0)=0$. It holds that $B-A$ is continuous and strictly increasing on $[c,0]$ with $B(c)-A(c)=a^+-a^-$ and $B(0)-A(0)=1$.
Let $r_{1}=c$, and define
\[
r_{n}=\inf\left\{  r>r_{n-1}:K(r)-K(r_{n-1})=\frac{1-(a^{+}-a^{-})}{2^{n-1}%
}\right\}  \qquad\forall n\geq2
\]
It is easy to see that the sequence $\{r_{n}\}$ is contained in $[c,0)$, and
increasing with $K(r_{n})-K(r_{n-1})=(1-a^{+}+a^{-})/{2^{n-1}}$ for $n\geq2$.
Moreover, $r_{n}\rightarrow0$ because $K$ is strictly increasing on $[c,0]$
with $K(0)=1$, and
\[
K(r_{n})=K(r_{1})+\sum_{i=2}^{n}K(r_{i})-K(r_{i-1})=1-\frac{1-a^{+}+a^{-}%
}{2^{n-1}}\rightarrow1
\]
Denote by $T_{n}=[A(r_{n}),B(r_{n})]$. Further, write $T_{0}=\emptyset$,
$T_{\infty}=\lim_{n\rightarrow\infty}T_{n}=(0,1)$. For each $n\in\mathbb{N}$,
define $\mu_{n}$ by%
\[
\mu_{n}(D)=P(f\in D\mid v\in T_{n}\setminus T_{n-1})\qquad\forall
D\in\mathcal{B}(\mathbb{R})
\]
Note that $\{r_{n}\}_{n\in\mathbb{N}}\subseteq\lbrack c,0)$, and $A$ and $B$
are both strictly monotone on $[c,0]$ satisfying $A\leq B$ and $A(0)=1-B(0)=0$%
. It holds that $A(r_{n}),B(r_{n})\in(0,1)$ for all $n\in\mathbb{N}$. Hence,
$\mu_{n}$ is a compactly supported Borel probability measure. Below we will
show $\int_{\mathbb{R}}x\mathrm{d}\mu_{n}(x)=0$ for $n\in\mathbb{N}$, and
\begin{equation}
\mu_{f}=\sum_{n\in\mathbb{N}}(K(r_{n})-K(r_{n-1}))\mu_{n} \label{eq-eqdf}%
\end{equation}
where $r_{0}$ is any number in $(-\infty,c)$ so that $K(r_{0})=0$. To show the
claim, using Lemma \ref{lm:change} and denoting by $m=K(r_{n})-K(r_{n-1})>0$,
we have
\begin{align*}
m\int_{\mathbb{R}}x\mathrm{d}\mu_{n}\left(  x\right)   &  =m\int_{\mathbb{R}%
}x\mathrm{d}\left(  P_{\left\{  v\in T_{n}\setminus T_{n-1}\right\}  }\circ
f_{v}^{-1}\right)  \left(  x\right)  =m\int f_{v}\mathrm{d}P_{\left\{  v\in
T_{n}\setminus T_{n-1}\right\}  }\\
&  =m\int q_{f}^{-}\mathrm{d}\left(  P\circ v^{-1}\right)  _{T_{n}\setminus
T_{n-1}}=m\int q_{f}^{-}\mathrm{d}\lambda_{T_{n}\setminus T_{n-1}}\\
&  ={\int_{T_{n}}q_{f}^{-}\mathrm{d}\lambda-\int_{T_{n-1}}q_{f}^{-}%
\mathrm{d}\lambda}\\
&  ={\int_{0}^{B(r_{n})}q_{f}^{-}\mathrm{d}\lambda-\int_{0}^{A(r_{n})}%
q_{f}^{-}\mathrm{d}\lambda-\int_{0}^{B(r_{n-1})}q_{f}^{-}\mathrm{d}%
\lambda+\int_{0}^{A(r_{n-1})}q_{f}^{-}\mathrm{d}\lambda}\\
&  ={H\circ B(r_{n})-H\circ A(r_{n})-H\circ B(r_{n-1})+H\circ A(r_{n-1})}=0
\end{align*}
where the last step follows from $H\circ A(r)=H\circ B(r)$ for all
$r\in\lbrack c,0]$. This implies that $\int_{\mathbb{R}}x\mathrm{d}\mu
_{n}(x)=0$ for $n\in\mathbb{N}$. To see \eqref{eq-eqdf}, note that $P(v\in
T_{n}\setminus T_{n-1})=\lambda(T_{n}\setminus T_{n-1})=\lambda(T_{n}%
)-\lambda(T_{n-1})=K(r_{n})-K(r_{n-1})$. Hence,
\begin{align*}
\sum_{n\in\mathbb{N}}(K(r_{n})-K(r_{n-1}))\mu_{n}(D)  &  =P\left(  f\in
D,\ v\in\bigcup_{n\in\mathbb{N}}\left(  T_{n}\setminus T_{n-1}\right)  \right)
\\
&  =P\left(  f\in D,\ v\in T_{\infty}\setminus T_{0}\right)  =\mu_{f}(D)
\end{align*}
Therefore, we have verified \eqref{eq-eqdf}.
%conclude that $\int_{\mathbb{R}}%
%x\mathrm{d}\mu_{n}(x)=0$ and @@\eqref{eq-eqdf} it seems to be missing@@ hold.
Take independent random variables $v:S\rightarrow\left(  0,1\right)  $ and
$w:S\rightarrow\mathbb{N}$ with $v\in\mathcal{U}$ and $w$ such that $P\left(
w=n\right)  =K(r_{n})-K(r_{n-1})$ for all $n\in\mathbb{N}$. By the
construction of $\{r_{n}\}$, we have $2P(w=n+1)=P(w=n)$ for all $n\geq2$.
Using the result of Theorem \ref{th:deus} on bounded random variables, on the
space $\left(  S,\sigma\left(  v\right)  ,P\right)  $ there exist identically
distributed random variables $g_{n}$ and $g_{n}^{\prime}$ such that
$g_{n}-g_{n}^{\prime}\overset{d}{\sim}\mu_{n}$ for each $n\in\mathbb{N}$.
Moreover, $q_{f}^{-}\circ A(r_{n})\leq g_{n},g_{n}^{\prime}\leq q_{f}^{-}\circ
B(r_{n})$ as the support of $\mu_{n}$ is contained in $[q_{f}^{-}\circ
A(r_{n}),q_{f}^{-}\circ B(r_{n})]$. Define the random variables $g$ and
$g^{\prime}$ by%
\[
g(s)=g_{w(s)}(s)\quad\text{and}\quad g^{\prime}(s)=g_{w(s)}^{\prime}%
(s)\qquad\forall s\in S
\]
First observe that, for all $D\in\mathcal{B}(\mathbb{R})$,%
\[
\left\{  g\in D\right\}  =\bigcup_{n\in\mathbb{N}}\left\{  g\in
D,\ w=n\right\}  =\bigcup_{n\in\mathbb{N}}\left\{  g_{n}\in D,\ w=n\right\}
\]
This shows that $g$ is measurable. Moreover, since $g_{n}$ and $w$ are
independent we have that, for all $D\in\mathcal{B}(\mathbb{R})$,
\[
\mu_{g}(D)=\sum_{n\in\mathbb{N}}P\left(  w=n\right)  \mu_{g_{n}}(D)=\sum
_{n\in\mathbb{N}}(K(r_{n})-K(r_{n-1}))\mu_{g_{n}}(D)
\]
The same argument for $g^{\prime}$ and the fact that $g_{n}\overset{d}{=}%
g_{n}^{\prime}$ for all $n\in\mathbb{N}$ show that $g\overset{d}{=}g^{\prime}%
$; the same argument for $g-g^{\prime}$ and the fact that $g_{n}-g_{n}%
^{\prime}\overset{d}{\sim}\mu_{n}$ for each $n\in\mathbb{N}$ combining with
\eqref{eq-eqdf} yield $g-g^{\prime}\overset{d}{\sim}\mu_{f}$. It remains to
verify that the constructed $g$ is in $L^{k-1}$. Recall the definition of
$\Phi=q_{f}^{-}\circ B-q_{f}^{-}\circ A$ in \eqref{eq-Phi}. We have
$|g_{n}|\leq\Phi(r_{n})$ because $q_{f}^{-}\circ A(r_{n})\leq g_{n}\leq
q_{f}^{-}\circ B(r_{n})$ and $q_{f}^{-}\circ A\leq0\leq q_{f}^{-}\circ B$.
Using \eqref{eq:WW15}, we obtain
\begin{align*}
\infty>\int_{c}^{0}(\Phi(r))^{k-1}{\mathrm{d}}K(r)  &  \geq\sum_{n\in
\mathbb{N}}\int_{(r_{n},r_{n+1}]}(\Phi(r))^{k-1}{\mathrm{d}}K(r)\\
&  \geq\sum_{n\in\mathbb{N}}P\left(  w=n+1\right)  (\Phi(r_{n}))^{k-1}%
\geq\frac{1}{2}\sum_{n=2}^{\infty}P\left(  w=n\right)  (\Phi(r_{n}))^{k-1}\\
&  \geq\frac{1}{2}\sum_{n=2}^{\infty}P\left(  w=n\right)  \mathbb{E}\left[
|g_{n}|^{k-1}\right]  =\frac{1}{2}\left(  \mathbb{E}\left[  |g|^{k-1}\right]
-K(r_{1})\mathbb{E}\left[  |g_{1}|^{k-1}\right]  \right)
\end{align*}
Noting that $\mathbb{E}\left[  |g_{1}|^{k-1}\right]  <\infty$ as $g_{1}$ is
bounded, we have $\mathbb{E}\left[  |g|^{k-1}\right]  <\infty$. This completes
the proof of the necessity statement. \hfill$\blacksquare$\bigskip

\subsection{Proof of Lemma \ref{lm:deus}}

It is a direct consequence of Theorem \ref{th:deus}, which we proved
above.\hfill$\blacksquare$

\subsection{Proof of Theorem \ref{th:wra}\label{app:wra}}

(i) $\implies$\ (ii). Let $w,f,g\in\mathcal{F}$ with $g\overset{d}{=}f$, if
$f\in\mathcal{I}^{\mathrm{fi}}\left(  w\right)  $, then $f=-w-\pi$ for some
$\pi\in\mathbb{R}$, then%
\[
w+f=-\pi=\mathbb{E}\left[  w\right]  +\mathbb{E}\left[  -w-\pi\right]
=\mathbb{E}\left[  w\right]  +\mathbb{E}\left[  g\right]  =\mathbb{E}\left[
w+g\right]  \succsim w+g
\]
where the third equality follows from $g\overset{d}{=}-w-\pi$, and the final
preference follows from weak risk aversion. Thus propensity to full insurance holds.

(ii) $\implies$ (i). For each $h\in\mathcal{F}$, by Lemma \ref{lm:deus}, there
exist $w,w^{\prime}\in\mathcal{F}$ such that $w\overset{d}{=}w^{\prime}$ and
$h-\mathbb{E}\left[  h\right]  \overset{d}{=}w-w^{\prime}$. Let
$f=-w+\mathbb{E}\left[  h\right]  $ and $g=-w^{\prime}+\mathbb{E}\left[
h\right]  $, clearly $f\overset{d}{=}g$ and $f\in\mathcal{I}^{\mathrm{fi}%
}\left(  w\right)  $. Propensity to full insurance implies that $w+f\succsim
w+g$, which gives $\mathbb{E}\left[  h\right]  =w+f\succsim w+g=w-w^{\prime
}+\mathbb{E}\left[  h\right]  \overset{d}{=}h$. Law invariance of $\succsim$
yields $w-w^{\prime}+\mathbb{E}\left[  h\right]  \sim h$, implying $\mathbb{E}%
\left[  h\right]  \succsim w-w^{\prime}+\mathbb{E}\left[  h\right]  \sim h$
and transitivity implies $\mathbb{E}\left[  h\right]  \succsim h$. Thus weak
risk aversion holds. \hfill$\blacksquare$

{\color{amethyst}
\subsection{On the relation between Lemma \ref{lm:deus} and Theorem \ref{th:wra}} \label{sec:sued}

We have just proved Theorem \ref{th:wra} by means of Lemma \ref{lm:deus}.
Here we show how, if Theorem \ref{th:wra} could be proved \emph{without}
relying on Lemma \ref{lm:deus}, the lemma would actually result as a
 corollary of the theorem.

(i) $\implies$\ (ii).\footnote{Of Lemma \ref{lm:deus}, assuming Theorem
	\ref{th:wra} to be true.} Consider, for each $c\in\mathbb{R}$, the set%
\[
\mathcal{G}_{c}=\{f\in\mathcal{F}:f\overset{d}{=}c+h-h^{\prime}\text{ for some
}h,h^{\prime}\in\mathcal{F}\text{ with }h\overset{d}{=}h^{\prime}%
\}\subseteq\{f\in\mathcal{F}:\mathbb{E}\left[  f\right]  =c\}
\]
Now define a relation $\sim$ on $\mathcal{F}$ by
\[
f\sim g\iff\text{either }f\overset{d}{=}g\text{ or }f,g\in\mathcal{G}%
_{c}\text{ for some }c\in\mathbb{R}%
\]
Clearly $\sim$ is law invariant (and symmetric).

Before proving transitivity note that%
\[
f,g\in\mathcal{G}_{c}\implies\mathbb{E}\left[  f\right]  =c=\mathbb{E}\left[
g\right]
\]
Now let $f_{1}\sim f_{2}$ and $f_{2}\sim f_{3}$, in order to prove $f_{1}\sim
f_{3}$, we consider the following four cases.

\begin{itemize}
	\item If $f_{1}\overset{d}{=}f_{2}$ and $f_{2}\overset{d}{=}f_{3}$, then
	$f_{1}\overset{d}{=}f_{3}$, and so $f_{1}\sim f_{3}$.
	
	\item If $f_{1}\overset{d}{=}f_{2}$ and [not $f_{2}\overset{d}{=}f_{3}$], then
	there exists $c\in\mathbb{R}$ such that $f_{2},f_{3}\in\mathcal{G}_{c}$, that
	is, $f_{2}\overset{d}{=}c+h_{2}-h_{2}^{\prime}$ for some $h_{2},h_{2}^{\prime
	}\in\mathcal{F}$ with $h_{2}\overset{d}{=}h_{2}^{\prime}$, and $f_{3}%
	\overset{d}{=}c+h_{3}-h_{3}^{\prime}$ for some $h_{3},h_{3}^{\prime}%
	\in\mathcal{F}$ with $h_{3}\overset{d}{=}h_{3}^{\prime}$. But
	\[
	f_{1}\overset{d}{=}f_{2}\overset{d}{=}c+h_{2}-h_{2}^{\prime}%
	\]
	thus $f_{1}\in\mathcal{G}_{c}$, and so $f_{1},f_{3}\in\mathcal{G}_{c}$, which implies
	$f_{1}\sim f_{3}$.
	
	\item If [not $f_{1}\overset{d}{=}f_{2}$] and $f_{2}\overset{d}{=}f_{3}$, the conclusion  	$f_{1}\sim f_{3}$ is obtained as in the previous case. 
%	then
%	there exists $c\in\mathbb{R}$ such that $f_{1},f_{2}\in\mathcal{G}_{c}$, that
%	is, $f_{1}\overset{d}{=}c+h_{1}-h_{1}^{\prime}$ for some $h_{1},h_{1}^{\prime
%	}\in\mathcal{F}$ with $h_{1}\overset{d}{=}h_{1}^{\prime}$, and $f_{2}%
%	\overset{d}{=}c+h_{2}-h_{2}^{\prime}$ for some $h_{2},h_{2}^{\prime}%
%	\in\mathcal{F}$ with $h_{2}\overset{d}{=}h_{2}^{\prime}$. But
%	\[
%	f_{3}\overset{d}{=}f_{2}\overset{d}{=}c+h_{2}-h_{2}^{\prime}%
%	\]
%	thus $f_{3}\in\mathcal{G}_{c}$, and so $f_{1},f_{3}\in\mathcal{G}_{c}$, which implies
%	$f_{1}\sim f_{3}$.
	
	\item If [not $f_{1}\overset{d}{=}f_{2}$] and [not $f_{2}\overset{d}{=}f_{3}%
	$], then there exist $c_{12},c_{23}\in\mathbb{R}$ such that $f_{1},f_{2}%
	\in\mathcal{G}_{c_{12}}$ and $f_{2},f_{3}\in\mathcal{G}_{c_{23}}$, but this
	implies $\mathbb{E}\left[  f_{2}\right]  =c_{12}$ and $\mathbb{E}\left[
	f_{2}\right]  =c_{23}$. Therefore, $c_{12}=c_{23}=c$, and  $f_{1},f_{3}%
	\in\mathcal{G}_{c}$ implies $f_{1}\sim f_{3}$.
\end{itemize}

Summing up, $\sim$ is a risk preference (indeed a law invariant equivalence
relation) on $\mathcal{F}$. Next we show that $\sim$ is propense to full
insurance. Take any $w,f,g\in\mathcal{F}$ such that $g\overset{d}{=}f$. If $f$
is a full insurance for $w$, then $f=-w-\pi$ for some $\pi\in\mathbb{R}$. It
follows that:

\begin{itemize}
	\item $w+f\in\mathcal{G}_{-\pi}$, because $w+f=-\pi=-\pi+0-0$ with
	$0\in\mathcal{F}$ and $0\overset{d}{=}0$;
	
	\item $w+g\in\mathcal{G}_{-\pi}$, because $w+g=-f-\pi+g=-\pi+g-f\ $with
	$g,f\in\mathcal{F}$ and $g\overset{d}{=}f$;
\end{itemize}

\noindent therefore (by definition of $\sim$) $w+f\sim w+g$. By Theorem
\ref{th:wra} (that we are \emph{assuming to be true}), $\sim$ is weakly risk averse,
that is, $\mathbb{E}\left[  f\right]  \sim f$ for all $f\in\mathcal{F}$.

We use the latter fact to show that, given any $f\in\mathcal{F}$, if $\mathbb{E}\left[  f\right]  =0$, then
$f\in\mathcal{G}_{0}$, that is, (i) $\implies$\ (ii). If $\mathbb{E}\left[
f\right]  =0$, since $f\sim\mathbb{E}\left[  f\right]  $, then $f\sim0$.

\begin{itemize}
	\item If $f$ is almost surely constant, then $\mathbb{E}\left[  f\right]  =0$
	implies that $f=0$ almost surely, and so $f\overset{d}{=}0=0+0-0$ with
	$0\in\mathcal{F}$ and $0\overset{d}{=}0$, thus $f\in\mathcal{G}_{0}$.
	
	\item Else $f$ is not almost surely constant, and so it cannot be the case
	that $f\overset{d}{=}0$. Then $f\sim0$ implies that there exists
	$c\in\mathbb{R}$ such that $f,0\in\mathcal{G}_{c}$, but as observed, it must
	then be the case that $c=\mathbb{E}\left[  f\right]  =0$, then $f\in
	\mathcal{G}_{0}$.
\end{itemize}

(ii) $\implies$ (i) of Lemma \ref{lm:deus} is trivial. \hfill$\blacksquare$

}

\subsection{Proof of Theorem \ref{th:sra}}

It is a direct consequence of Theorem \ref{th:omni}, which we prove
below.\hfill$\blacksquare$

\subsection{Proof of Proposition \ref{prop:counter}}

The $\subseteq$ inclusion. As observed, $\mathcal{I}^{\mathrm{cs}}\left(  w\right)  $ is the
set of all elements of $\mathcal{F}$ that are counter-monotonic with $w$, that
is, such that%
\[
\left[  f\left(  s\right)  -f\left(  s^{\prime}\right)  \right]  \left[
w\left(  s\right)  -w\left(  s^{\prime}\right)  \right]  \leq0
\]
$P\times P$ almost surely. Thus, by Theorem 2.14 of R\"{u}schendorf
(2013),\footnote{There is a typo in both relation (2.39) and the last line of
the mentioned theorem of R\"{u}schendorf: the inequality on the left-hand side
of the implication should be strict, in both cases.} if $f\in\mathcal{I}%
^{\mathrm{cs}}\left(  w\right)  $, then%
\[
F_{f,w}\leq G
\]
for all joint distributions with marginals $F_{f}$ and $F_{w}$. In particular,
if $g\overset{d}{=}f$, then $F_{f,w}\leq F_{g,w}$ which is equivalent to
$f\trianglerighteq_{w}g$.

The $\supseteq$ inclusion. Assume that $f\trianglerighteq_{w}g$ for all $g\overset{d}{=}f$, that is,
$F_{f,w}\leq F_{g,w}$ for all $g\overset{d}{=}f$. We want to show that $f$ is
counter-monotonic with $w$. By Theorem 3.1 of Puccetti and Wang (2015), it
suffices to show that
\[
F_{f,w}\left(  x,y\right)  \leq\left(  F_{f}\left(  x\right)  +F_{w}\left(
y\right)  -1\right)  ^{+}\qquad\forall(x,y)\in\mathbb{R}^{2}%
\]
since the opposite inequality is true for all joint distributions with
marginals $F_{f}$ and $F_{w}$.

Let $g\in\mathcal{F}$ be such that $g\overset{d}{=}f$ and $g$ is
counter-monotonic with $w$. If $\Sigma$ is generated by a finite partition of
equiprobable cells, then such a $g$ can be constructed by rearranging the
values of $f$ over the cells. Else, we can take $ v\in\mathcal{U}$ such
that a.s.~$w=w_{ v}=F_{w}^{-1}\left(   v\right)  $ and define
$g=F_{f}^{-1}\left(  1- v\right)  $, now $g\overset{d}{=}f$ because
$1- v\in\mathcal{U}$, and it is counter-monotonic with $w$ because%
\[
\left(  w,g\right)  =\left(  F_{w}^{-1}\left(   v\right)  ,F_{f}%
^{-1}\left(  1- v\right)  \right)
\]
$P\times P$ almost surely.

With this, for all $x,y\in\mathbb{R}^{2}$,
\[
F_{f,w}\left(  x,y\right)  \leq F_{g,w}\left(  x,y\right)  =\left(
F_{g}\left(  x\right)  +F_{w}\left(  y\right)  -1\right)  ^{+}=\left(
F_{f}\left(  x\right)  +F_{w}\left(  y\right)  -1\right)  ^{+}%
\]
where the first equality follows from Theorem 3.1 of Puccetti and Wang (2015)
and the counter-monotonicity of $g$ and $w$, the second from the fact that
$g\overset{d}{=}f$.\hfill$\blacksquare$

\subsection{Proof of Theorem \ref{th:omni}}

\label{proof:thomni}(i) $\implies$ (vi). Let $w,f,g\in\mathcal{F}$ with
$g\overset{d}{=}f$ and $f\trianglerighteq_{w}g$. By Lemma \ref{lm:com-cv}, $w+f\geq
_{\mathrm{cv}}w+g$, and strong risk aversion implies $w+f\succsim w+g$. Thus
$\succsim$ is propense to hedging.

(vi) $\implies$ (v). Let $w,f,g\in\mathcal{F}$ with $g\overset{d}{=}f$ and
$f\in\mathcal{I}^{\mathrm{cs}}\left(  w\right)  $, by Proposition
\ref{prop:counter}, it follows that $f\trianglerighteq_{w}g$, and propensity to hedging
implies $w+f\succsim w+g$. Thus $\succsim$ is propense to contingency-schedule insurance.

(v) $\implies$ (iv) because $\mathcal{I}^{\mathrm{is}}\left(  w\right)
\subseteq\mathcal{I}^{\mathrm{cs}}\left(  w\right)  $ for all $w\in
\mathcal{F}$.

(iv) $\implies$ (iii) and (iv) $\implies$ (ii) because $\mathcal{I}%
^{\mathrm{dl}}\left(  w\right)  ,\mathcal{I}^{\mathrm{pr}}\left(  w\right)
\subseteq\mathcal{I}^{\mathrm{is}}\left(  w\right)  $ for all $w\in
\mathcal{F}$.

(iii) $\implies$ (i) and (ii) $\implies$ (i). The case in which $\Sigma$ is
generated by a finite partition of equiprobable events follows from Lemma
\ref{lem:cv}. Now, let $P$ be nonatomic. Let $f,g\in\mathcal{F}$ be such that
$f\geq_{\mathrm{cv}}g$. We want to show that $f\succsim g$. Let $ v
\in\mathcal{U}$. By Lemma \ref{lm:fs}-(i), $f_{ v}\overset{d}{=}f$ and
$g_{ v}\overset{d}{=}g$. Consider the filtration $\left\{  \Sigma
_{n}^{ v}:n\in\mathbb{N}\right\}  $ that we built for Lemma \ref{lm:sti}
and note that
\[
f_{n}:=\mathbb{E}\left[  f_{ v}\mid\Sigma_{n}^{ v}\right]
\rightarrow f_{ v}\quad\text{and\quad}g_{n}:=\mathbb{E}\left[
g_{ v}\mid\Sigma_{n}^{ v}\right]  \rightarrow g_{ v}%
\]
in $\mathcal{L}^{\infty}$ with respect to bounded a.s.~convergence if
$\mathcal{F}=\mathcal{L}^{\infty}$ and in $\mathcal{M}^{\infty}$ if
$\mathcal{F}=\mathcal{M}^{\infty}$. We want to show that, for each
$n\in\mathbb{N}$, $f_{n}\geq_{\mathrm{cv}}g_{n}$. To this end, let $F,G$ and
$F_{n},G_{n}$ be the distribution functions of $f,g$ and $f_{n},g_{n}$,
respectively. Define $\varphi,\gamma:[0,1]\rightarrow\mathbb{R}$ by%
\[
\varphi\left(  p\right)  =\int_{0}^{p}F^{-1}(t)\mathrm{d}\lambda
\quad\text{and}\quad\gamma\left(  p\right)  =\int_{0}^{p}G^{-1}(t)\mathrm{d}%
\lambda
\]
As well-known,\footnote{See, e.g., Theorem 3.A.5 of Shaked and Shanthikumar
(2007).} $f\geq_{\mathrm{cv}}g$ is equivalent to $\varphi\geq\gamma$ with
$\varphi(1)=\gamma(1)$. Arbitrarily choose $n\in\mathbb{N}$ and define
$\varphi_{n}$ and $\gamma_{n}$ in a similar way. Now note that, by Lemma
\ref{lm:sti}, we have $\lambda$-a.s.%
\[
F_{n}^{-1}=\mathbb{E}_{\lambda}\left[  F^{-1}\mid\sigma\left(  \Psi
_{n}\right)  \right]
\]
Therefore, for each $i=1,\dots,2^{n}$,%
\begin{equation}
\varphi_{n}\left(  \frac{i}{2^{n}}\right)  =\int_{0}^{\frac{i}{2^{n}}}%
F_{n}^{-1}(t)\mathrm{d}\lambda=\int_{0}^{\frac{i}{2^{n}}}F^{-1}(t)\mathrm{d}%
\lambda=\varphi\left(  \frac{i}{2^{n}}\right)  \label{eq:fn}%
\end{equation}
A similar argument holds for $g$ and $g_{n}$. Thus,
\[
\varphi\geq\gamma\Longrightarrow\varphi_{n}\left(  \frac{i}{2^{n}}\right)
\geq\gamma_{n}\left(  \frac{i}{2^{n}}\right)  \qquad\forall i=1,\dots,2^{n}%
\]
By definition $\varphi_{n}(0)=\gamma_{n}(0)=0$. The functions $\varphi_{n}$
and $\gamma_{n}$ are absolutely continuous on $\left[  0,1\right]  $.
Moreover, on each segment $[(i-1)/2^{n},i/2^{n}]$, for each $p\in
\lbrack(i-1)/2^{n},i/2^{n}]$, we have%
\begin{align*}
\varphi_{n}\left(  p\right)   &  =\int_{0}^{p}F_{n}^{-1}(t)\mathrm{d}%
\lambda=\int_{0}^{\frac{i-1}{2^{n}}}F_{n}^{-1}(t)\mathrm{d}\lambda+\int
_{\frac{i-1}{2^{n}}}^{p}\underset{=c_{i,n} \: \lambda
\mbox{\scriptsize{-a.s.}}}{\underbrace{F_{n}^{-1}(t)}}\mathrm{d}\lambda\\
&  =\varphi_{n}\left(  \frac{i-1}{2^{n}}\right)  +c_{i,n}\left(  p-\frac
{i-1}{2^{n}}\right)
\end{align*}
because $F_{n}^{-1}(t)$ is $\lambda$-a.s.~constant on $\left(  (i-1)/2^{n}%
,i/2^{n}\right)  $. But then $\varphi_{n}\ $is affine on $[(i-1)/2^{n}%
,i/2^{n}]$, and the same is true for $\gamma_{n}$. Therefore, the inequality
$\varphi_{n}\geq\gamma_{n}$ on the points $\{i/2^{n}:i=0,\dots,2^{n}\}$
implies $\varphi_{n}\geq\gamma_{n}$ on $\left[  0,1\right]  $. As the equality
$\varphi_{n}(1)=\gamma_{n}(1)$ follows from $\varphi(1)=\gamma(1)$, this
proves that $f_{n}\geq_{\mathrm{cv}}g_{n}$ in $\mathcal{L}^{\infty}\left(
S,\Sigma,P\right)  $, but then $f_{n}\geq_{\mathrm{cv}}g_{n}$ in
$\mathcal{L}^{\infty}\left(  S,\Sigma_{n}^{ v},P_{\mid\Sigma
_{n}^{ v}}\right)  $.\footnote{Since $\Sigma_{n}^{ v}$ is finite,
then $\mathcal{L}^{\infty}\left(  S,\Sigma_{n}^{ v},P_{\mid\Sigma
_{n}^{ v}}\right)  =\mathcal{M}^{\infty}\left(  S,\Sigma_{n}^{ v
},P_{\mid\Sigma_{n}^{ v}}\right)  $.} As $n$ was chosen arbitrarily in
$\mathbb{N}$, we conclude that, for each $n\in\mathbb{N}$, $f_{n}%
\geq_{\mathrm{cv}}g_{n}$ in $\mathcal{L}^{\infty}\left(  S,\Sigma
_{n}^{ v},P_{\mid\Sigma_{n}^{ v}}\right)  $. Now the restriction
of $\succsim$ to $\mathcal{L}^{\infty}\left(  S,\Sigma_{n}^{ v}%
,P_{\mid\Sigma_{n}^{ v}}\right)  $ is either propense to
deductible-limit insurance or propense to proportional insurance because
$\succsim$ satisfies either (iii) or (ii) on $\mathcal{F}$, and we can apply
Lemma \ref{lem:cv} to conclude that%
\begin{equation}
f_{n}\succsim g_{n}\qquad\forall n\in\mathbb{N} \label{eq:pref-n}%
\end{equation}
But, as observed, $f_{n}\rightarrow f_{ v}$ and $g_{n}\rightarrow
g_{ v}$, thus the continuity of $\succsim$ guarantees that $f_{ v
}\succsim g_{ v}$, and law invariance delivers $f\succsim g$, as
wanted.\hfill$\blacksquare$ \bigskip

The conclusions of Theorem \ref{th:omni} hold also for risk preferences on
$\mathcal{L}^{p}$ for $p\in\lbrack1,\infty)$ if continuity is formulated with
respect to convergence in $\mathcal{L}^{p}$. This is because in Lemma
\ref{lm:sti}, we proved that the convergence of $f_{n}\rightarrow f_{ v
}$ and $g_{n}\rightarrow g_{ v}$ is in the corresponding sense.

\subsection{Weak monotonicity and weak secularity}

Next we introduce weaker notions of monotonicity and secularity that are sufficient for some of the results that follow. 

\label{app:weak}
\begin{definition}
A risk preference $\succsim$ is:

\begin{itemize}
\item \emph{weakly monotone} when, for all $\eta,\gamma\in\mathbb{R}$,%
\[
\eta>\gamma\implies\eta\succ\gamma
\]

\item \emph{weakly secular} (or \emph{solvable}) when, for all $g\in\mathcal{F}$, there exists
$\gamma\in\mathbb{R}$, such that 
$
g\sim\gamma
$.

\end{itemize}
\end{definition}

%	Note that if $\succsim$ is monotone and $\eta>\gamma$, then $\eta
%	=\gamma+\left(  \eta-\gamma\right)  $ with $\varepsilon=\eta-\gamma>0$,
%	implies $\eta\succ\gamma$, then $\succsim$ is wealky monotone. If $\succsim$
%	is secular, then for all $g\in\mathcal{F}$, there exists $\rho\in\mathbb{R}$,
%	such that%
%	\[
%	g\sim0-\rho=-\rho
%	\]
%	then $\succsim$ is weakly secular.
%	
	As for the interpretation, weak monotonicity just requires that larger sure
	payoffs are preferred to smaller ones, weak secularity that every random
	payoff has a certainty equivalent.

\subsection{Proof of Proposition \ref{prop:rn}}

This proof only requires weak monotonicity.

Clearly, (iv) $\implies$ (iii) $\implies$ (ii) $\implies$ (i). For the sake of
brevity, call (v) the property
\[
f\succsim g\iff\mathbb{E}\left[  f\right]  \geq\mathbb{E}\left[  g\right]
\]
for all $f,g\in\mathcal{F}$. Let $w,f,g\in\mathcal{F}$.

(i) $\implies$ (iv) If $g\overset{d}{=}f$, then $\mathbb{E}\left[  w+f\right]
=\mathbb{E}\left[  w\right]  +\mathbb{E}\left[  f\right]  =\mathbb{E}\left[
w\right]  +\mathbb{E}\left[  g\right]  =\mathbb{E}\left[  w+g\right]  $. Risk
neutrality delivers
\[
w+f\sim\mathbb{E}\left[  w+f\right]  =\mathbb{E}\left[  w+g\right]  \sim w+g
\]
and transitivity implies $w+f\sim w+g$. Thus dependence neutrality holds.

(v) $\implies$ (i) Since $\mathbb{E}\left[  f\right]  =\mathbb{E}\left[
\mathbb{E}\left[  f\right]  \right]  $, condition (v) implies $f\sim
\mathbb{E}\left[  f\right]  $. Thus risk neutrality holds.

(i) $\implies$ (v) If $f\succsim g$, then risk neutrality yields
$\mathbb{E}\left[  f\right]  \sim f\succsim g\sim\mathbb{E}\left[  g\right]
$, and transitivity implies $\mathbb{E}\left[  f\right]  \succsim
\mathbb{E}\left[  g\right]  $. If $\mathbb{E}\left[  f\right]  <\mathbb{E}%
\left[  g\right]  $, weak monotonicity would imply $\mathbb{E}\left[
f\right]  \prec\mathbb{E}\left[  g\right]  $, a contradiction, therefore it
must be the case that $\mathbb{E}\left[  f\right]  \geq\mathbb{E}\left[
g\right]  $. Summing up: $f\succsim g\implies\mathbb{E}\left[  f\right]
\geq\mathbb{E}\left[  g\right]  $.

Conversely, if $\mathbb{E}\left[  f\right]  \geq\mathbb{E}\left[  g\right]  $, then:

\begin{itemize}
\item either $\mathbb{E}\left[  f\right]  =\mathbb{E}\left[  g\right]  $, then
risk neutrality and reflexivity yield $f\sim\mathbb{E}\left[  f\right]
\sim\mathbb{E}\left[  g\right]  \sim g$, and transitivity implies $f\succsim
g$;

\item or $\mathbb{E}\left[  f\right]  >\mathbb{E}\left[  g\right]  $, then
risk neutrality and weak monotonicity yield $f\sim\mathbb{E}\left[  f\right]
\succ\mathbb{E}\left[  g\right]  \sim g$, and transitivity implies $f\succsim
g$.
\end{itemize}

Summing up: $\mathbb{E}\left[  f\right]  \geq\mathbb{E}\left[  g\right]
\implies f\succsim g$. Thus, (v) holds.\hfill$\blacksquare$

\subsection{Proof of Proposition \ref{th:PST}}

This proof only requires weak monotonicity.

Let $w,f,g\in\mathcal{F}$.

(i) $\implies$ (ii). If $f\geq_{\mathrm{fsd}}g$, then $\mathbb{E}\left[
w+f\right]  =\mathbb{E}\left[  w\right]  +\mathbb{E}\left[  f\right]
\geq\mathbb{E}\left[  w\right]  +\mathbb{E}\left[  g\right]  =\mathbb{E}%
\left[  w+g\right]  $, it follows that $\mathbb{E}\left[  w+f\right]
\geq\mathbb{E}\left[  w+g\right]  $ and, by (i), $w+f\succsim w+g$.

(ii) $\implies$ (i). If $f\overset{d}{=}g$, then $f\geq_{\mathrm{fsd}}%
g\geq_{\mathrm{fsd}}f$. By (ii), we have $w+f\succsim w+g\succsim w+f$ and so
$w+f\sim w+g$. Hence, $\succsim$ is dependence neutral, and Proposition
\ref{prop:rn} implies that $\succsim$ admits an expected-value representation.

(i) $\implies$ (iii). Since $\succsim$ is represented by the expected value,
(iii) follows immediately.

(iii) $\implies$ (i) If $f\overset{d}{=}g$, by law invariance, $f\sim g$, by
(iii), $w+f\sim w+g$. Thus, (iii) yields dependence neutrality, and
Proposition \ref{prop:rn} implies that $\succsim$ admits an expected-value representation.

(i) $\implies$ (iv). Since $\succsim$ is represented by the expected value, it
is complete. Also, if $f\succ g$, then $\mathbb{E}\left[  f\right]
>\mathbb{E}\left[  g\right]  $; Theorem 1 of Pomatto, Strack, and Tamuz (2020) implies
that $w+\tilde{f}>_{\mathrm{fsd}}w+\tilde{g}$ for some $w,\tilde{f},\tilde
{g}\in\mathcal{F}$ such that $f\overset{d}{=}\tilde{f}$, $g\overset{d}%
{=}\tilde{g}$, and $w$ is independent of both $\tilde{f}$ and $\tilde{g}$.

(iv) $\implies$ (i). If $f\succ g$, then $w+\tilde{f}>_{\mathrm{fsd}}%
w+\tilde{g}$ for some $w,\tilde{f},\tilde{g}\in\mathcal{F}$ such that
$f\overset{d}{=}\tilde{f}$, $g\overset{d}{=}\tilde{g}$, and $w$ is independent
of both $\tilde{f}$ and $\tilde{g}$. Thus, $f\succ g$ implies $\mathbb{E}%
[w+\tilde{f}]>\mathbb{E}[w+\tilde{g}]$, whence $\mathbb{E}\left[  f\right]
=\mathbb{E}[\tilde{f}]>\mathbb{E}[\tilde{g}]=\mathbb{E}\left[  g\right]  $ and
$\mathbb{E}\left[  f\right]  >\mathbb{E}\left[  g\right]  $. Since $\succsim$
is complete, by contraposition, it follows that $\mathbb{E}\left[  f\right]
\leq\mathbb{E}\left[  g\right]  $ implies $f\precsim g$. In particular,
$\mathbb{E}\left[  f\right]  =\mathbb{E}\left[  g\right]  $ implies $f\sim g$.
Finally, $\mathbb{E}\left[  f\right]  =\mathbb{E}\left[  \mathbb{E}\left[
f\right]  \right]  $ implies $f\sim\mathbb{E}\left[  f\right]  $. Thus risk
neutrality holds. Since $\succsim$ is a (weakly) monotone risk preference, by
Proposition \ref{prop:rn}, it admits an expected-value representation.\hfill
$\blacksquare$

%\newpage

\subsection{Proofs of the results of Section \ref{sect:compa}}
\label{app:compa}
\noindent\textbf{Proof of Lemma \ref{prop:Ya}. }This proof only requires weak
monotonicity and weak secularity.

Let $f,g\in\mathcal{F}$, $\rho_{\mathrm{A}},\rho_{\mathrm{B}},\gamma
\in\mathbb{R}$.

(i) $\implies$ (ii). If $f=\mathbb{E}\left[  g\right]  $, then both
$f-\rho_{\mathrm{A}}$ and $f-\rho_{\mathrm{B}}$ are sure payoffs. Since
$f-\rho_{\mathrm{A}}\succsim_{\mathrm{A}}g$, by (i), $f-\rho_{\mathrm{A}%
}\succsim_{\mathrm{B}}g\sim_{\mathrm{B}}f-\rho_{\mathrm{B}}$. By weak
monotonicity, $\rho_{\mathrm{A}}>\rho_{\mathrm{B}}$ would lead to the
contradiction $f-\rho_{\mathrm{B}}\succ_{\mathrm{B}}f-\rho_{\mathrm{A}}$, then
it must be the case that $\rho_{\mathrm{B}}\geq\rho_{\mathrm{A}}$.

(ii) $\implies$ (i). If $\gamma\succsim_{\mathrm{A}}g$, then $\mathbb{E}%
\left[  g\right]  -\left(  \mathbb{E}\left[  g\right]  -\gamma\right)
\succsim_{\mathrm{A}}g$. Now, let $f=\mathbb{E}\left[  g\right]  $, if
$g\sim_{\mathrm{A}}f-\rho_{\mathrm{A}}$ (and such a $\rho_{\mathrm{A}}$ exists
by weak secularity), we have
\[
f-\left(  \mathbb{E}\left[  g\right]  -\gamma\right)  =\mathbb{E}\left[
g\right]  -\left(  \mathbb{E}\left[  g\right]  -\gamma\right)  \succsim
_{\mathrm{A}}g\sim_{\mathrm{A}}f-\rho_{\mathrm{A}}%
\]
By weak monotonicity, $\mathbb{E}\left[  g\right]  -\gamma>\rho_{\mathrm{A}}$
would lead to the contradiction $f-\rho_{\mathrm{A}}\succ_{\mathrm{A}%
}f-\left(  \mathbb{E}\left[  g\right]  -\gamma\right)  $, then it must be the
case that $\mathbb{E}\left[  g\right]  -\gamma\leq\rho_{\mathrm{A}}$. Now let
$\rho_{\mathrm{B}}$ be such that $g\sim_{\mathrm{B}}f-\rho_{\mathrm{B}}$ (and
such a $\rho_{\mathrm{B}}$ exists by weak secularity). By (ii) and what we
have just observed, $\rho_{\mathrm{B}}\geq\rho_{\mathrm{A}}\geq\mathbb{E}%
\left[  g\right]  -\gamma$, and $f-\left(  \mathbb{E}\left[  g\right]
-\gamma\right)  \geq f-\rho_{\mathrm{B}}$, by weak monotonicity (and
reflexivity for the equality case)%
\[
f-\left(  \mathbb{E}\left[  g\right]  -\gamma\right)  \succsim_{\mathrm{B}%
}f-\rho_{\mathrm{B}}%
\]
but then $\gamma=f-\left(  \mathbb{E}\left[  g\right]  -\gamma\right)
\succsim_{\mathrm{B}}f-\rho_{\mathrm{B}}\sim_{\mathrm{B}}g$, so that
$\gamma\succsim_{\mathrm{B}}g$.

The final part of the statement is a consequence of the fact that
$\mathbb{E}\left[  g\right]  $ dominates any random payoff $g$ according to
$\geq_{\mathrm{cv}}$.\hfill$\blacksquare$

\begin{lemma}
\label{lem:strength}Let $\succsim$ be a monotone and secular risk preference
on $\mathcal{F}$. Then:

\begin{enumerate}
\item for all $f,g\in\mathcal{F}$, $f\succsim g\iff\rho\left(  g,f\right)
\geq0$;

\item the certainty equivalent map $g\mapsto-\rho\left(  g,0\right)  $
represents $\succsim$ on $\mathcal{F}$;

\item if $f,f^{\prime},g,g^{\prime}\in\mathcal{F}$, $f\overset{d}{=}f^{\prime
}$, and $g\overset{d}{=}g^{\prime}$, then $\rho\left(  g,f\right)
=\rho\left(  g^{\prime},f^{\prime}\right)  $.
\end{enumerate}

If moreover $\succsim$ is continuous, then $\rho:\mathcal{F}\times
\mathcal{F}\rightarrow\mathbb{R}$ is (jointly) sequentially continuous.
\end{lemma}

\noindent\textbf{Proof.} Let $f,g\in\mathcal{F}$.

1. By definition of $\rho:\mathcal{F}\times\mathcal{F}\rightarrow\mathbb{R}$,
$g\sim f-\rho\left(  g,f\right)  $. If $f\succsim g$, by transitivity
$f\succsim f-\rho\left(  g,f\right)  $, monotonicity then excludes the case
$\rho\left(  g,f\right)  <0$. Conversely, if $\rho\left(  g,f\right)  \geq0$,
monotonicity and reflexivity imply%
\[
f=\left(  f-\rho\left(  g,f\right)  \right)  +\rho\left(  g,f\right)  \succsim
f-\rho\left(  g,f\right)  +0\sim g
\]
transitivity allows to conclude $f\succsim g$.

2. By definition of $\rho:\mathcal{F}\times\mathcal{F}\rightarrow\mathbb{R}$,
$g\sim0-\rho\left(  g,0\right)  =-\rho\left(  g,0\right)  $, then
$-\rho\left(  g,0\right)  $ is the certainty equivalent of $\mathcal{F}$. With
this, for all $f,g\in\mathcal{F}$%
\[
f\succsim g\iff-\rho\left(  f,0\right)  \succsim-\rho\left(  g,0\right)
\iff-\rho\left(  f,0\right)  \geq-\rho\left(  g,0\right)
\]
where the latter relation follows by monotonicity.

3. Note $f\overset{d}{=}f^{\prime}$ implies $f-\rho\left(  g,f\right)
\overset{d}{=}f^{\prime}-\rho\left(  g,f\right)  $, repeated application of
law invariance yield%
\[
g^{\prime}\sim g\sim f-\rho\left(  g,f\right)  \sim f^{\prime}-\rho\left(
g,f\right)
\]
transitivity and the definition of $\rho$ yield $\rho\left(  g,f\right)
=\rho\left(  g^{\prime},f^{\prime}\right)  $.

Finally, assume that $\succsim$ is continuous.
Next we show
that, if $k\in\mathbb{R}$, $f_{n}\rightarrow f$ in $\mathcal{F}$, $g_{n}\rightarrow g$ in
$\mathcal{F}$, and $\rho\left(  g_{n},f_{n}\right)  \leq k$ (resp.~$\geq k$)
for all $n\in\mathbb{N}$, then $\rho\left(  g,f\right)  \leq k$ (resp.~$\geq
k$). Indeed, for all
$n\in\mathbb{N}$, $\rho\left(  g_{n},f_{n}\right)  \leq k$ implies
$-\rho\left(  g_{n},f_{n}\right)  \geq-k$, by monotonicity,
\[
g_{n}\sim f_{n}-\rho\left(  g_{n},f_{n}\right)  \succsim f_{n}-k
\]
by continuity
\[
g\succsim f-k
\]
but then $f-\rho\left(  g,f\right)  \sim g\succsim f-k$, and monotonicity
again yields $\rho\left(  g,f\right)  \leq k$. Analogously, for all
$n\in\mathbb{N}$, $\rho\left(  g_{n},f_{n}\right)  \geq k$ implies
$\rho\left(  g,f\right)  \geq k$.

Now assume that $f_{n}\rightarrow f$ in $\mathcal{F}$, and $g_{n}\rightarrow
g$ in $\mathcal{F}$, and, per contra $\rho\left(  g_{n},f_{n}\right)
\nrightarrow\rho\left(  g,f\right)  $. Then there exists $\eta>0$ such that
for all $m\in\mathbb{N}$ there exists $n_{m}>m$ such that $\rho\left(
g_{n_{m}},f_{n_{m}}\right)  \notin\left(  \rho\left(  g,f\right)  -\eta
,\rho\left(  g,f\right)  +\eta\right)  $. Therefore there exists a subsequence
$\left\{  \left(  g_{n_{l}},f_{n_{l}}\right)  \right\}  _{l\in\mathbb{N}}$ of
$\left\{  \left(  g_{n},f_{n}\right)  \right\}  _{n\in\mathbb{N}}$ such that
$\rho\left(  g_{n_{l}},f_{n_{l}}\right)  \notin\left(  \rho\left(  g,f\right)
-\eta,\rho\left(  g,f\right)  +\eta\right)  $ for all $l\in\mathbb{N}$. But
then, either $\rho\left(  g_{n_{l}},f_{n_{l}}\right)  \leq\rho\left(
g,f\right)  -\eta$ for infinitely many $l$, or $\rho\left(  g_{n_{l}}%
,f_{n_{l}}\right)  \geq\rho\left(  g,f\right)  +\eta$ for infinitely many
$l\in\mathbb{N}$. In the first case, there exists a subsequence $\left\{
\left(  g_{n_{i}},f_{n_{i}}\right)  \right\}  _{i\in\mathbb{N}}$ of $\left\{
\left(  g_{n_{l}},f_{n_{l}}\right)  \right\}  _{l\in\mathbb{N}}$ such that
$\rho\left(  g_{n_{i}},f_{n_{i}}\right)  \leq\rho\left(  g,f\right)  -\eta$
for all $i\in\mathbb{N}$, and by the previous observation $\rho\left(
g,f\right)  \leq\rho\left(  g,f\right)  -\eta$, a  contradiction. In the second
case, the contradiction $\rho\left(  g,f\right)  \geq\rho\left(  g,f\right)
+\eta$ is obtained. This yields the desired joint sequential continuity.\hfill
$\blacksquare\bigskip$

\noindent\textbf{Proof of Lemma \ref{prop:Ro}.} Let $\mathrm{A}$ be risk
neutral. Note that for $\succsim_{\mathrm{A}}$ the assumptions of monotonicity
and secularity are implied by weak monotonicity. In fact, by Proposition
\ref{prop:rn}, weak monotone and risk neutral risk preferences are represented
by the expected value, so they are monotone. As to secularity, for all
$f,g\in\mathcal{F}$,%
\[
f-\left(  \mathbb{E}\left[  f\right]  -\mathbb{E}\left[  g\right]  \right)
\sim_{\mathrm{A}}\mathbb{E}\left[  f-\left(  \mathbb{E}\left[  f\right]
-\mathbb{E}\left[  g\right]  \right)  \right]  =\mathbb{E}\left[  g\right]
\sim_{\mathrm{A}}g
\]
that is, $\rho_{\mathrm{A}}\left(  g,f\right)  =\mathbb{E}\left[  f\right]
-\mathbb{E}\left[  g\right]  $.

We only prove point 2 because point 1 is well known.

\begin{Hidden}
This proof only requires weak monotonicity of the risk preferences
$\succsim_{\mathrm{A}}$ and $\succsim_{\mathrm{B}}$.

1. Let $\mathrm{B}$ be weakly more risk averse than $\mathrm{A}$. For all
$f\in\mathcal{F}$, since $\mathrm{A}$ is risk neutral, $\mathbb{E}\left[
f\right]  \sim_{\mathrm{A}}f$, since $\mathrm{B}$ is weakly more risk averse
than $\mathrm{A}$, then $\mathbb{E}\left[  f\right]  \succsim_{\mathrm{B}}f$,
and so $\mathrm{B}$ is weakly risk averse.

Conversely, let $\mathrm{B}$ be weakly risk averse. For all $f\in\mathcal{F}$
and $\gamma\in\mathbb{R}$, if $\gamma\succsim_{\mathrm{A}}f$, since
$\mathrm{A}$ is risk neutral, $f\sim_{\mathrm{A}}\mathbb{E}\left[  f\right]
$, and by transitivity, we have $\gamma\succsim_{\mathrm{A}}\mathbb{E}\left[
f\right]  $. Weak monotonicity of $\succsim_{\mathrm{A}}$ implies $\gamma
\geq\mathbb{E}\left[  f\right]  $, monotonicity of $\succsim_{\mathrm{B}}$
implies $\gamma\succsim_{\mathrm{B}}\mathbb{E}\left[  f\right]  $, and since
$\mathrm{B}$ is weakly risk averse, $\mathbb{E}\left[  f\right]
\succsim_{\mathrm{B}}f$. Transitivity allows to conclude $\gamma
\succsim_{\mathrm{B}}f$, and so $\mathrm{B}$ is weakly more risk averse than
$\mathrm{A}$.
\end{Hidden}

2. Let $\mathrm{B}$ be strongly more risk averse than $\mathrm{A}$. If
$f\geq_{\mathrm{cv}}g$, then $\mathbb{E}\left[  f\right]  =\mathbb{E}\left[
g\right]  $. Since $\mathrm{A}$ is risk neutral, as observed, $\rho
_{\mathrm{A}}\left(  g,f\right)  =\mathbb{E}\left[  f\right]  -\mathbb{E}%
\left[  g\right]  =0$. Since $\mathrm{B}$ is strongly more risk averse than
$\mathrm{A}$, then
\[
\rho_{\mathrm{B}}\left(  g,f\right)  \geq\rho_{\mathrm{A}}\left(  g,f\right)
=0
\]
Lemma \ref{lem:strength} yields $f\succsim_{\mathrm{B}}g$, and so $\mathrm{B}$
is strongly risk averse.

Conversely, if $\mathrm{B}$ is strongly risk averse, then
\[
f\geq_{\mathrm{cv}}g\implies f\succsim_{\mathrm{B}}g
\]
Lemma \ref{lem:strength} yields $\rho_{\mathrm{B}}\left(  g,f\right)  \geq0$.
But, as observed, since $\mathrm{A}$ is risk neutral, $\rho_{\mathrm{A}%
}\left(  g,f\right)  =\mathbb{E}\left[  f\right]  -\mathbb{E}\left[  g\right]
=0$, and so $\rho_{\mathrm{B}}\left(  g,f\right)  \geq\rho_{\mathrm{A}}\left(
g,f\right)  $ which shows that $\mathrm{B}$ is strongly more risk averse than
$\mathrm{A}$.\hfill$\blacksquare\bigskip$

\noindent\textbf{Proof of Theorem \ref{th:comp-wra}}. This proof only requires
weak monotonicity and weak secularity.

(i) $\implies$ (ii). Let $w,f,g\in\mathcal{F}$, with $f\overset{d}{=}g$, if
$f\in\mathcal{I}^{\mathrm{fi}}\left(  w\right)  $, then
\[
\left(  w+f\right)  -\rho_{\mathrm{A}}\left(  w+g,w+f\right)  \sim
_{\mathrm{A}}w+g\quad\text{and\quad}\left(  w+f\right)  -\rho_{\mathrm{B}%
}\left(  w+g,w+f\right)  \sim_{\mathrm{B}}w+g
\]
but $\gamma=\left(  w+f\right)  -\rho_{\mathrm{A}}\left(  w+g,w+f\right)
\in\mathbb{R}$, because $f\in\mathcal{I}^{\mathrm{fi}}\left(  w\right)  $. By
(i), $\left(  w+f\right)  -\rho_{\mathrm{A}}\left(  w+g,w+f\right)
\succsim_{\mathrm{B}}w+g\sim_{\mathrm{B}}\left(  w+f\right)  -\rho
_{\mathrm{B}}\left(  w+g,w+f\right)  $, by weak monotonicity, $\rho
_{\mathrm{A}}\left(  w+g,w+f\right)  \leq\rho_{\mathrm{B}}\left(
w+g,w+f\right)  $. This shows that $\mathrm{B}$ is more propense to full
insurance than $\mathrm{A}$.

(ii) $\implies$ (i). Let $h\in\mathcal{F}$ and $\gamma\in\mathbb{R}$ be such that
$\gamma\succsim_{\mathrm{A}}h$. By weak secularity there exists $\eta
\in\mathbb{R}$ such that $\gamma\succsim_{\mathrm{A}}h\sim_{\mathrm{A}}\eta$,
and by weak monotonicity $\gamma\geq\eta$. By Lemma \ref{lm:deus}, there exist
$w,w^{\prime}\in\mathcal{F}$ such that $w\overset{d}{=}w^{\prime}$ and
$h-\mathbb{E}\left[  h\right]  \overset{d}{=}w-w^{\prime}$. Let
$f=-w+\mathbb{E}\left[  h\right]  $ and $g=-w^{\prime}+\mathbb{E}\left[
h\right]  $, clearly $f\overset{d}{=}g$ and $f\in\mathcal{I}^{\mathrm{fi}%
}\left(  w\right)  $. By (ii), $\rho_{\mathrm{A}}\left(  w+g,w+f\right)
\leq\rho_{\mathrm{B}}\left(  w+g,w+f\right)  $, and by definition of $\rho$,%
\[
\underset{=\mathbb{E}\left[  h\right]  -\rho_{\mathrm{A}}\left(
w+g,w+f\right)  }{\underbrace{\left(  w+f\right)  -\rho_{\mathrm{A}}\left(
w+g,w+f\right)  }}\sim_{\mathrm{A}}\underset{\overset{d}{=}h}{\underbrace
{w+g}}\quad\text{and\quad}\left(  w+f\right)  -\rho_{\mathrm{B}}\left(
w+g,w+f\right)  \sim_{\mathrm{B}}w+g
\]
law invariance yields $\left(  w+f\right)  -\rho_{\mathrm{A}}\left(
w+g,w+f\right)  \sim_{\mathrm{A}}h$, but since $\left(  w+f\right)
-\rho_{\mathrm{A}}\left(  w+g,w+f\right)  $ is constant, then%
\[
\eta=\left(  w+f\right)  -\rho_{\mathrm{A}}\left(  w+g,w+f\right)  \geq\left(
w+f\right)  -\rho_{\mathrm{B}}\left(  w+g,w+f\right)  \sim_{\mathrm{B}%
}w+g\overset{d}{=}h
\]
Weak monotonicity and law invariance yield $\eta\succsim_{\mathrm{B}}h$, and
weak monotonicity again yields $\gamma\succsim_{\mathrm{B}}h$. This shows that
$\mathrm{B}$ is weakly more risk averse than $\mathrm{A}$. \hfill
$\blacksquare$\bigskip

In the following two lemmas, analogous to those of Appendix \ref{sec:mps}, we assume that
$\Sigma$ is generated by a partition $\mathcal{S}$ of equiprobable events
(called cells), and we fix two continuous, monotone, and secular risk
preferences $\succsim_{\mathrm{A}}$ and $\succsim_{\mathrm{B}}$ on
$\mathcal{F}$.

\begin{lemma}
Let $f,g\in\mathcal{F}$ be such that $g$ is a mean preserving spread of $f$.
If either (ii) or (iii) of Theorem \ref{th:omni-comp} holds, then
$\rho_{\mathrm{B}}(g,f)\geq\rho_{\mathrm{A}}(g,f)$.
\end{lemma}

\noindent\textbf{Proof.} When (iii) of Theorem \ref{th:omni-comp} holds, the
results follows immediately from Lemma \ref{lm-MPSdl}. Suppose now that (ii)
of Theorem \ref{th:omni-comp} holds. Let
\[
g=f-\delta1_{S_{1}}+\delta1_{S_{2}}%
\]
with $\delta\geq0$ and $S_{1}, S_{2}$ two distinct cells  in
$\mathcal{S}$ such that $f\left(  S_{1}\right)  \leq f\left(  S_{2}\right)  $. If
$\delta=0$, then $f=g$, and $\rho_{\mathrm{A}}(g,f)=\rho_{\mathrm{B}}(g,f)=0$.
If $\delta>0$ and $f(S_{1})<f(S_{2})$, it follows from Lemma \ref{lm-MPSpr}
that $\rho_{\mathrm{B}}(g,f)\geq\rho_{\mathrm{A}}(g,f)$. If $\delta>0$ and
$f(S_{1})=f(S_{2})$, define $f_{\varepsilon}=f-\varepsilon1_{S_{1}%
}+\varepsilon1_{S_{2}}$ with $\varepsilon\in(0,\delta)$. Note that %$f_{\varepsilon}     \left(  S_{1}\right)  <f_{\varepsilon}\left(  S_{2}\right)$ and
\begin{align*}
f_{\varepsilon}  &  =f-\varepsilon1_{S_{1}}+\varepsilon1_{S_{2}}%
=f1_{S\setminus\{S_{1},S_{2}\}}+\left(  f(S_{1})-\varepsilon\right)  1_{S_{1}%
}+\left(  f(S_{2})+\varepsilon\right)  1_{S_{2}} \\g  &  =f-\delta1_{S_{1}}+\delta1_{S_{2}}=f-\left(  \varepsilon+\left(
\delta-\varepsilon\right)  \right)  1_{S_{1}}+\left(  \varepsilon+\left(
\delta-\varepsilon\right)  \right)  1_{S_{2}}  =f_{\varepsilon}-\left(  \delta-\varepsilon\right)  1_{S_{1}}+\left(
\delta-\varepsilon\right)  1_{S_{2}}%
\end{align*}
Thus $g$ is a mean preserving spread of $f_{\varepsilon}$ with $f_{\varepsilon
}(S_{1})<f_{\varepsilon}(S_{2})$ and $\delta-\varepsilon>0$. By the previous
argument $\rho_{\mathrm{B}}(g,f_{\varepsilon})\geq\rho_{\mathrm{A}%
}(g,f_{\varepsilon})$ for all $\varepsilon\in(0,\delta)$. Let $\{\varepsilon
_{n}\}_{n\in\mathbb{N}}\subseteq(0,\delta)$ be such that $\lim_{n\rightarrow
\infty}\varepsilon_{n}=0$. By Lemma \ref{lem:strength} and continuity of both
$\succsim_{\mathrm{A}}$ and $\succsim_{\mathrm{B}}$, it follows that
$\rho_{\mathrm{A}}(g,f_{\varepsilon_{n}})\rightarrow\rho_{\mathrm{A}}(g,f)$
and $\rho_{\mathrm{B}}(g,f_{\varepsilon_{n}})\rightarrow\rho_{\mathrm{B}%
}(g,f)$, and so $\rho_{\mathrm{B}}(g,f)\geq\rho_{\mathrm{A}}(g,f)$. This
completes the proof. \hfill$\blacksquare$

\begin{lemma}
\label{lm-comfinte}Let $f,g\in\mathcal{F}$ be such that $f\geq_{\mathrm{cv}}%
g$. If either (ii) or (iii) of Theorem \ref{th:omni-comp} holds, then
$\rho_{\mathrm{B}}(g,f)\geq\rho_{\mathrm{A}}(g,f)$.
\end{lemma}

\noindent\textbf{Proof.} If $f\geq_{\mathrm{cv}}g$ in $\mathcal{F}$, then
there exists a sequence $h_{0},h_{1},\dots,h_{m}$ such that $f=h_{0}$,
$g=h_{m}$ and each $h_{k+1}$ is either a mean preserving spread of $h_{k}$ or
it is obtained by $h_{k}$ through the permutation of the values that $h_{k}$
takes on two cells. By the previous lemma, we have $\rho_{\mathrm{B}}%
(h_{k+1}-x,h_{k}-x)\geq\rho_{\mathrm{A}}(h_{k+1}-x,h_{k}-x)$ for all
$x\in\mathbb{R}$ and $k=0,1,\dots,m-1$ as either $h_{k+1}-x$ is  a mean preserving
spread of $h_{k}-x$ or $h_{k+1}-x\overset{d}{=}h_{k}-x$. Next, we prove by induction that, for all $x\in\mathbb{R}$ and $j=1,2,\dots,m$,
\[
\rho_{\mathrm{B}}(h_{j}-x,h_{0}-x)\geq\rho_{\mathrm{A}}(h_{j}-x,h_{0}-x)
\]
As we have just observed, for $j=1$, we have $\rho_{\mathrm{B}}(h_{1}%
-x,h_{0}-x)\geq\rho_{\mathrm{A}}(h_{1}-x,h_{0}-x)$ for all $x\in\mathbb{R}$.
Suppose that, for $j=k$, $\rho_{\mathrm{B}}(h_{k}-x,h_{0}-x)\geq\rho_{\mathrm{A}}%
(h_{k}-x,h_{0}-x)$ for all $x\in\mathbb{R}$; it then
suffices to verify that $\rho_{\mathrm{B}}(h_{k+1}-x,h_{0}-x)\geq
\rho_{\mathrm{A}}(h_{k+1}-x,h_{0}-x)$ for all $x\in\mathbb{R}$. To see this,
denote by $\eta_{\mathrm{A}}=\rho_{\mathrm{A}}(h_{k+1}-x,h_{k}-x)$ and
$\eta_{\mathrm{B}}=\rho_{\mathrm{B}}(h_{k+1}-x,h_{k}-x)$. It holds that
\[
h_{k}-x-\eta_{\mathrm{A}}\sim_{\mathrm{A}}h_{k+1}-x\quad\text{and}\quad
h_{k}-x-\eta_{\mathrm{B}}\sim_{\mathrm{B}}h_{k+1}-x
\]
As we have observed above, $\eta_{\mathrm{B}}\geq\eta_{\mathrm{A}}$ and since
$\succsim_{\mathrm{A}}$ is monotone we have $h_{k}-x-\eta_{\mathrm{B}}%
\precsim_{\text{\textrm{A}}}h_{k+1}-x$. Therefore
\[
h_{0}-x-\eta_{\mathrm{B}}-\rho_{\mathrm{A}}(h_{k}-x-\eta_{\mathrm{B}}%
,h_{0}-x-\eta_{\mathrm{B}})\sim_{\mathrm{A}}h_{k}-x-\eta_{\mathrm{B}}%
\precsim_{\text{\textrm{A}}}h_{k+1}-x\sim_{\mathrm{A}}h_{0}-x-\rho
_{\mathrm{A}}(h_{k+1}-x,h_{0}-x)
\]
and, by monotonicity, $\rho_{\mathrm{A}}(h_{k+1}-x,h_{0}-x)\leq\eta
_{\mathrm{B}}+\rho_{\mathrm{A}}(h_{k}-x-\eta_{\mathrm{B}},h_{0}-x-\eta
_{\mathrm{B}})$. Moreover,%
\[
h_{0}-x-\eta_{\mathrm{B}}-\rho_{\mathrm{B}}(h_{k}-x-\eta_{\mathrm{B}}%
,h_{0}-x-\eta_{\mathrm{B}})\sim_{\mathrm{B}}h_{k}-x-\eta_{\mathrm{B}}%
\sim_{\mathrm{B}}h_{k+1}-x
\]
and so $\rho_{\mathrm{B}}(h_{k+1}-x,h_{0}-x)=\eta_{\mathrm{B}}+\rho
_{\mathrm{B}}(h_{k}-x-\eta_{\mathrm{B}},h_{0}-x-\eta_{\mathrm{B}})$. By
induction $\rho_{\mathrm{B}}(h_{k}-x-\eta_{\mathrm{B}},h_{0}-x-\eta
_{\mathrm{B}})\geq\rho_{\mathrm{A}}(h_{k}-x-\eta_{\mathrm{B}},h_{0}%
-x-\eta_{\mathrm{B}})$, and so%
\begin{align*}
\rho_{\mathrm{A}}(h_{k+1}-x,h_{0}-x)  &  \leq\eta_{\mathrm{B}}+\rho
_{\mathrm{A}}(h_{k}-x-\eta_{\mathrm{B}},h_{0}-x-\eta_{\mathrm{B}})\\
&  \leq\eta_{\mathrm{B}}+\rho_{\mathrm{B}}(h_{k}-x-\eta_{\mathrm{B}}%
,h_{0}-x-\eta_{\mathrm{B}})\\
&  =\rho_{\mathrm{B}}(h_{k+1}-x,h_{0}-x)
\end{align*}
as wanted. \hfill$\blacksquare\bigskip$

\noindent\textbf{Proof of Theorem \ref{th:omni-comp}}. (i) $\implies$ (vi).
Let $w,f,g\in\mathcal{F}$ with $g\overset{d}{=}f$ and $f\trianglerighteq_{w}g$. By Lemma
\ref{lm:com-cv}, $w+f\geq_{\mathrm{cv}}w+g$, and (i) implies $\rho
_{\mathrm{B}}(w+g,w+f)\geq\rho_{\mathrm{A}}(w+g,w+f)$. Thus (vi) holds.

(vi) $\implies$ (v). Let $w,f,g\in\mathcal{F}$ with $g\overset{d}{=}f$ and
$f\in\mathcal{I}^{\mathrm{cs}}\left(  w\right)  $, by Proposition
\ref{prop:counter}, it follows that $f\trianglerighteq_{w}g$, and (vi) implies
$\rho_{\mathrm{B}}(w+g,w+f)\geq\rho_{\mathrm{A}}(w+g,w+f)$. Thus (v) holds.

(v) $\implies$ (iv) because $\mathcal{I}^{\mathrm{is}}\left(  w\right)
\subseteq\mathcal{I}^{\mathrm{cs}}\left(  w\right)  $ for all $w\in
\mathcal{F}$.

(iv) $\implies$ (iii) and (iv) $\implies$ (ii) because $\mathcal{I}%
^{\mathrm{dl}}\left(  w\right)  ,\mathcal{I}^{\mathrm{pr}}\left(  w\right)
\subseteq\mathcal{I}^{\mathrm{is}}\left(  w\right)  $ for all $w\in
\mathcal{F}$.

(iii) $\implies$ (i) and (ii) $\implies$ (i). The case in which $\Sigma$ is
generated by a finite partition of equiprobable events follows from Lemma
\ref{lm-comfinte}. Now, let $P$ be nonatomic. Let $f,g\in\mathcal{F}$ be such
that $f\geq_{\mathrm{cv}}g$. We want to show that $\rho_{\mathrm{B}}%
(g,f)\geq\rho_{\mathrm{A}}(g,f)$.

The sequences $\{f_{n}\}_{n\in\mathbb{N}}$ and $\{g_{n}\}_{n\in\mathbb{N}}$
 introduced in the
proof of Theorem \ref{th:omni} in Appendix \ref{proof:thomni} have the following properties:

\begin{itemize}
\item $f_{n},g_{n}\in\mathcal{L}^{\infty}\left(  S,\Sigma_{n}^{ v
},P_{\mid\Sigma_{n}^{ v}}\right)  $ where $\left\{  \Sigma_{n}%
^{ v}:n\in\mathbb{N}\right\}  $ is the filtration that we built for Lemma
\ref{lm:sti};

\item $f_{n}\geq_{\mathrm{cv}} g_{n}$ for all $n\in\mathbb{N}$;

\item $f_{n}\rightarrow f_{ v}$ and $g_{n}\rightarrow g_{ v}$
in $\mathcal{F}$, with $f_{ v}\overset{d}{=}f$ and $g_{ v}\overset{d}{=}g$.
\end{itemize}

The restrictions of $\succsim_{\mathrm{A}}$ and $\succsim_{\mathrm{B}}$ to
$\mathcal{L}^{\infty}\left(  S,\Sigma_{n}^{ v},P_{\mid\Sigma
_{n}^{ v}}\right)  $ are continuous, monotone, and secular risk preferences that
 satisfy either (ii) or (iii) in this theorem, and
we can apply Lemma \ref{lm-comfinte} to conclude  
\begin{equation}
\rho_{\mathrm{B}}(g_{n},f_{n})\geq\rho_{\mathrm{A}}(g_{n},f_{n})\qquad\forall
n\in\mathbb{N} \label{eq:comfin}%
\end{equation}
but, by Lemma \ref{lem:strength}, both $\rho_{\mathrm{A}}$ and $\rho
_{\mathrm{B}}$ are law invariant and continuous, and hence
\[
\rho_{\mathrm{B}}(g,f)=\rho_{\mathrm{B}}(g_{ v},f_{ v})\geq
\rho_{\mathrm{A}}(g_{ v},f_{ v})=\rho_{\mathrm{A}}(g,f)
\]
as wanted. \hfill$\blacksquare$

\begin{proposition}
\label{prop:Ro-insurance}Let $\succsim_{\mathrm{A}}$ and $\succsim
_{\mathrm{B}}$ be monotone and secular risk preferences.

\begin{enumerate}
\item If $\succsim_{\mathrm{A}}$ is neutral to full insurance, then
$\mathrm{B}$ is more propense to full insurance than $\mathrm{A}$ if and only
if $\mathrm{B}$ is propense to full insurance.

\item If $\succsim_{\mathrm{A}}$ is neutral to hedging, then $\mathrm{B}$ is
more propense to hedging than $\mathrm{A}$ if and only if $\mathrm{B}$ is
propense to hedging.
\end{enumerate}
\end{proposition}

\noindent\textbf{Proof.} Note that by Proposition \ref{prop:rn}, $\mathrm{A}$
is neutral to full insurance if and only if she is neutral to hedging if and
only if she is risk neutral.

1. By Theorem \ref{th:comp-wra}, $\mathrm{B}$ is more propense to full
insurance than $\mathrm{A}$ if and only if $\mathrm{B}$ is weakly more risk
averse than $\mathrm{A}$. By Lemma \ref{prop:Ro}, $\mathrm{B}$ is weakly
more risk averse than $\mathrm{A}$ if and only if $\mathrm{B}$ is weakly risk
averse. By Theorem \ref{th:wra}, $\mathrm{B}$ is weakly risk averse if and
only if $\mathrm{B}$ is propense to full insurance.

2. Let $w,f,g\in\mathcal{F}$ with $g\overset{d}{=}f$. Assume that $\mathrm{B}$
is more propense to hedging than $\mathrm{A}$. If $f\trianglerighteq_{w}g$, then
$\mathbb{E}\left[  w+f\right]  =\mathbb{E}\left[  w+g\right]  $ (because
$f\overset{d}{=}g$). Since $\mathrm{A}$ is risk neutral, as observed in the
proof of Lemma \ref{prop:Ro}, $\rho_{\mathrm{A}}\left(  w+g,w+f\right)
=\mathbb{E}\left[  w+f\right]  -\mathbb{E}\left[  w+g\right]  =0$. Since
$\mathrm{B}$ is more propense to hedging than $\mathrm{A}$, then
\[
\rho_{\mathrm{B}}\left(  w+g,w+f\right)  \geq\rho_{\mathrm{A}}\left(
w+g,w+f\right)  =0
\]
Lemma \ref{lem:strength} yields $w+f\succsim_{\mathrm{B}}w+g$, and so
$\mathrm{B}$ is propense to hedging.

Conversely, if $\mathrm{B}$ is propense to hedging, then
\[
f\trianglerighteq_{w}g\implies w+f\succsim_{\mathrm{B}}w+g
\]
Lemma \ref{lem:strength} yields $\rho_{\mathrm{B}}\left(  w+g,w+f\right)
\geq0$. But, as observed in the proof of Lemma \ref{prop:Ro}, since
$\mathrm{A}$ is risk neutral, $\rho_{\mathrm{A}}\left(  w+g,w+f\right)
=\mathbb{E}\left[  w+f\right]  -\mathbb{E}\left[  w+g\right]  =0$, and so
$\rho_{\mathrm{B}}\left(  w+g,w+f\right)  \geq\rho_{\mathrm{A}}\left(
w+g,w+f\right)  $, which shows that $\mathrm{B}$ is more propense to hedging
than $\mathrm{A}$.\hfill$\blacksquare$

{\color{amethyst}
	
\section{Additional results and considerations}\label{app:miscellanea}

\subsection{Total wealth and wealth changes} \label{app:vs}
In choice under risk, to each risk preference $\succsim$ on $\mathcal{F}$ and
each initial wealth $w_{0}\in\mathbb{R}$, another risk preference
\begin{equation}
	f\succsim^{w_{0}}g\iff w_{0}+f\succsim w_{0}+g\label{eq:cons}%
\end{equation}
is associated. In this perspective, random payoffs are
interpreted as risks -- that is, changes in wealth -- relative to an initial
wealth $w_{0}$. Accordingly, the ranking%
\[
f\succsim^{w_{0}}g
\]
is interpreted as `$f$ is preferred to $g$, given $w_{0}$'. Obviously, the risk preference $\succsim^{0}$, corresponding to $w_{0}=0$,
is nothing but $\succsim$ itself. This appendix
shows that the study of risk attitudes -- in its traditional form as well as
in the insurance-based one of the current paper -- is independent of whether
we consider either a preference relation $\succsim$ over random final wealth
levels or any preference relation $\succsim^{w_{0}}$ over risks.

\begin{proposition}
	\label{prop:w-change}The following properties are equivalent for a risk
	preference $\succsim$:
	
	\begin{enumerate}
		\item[(i)] $\succsim$ is propense to full insurance (weakly risk averse);
		
		\item[(ii)] for some $w_{0}\in\mathbb{R}$, $\succsim^{w_{0}}$ is propense to
		full insurance (weakly risk averse);
		
		\item[(iii)] for every $w_{0}\in\mathbb{R}$, $\succsim^{w_{0}}$ is propense
		to full insurance (weakly risk averse).
	\end{enumerate}
\end{proposition}

\noindent\textbf{Proof.} We only prove that (ii) $\implies$ (iii), the rest
being obvious. Assume that (ii) holds for a given $w_{0}$ and arbitrarily
choose $w_{0}^{\prime}\in\mathbb{R}$. For all $w,f,g\in\mathcal{F}$ with
$g\overset{d}{=}f$, if $f$ is full insurance for $w$, then $f=-w-\pi$ for some
$\pi\in\mathbb{R}$. But then $f$ is full insurance also for $y=w+w_{0}%
^{\prime}-w_{0}$, in fact $f=-\left(  w+w_{0}^{\prime}-w_{0}\right)  -\left(
\pi-w_{0}^{\prime}+w_{0}\right)  $. Since $\succsim^{w_{0}}$ is propense to
full insurance, then $y+f\succsim^{w_{0}}y+g$, explicitly%
\[
w_{0}+\underset{y}{\underbrace{w+w_{0}^{\prime}-w_{0}}}+f\succsim
w_{0}+\underset{y}{\underbrace{w+w_{0}^{\prime}-w_{0}}}+g
\]
and so $w_{0}^{\prime}+\left(  w+f\right)  \succsim w_{0}^{\prime}+\left(
w+g\right)  $, that is, $w+f\succsim^{w_{0}^{\prime}}w+g$, as wanted.\hfill
$\blacksquare$\bigskip

In words, when a preference relation is propense to full insurance (weakly
risk averse) at some initial wealth level, it remains so at any other initial
level. Intuitively, propensity to full insurance (weak risk aversion) per se
is a feature of a preference relation that depends only on the variability of
payoffs and as such it is unaffected by the addition of a constant (i.e., by
an initial wealth $w_{0}$). In contrast, the degree of propensity to full
insurance (weak risk aversion)\emph{\ }may well change with the initial wealth
level as risk preferences are, in general, not invariant under the addition of
constants (in the jargon, they are not translation invariant).

This intuition is confirmed by the main idea of the proof above: $f$ is full
insurance for a risk $w$ if and only if it is full insurance for the corresponding final
wealth $w_{0}+w$, i.e., $\mathcal{I}^{\mathrm{fi}}\left(  w\right)
=\mathcal{I}^{\mathrm{fi}}\left(  w_{0}+w\right)  $. This invariance is easily
seen to hold for partial insurance as well, that is, $\mathcal{I}%
^{\mathrm{pi}}\left(  w\right)  =\mathcal{I}^{\mathrm{pi}}\left(
w_{0}+w\right)  $ for each $\mathrm{pi}\in\left\{  \mathrm{pr},\mathrm{dl}%
,\mathrm{is},\mathrm{cs}\right\}  $. Accordingly the last proposition
continues to hold with `partial' and `strongly' in place of `full' and `weakly'.	

%\newpage

\subsection{A generalization of Theorem \ref{th:wra}\label{sec-gen-th1}}

A possible issue in testing empirically our notion of propensity to full
insurance is the universal quantification `for all $\pi\in\mathbb{R}$'
regarding insurance prices. Typically, in insurance markets there is only a
finite number of insurance providers, each adopting a premium principle $\Pi$
that associates a premium $\pi=\Pi\left(  h\right)  $ to each insurance payoff
$h$ in $\mathcal{F}$. In this appendix, we address this issue by providing a
weaker version of our notion that still characterizes weak risk aversion. To
this end, next we introduce a general class of pricing rules for insurance markets.

\begin{definition}
	A function $\Pi:\mathcal{F}\rightarrow\mathbb{R}$ is a\emph{ premium
		calculation principle} when there exists $\theta>0$ such that
	\[
	\Pi\left(  h+\gamma\right)  =\Pi\left(  h\right)  +\theta\gamma
	\]
	for all $h\in\mathcal{F}$ and all $\gamma\in\mathbb{R}$.
\end{definition}

This notion includes most pricing rules used in the insurance industry,
such as those presented by Dickson (2017, Chapter 3), and in particular
the fair premium principle for which $\Pi$ is the expected value.

With a prespecified premium calculation principle $\Pi$ replacing the
arbitrary premium $\pi\in\mathbb{R}$, propensity to full insurance takes the
following weaker form.

\begin{definition}
	\label{def:iiciente}A risk preference $\succsim$ is \emph{propense to full
		insurance at price }$\Pi$ when, for all $w,f,g\in\mathcal{F}$ with
	$g\overset{d}{=}f$,
	\begin{equation}
		f=-w-\Pi(-w)\implies w+f\succsim w+g\label{eq:pcp}%
	\end{equation}
	
\end{definition}

The interpretation is analogous to the original Definition
\ref{def:iciente}-(i), but without the quantification `for all $\pi\in\mathbb{R}%
$' previously mentioned. With this amended definition, the conclusion of Theorem \ref{th:wra}
continues to hold.

\begin{theorem}
	\label{th:wrap}Let $\Pi:\mathcal{F}\rightarrow\mathbb{R}$ be a\emph{ }premium
	calculation  principle. The following properties are equivalent for a risk preference:
	
	\begin{itemize}
		\item[(i)] weak risk aversion;
		
		\item[(ii)] propensity to full insurance;
		
		\item[(iii)] propensity to full insurance at price $\Pi$. 
	\end{itemize}
\end{theorem}

\textbf{Proof. }The implication (i) $\implies$ (ii) is part of Theorem
\ref{th:wra} and (ii) $\implies$ (iii) is immediate. As for (iii) $\implies$
(i) observe that for each $h\in\mathcal{F}$ there exist, by Lemma
\ref{lm:deus}, elements $z,z^{\prime}\in\mathcal{F}$ such that $z\overset
{d}{=}z^{\prime}$ and $h-\mathbb{E}\left[  h\right]  \overset{d}{=}%
z-z^{\prime}$. By (\ref{eq:pcp}), for
\[
\gamma=\frac{\Pi(-z)+\mathbb{E}[h]}{\theta}%
\]
we have that $\Pi(-z-\gamma)=-\mathbb{E}[h]$. Let $w=z+\gamma$ and $w^{\prime
}=z^{\prime}+\gamma$. Clearly, $w\overset{d}{=}w^{\prime}$, $h-\mathbb{E}%
\left[  h\right]  \overset{d}{=}w-w^{\prime}$, and $\mathbb{E}[h]=-\Pi
(-z-\gamma)=-\Pi(-w)$. Let
\[
f=-w+\mathbb{E}\left[  h\right]  =-w-\Pi(-w)\quad\text{and}\quad g=-w^{\prime
}+\mathbb{E}\left[  h\right]  =-w^{\prime}-\Pi(-w)
\]
Clearly, $f\overset{d}{=}g$. Propensity to full insurance at price $\Pi$,
yields $w+f\succsim w+g$, that is,%
\[
\mathbb{E}\left[  h\right]  =w+f\succsim w+g=w-w^{\prime}+\mathbb{E}\left[
h\right]  \overset{d}{=}h
\]
Law invariance of $\succsim$ guarantees that $w-w^{\prime}+\mathbb{E}\left[
h\right]  \sim h$, and so $\mathbb{E}\left[  h\right]  \succsim w-w^{\prime
}+\mathbb{E}\left[  h\right]  \sim h$. Transitivity then implies
$\mathbb{E}\left[  h\right]  \succsim h$, showing that (i) holds.\hfill
$\blacksquare\bigskip$

Also the definitions of propensity to proportional and deductible-limit
insurance can be weakened in the same manner, with the conclusions of Theorem
\ref{th:sra} still holding true. The details are omitted for brevity.

We close by observing that another common feature of most pricing rules for
insurance markets is \emph{law invariance},\footnote{Law invariance of $\Pi$ means that $h\overset{d}%
	{=}h^{\prime}$ implies $\Pi(h)=\Pi(h^{\prime})$.} with this (\ref{eq:pcp})
becomes
\begin{equation}
	w+\underset{\text{full insurance at its own price }}{\underbrace{\left(
			-w-\Pi(-w)\right)  }}\succsim w+\underset{\text{payoff }h\text{ at
			its own price}}{\underbrace{\left(  h-\Pi(h)\right)  }}\label{eq:wwra}%
\end{equation}
for all $h\overset{d}{=}-w$, because law invariance guarantees
$\Pi(h)=\Pi(-w)$.
%Indeed, if $g\overset{d}{=}f=-w-\Pi(-w)$, then $h=g+\Pi(-w)\overset{d}{=}-w$
%and by law invariance of $\Pi$ we have $\Pi(h)=\Pi(-w)$, thus (\ref{eq:wwra})
%implies%
%\[
%w+f\succsim w+\left(  h-\Pi(h)\right)  =w+\left(  h-\Pi(-w)\right)  =w+g
%\]
%Conversely, under (\ref{eq:pcp}), if $h\overset{d}{=}-w$, then $g=h-\Pi
%(h)\overset{d}{=}-w-\Pi(h)$, morever law invariance gives $\Pi(h)=\Pi(-w)$,
%thus $g\overset{d}{=}-w-\Pi(-w)$, finally setting $f=-w-\Pi(-w)$, we have%
%\[
%w+\left(  -w-\Pi(-w)\right)  =w+f\succsim w+g=w+\left(  h-\Pi(h)\right)
%\]
%
Note that (\ref{eq:wwra}) is analogous to condition (\ref{eq:wra}) in the
introduction, again without the universal quantification.
	
\subsection{Extension to $\mathcal{L}^{p}$ spaces and to $\mathcal{F}_{0}$ } \label{app:arente}

In the following proofs, continuity for risk preferences on $\mathcal{L}^{p}$
spaces is with respect to $p$-norm convergence, continuity for risk
preferences on the space
$\mathcal{F}_{0}$ of \emph{simple random payoffs} -- those that take, almost surely,
only finitely many values -- is with respect to bounded a.s.~convergence.

As discussed in the main text only the proofs of the results concerning
propensity to full insurance (Theorems \ref{th:wra} and \ref{th:comp-wra})
need to be modified by adding the assumption of continuity,  the ones
regarding propensity about partial insurance remain unchanged.\bigskip

\noindent\textbf{Proof of Theorem \ref{th:wra} for continuous risk preferences
	on }$\mathcal{F}_{0}$\textbf{ and }$\mathcal{L}^{p}$\textbf{, with }$p\in\left[
1,\infty\right)  $\textbf{. }

(i) $\implies$\ (ii). The proof is the same as that in Appendix \ref{app:wra}.

(ii) $\implies$ (i). Let $f\in\mathcal{F}_{0}$ (resp.~$\mathcal{L}^{p}$).
Choosing $ v$ and $\Sigma_{n}^{ v}$ as in Lemma \ref{lm:sti},
\begin{equation}
	f_{n}:=\mathbb{E}\left[  f\mid\Sigma_{n}^{ v}\right]  \rightarrow
	f\label{eq-Scon}%
\end{equation}
in bounded a.s.~convergence (resp.~in $\mathcal{L}^{p}$). It is obvious to see
that $f_{n}\in\mathcal{F}_{0}\subseteq\mathcal{L}^{p}$ for all $n\in
\mathbb{N}$. Theorem \ref{th:wra}, applied to the restriction of $\succsim$ to
$\mathcal{F}_{0}\left(  S,\Sigma_{n}^{ v},P_{\mid\Sigma_{n}^{ v}%
}\right)  =\mathcal{L}^{p}\left(  S,\Sigma_{n}^{ v},P_{\mid\Sigma
	_{n}^{ v}}\right)  =\mathcal{L}^{\infty}\left(  S,\Sigma_{n}^{ v
},P_{\mid\Sigma_{n}^{ v}}\right)  $, yields $\mathbb{E}[f_{n}]\succsim
f_{n}$ for all $n\in\mathbb{N}$. But $\mathbb{E}[f_{n}]\rightarrow
\mathbb{E}[f]$ and $f_{n}\rightarrow f$, and the continuity of $\succsim$
implies $\mathbb{E}[f]\succsim f$. Thus weak risk aversion holds.\hfill
$\blacksquare\bigskip$

\noindent\textbf{Proof of Theorem \ref{th:comp-wra} for continuous risk
	preferences on }$\mathcal{F}_{0}$\textbf{ and }$\mathcal{L}^{p}$\textbf{,
with }$p\in\left[  1,\infty\right)  $\textbf{. }

(i) $\implies$\ (ii). The proof is the same as that in Appendix
\ref{app:compa}.

(ii) $\implies$ (i). Let $f\in\mathcal{F}_{0}$ (resp.~$\mathcal{L}^{p}$) and
$\gamma\in\mathbb{R}$ be such that $\gamma\succsim_{\mathrm{A}}f$. We want to
show that $\gamma\succsim_{\mathrm{B}}f$. Define $\{f_{n}\}$ as in
\eqref{eq-Scon}. Note that $\gamma\sim_{\mathrm{A}}f_{n}-\rho_{\mathrm{A}%
}(\gamma,f_{n})$ for all $n\in\mathbb{N}$. Theorem \ref{th:comp-wra}, applied
to the restriction of $\succsim$ to $\mathcal{F}_{0}\left(  S,\Sigma
_{n}^{ v},P_{\mid\Sigma_{n}^{ v}}\right)  =\mathcal{L}^{p}\left(
S,\Sigma_{n}^{ v},P_{\mid\Sigma_{n}^{ v}}\right)  =\mathcal{L}%
^{\infty}\left(  S,\Sigma_{n}^{ v},P_{\mid\Sigma_{n}^{ v}}\right)
$, yields $\gamma\succsim_{\mathrm{B}}f_{n}-\rho_{\mathrm{A}}(\gamma,f_{n})$
for all $n\in\mathbb{N}$. But then
\[
f_{n}-\rho_{\mathrm{B}}(\gamma,f_{n})\sim_{\mathrm{B}}\gamma\succsim
_{\mathrm{B}}f_{n}-\rho_{\mathrm{A}}(\gamma,f_{n})
\]
and, together with transitivity, monotonicity implies that $\rho_{\mathrm{B}%
}(\gamma,f_{n})\leq\rho_{\mathrm{A}}(\gamma,f_{n})$ for all $n\in\mathbb{N}%
$.\footnote{Monotonicity is equivalent to
	\[
	f-\zeta\succsim f-\xi\iff\zeta\leq\xi
	\]
	whenever $f\in\mathcal{F}_{0}$ and $\zeta,\xi\in\mathbb{R}$.} Since $\succsim$
is continuous and $f_{n}\rightarrow f$ suitably, it follows, by Lemma
\ref{lem:strength},\footnote{Adjusted for the corresponding type of
	convergence of sequences of random payoffs.} that $\rho_{\mathrm{A}}%
(\gamma,f_{n})\rightarrow\rho_{\mathrm{A}}(\gamma,f)$ and $\rho_{\mathrm{B}%
}(\gamma,f_{n})\rightarrow\rho_{\mathrm{B}}(\gamma,f)$, and so $\rho
_{\mathrm{B}}(\gamma,f)\leq\rho_{\mathrm{A}}(\gamma,f)$. But
\[
f-\rho_{\mathrm{A}}(\gamma,f)\sim_{\mathrm{A}}\gamma\succsim_{\mathrm{A}}f=f-0
\]
and another application of transitivity and monotonicity yields $\rho
_{\mathrm{A}}(\gamma,f)\leq0$, and so $\rho_{\mathrm{B}}(\gamma,f)\leq
\rho_{\mathrm{A}}(\gamma,f)\leq0$. With this (monotonicity again)%
\[
\gamma\sim_{\mathrm{B}}f-\rho_{\mathrm{B}}(\gamma,f)\succsim_{\mathrm{B}}f-0=f
\]
and (transitivity again) $\gamma\succsim_{\mathrm{B}}f$, as desired.
\hfill$\blacksquare\bigskip$
}

\end{document}